# Easily testable logical networks based on a «widened long flip-flop»*

**Nick Stukach, Kyiv (Ukraine)**

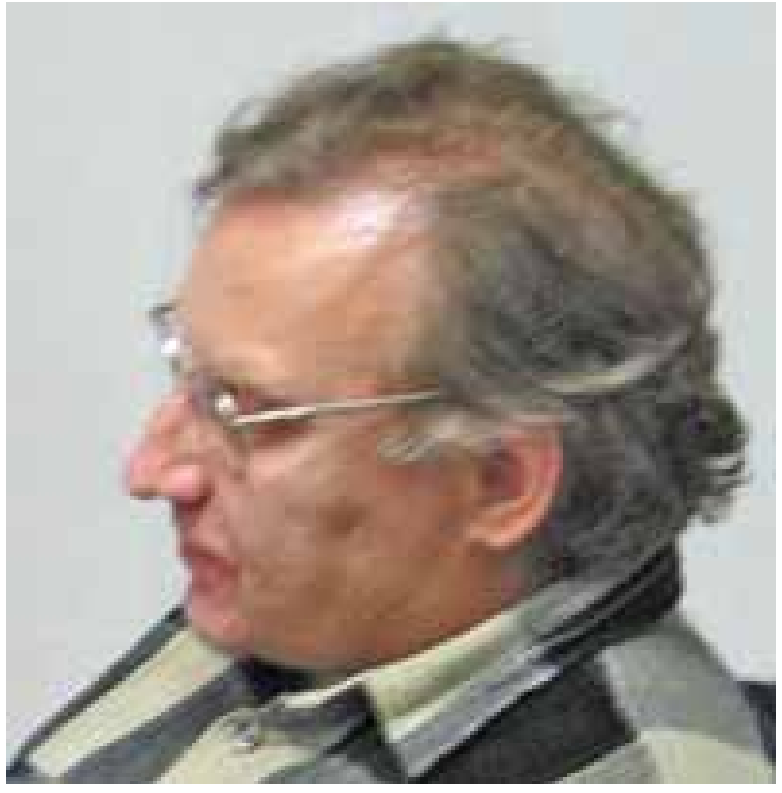

## Content

* See the hypertext variant of the article at http://stoo.co.ua

# Introduction

The article can be useful for solving problems arising for mission-critical digital systems.

The author's previous publications relating to the theme touched on are [1—11].

The author uses the model of multiple stuck-at-faults for logical networks. This model was defined more exactly with a view to be maximally strict and complete (see chapter I).

The beginning of the theory being set out in the article is caused by appearance of the «long flip-flop» which is nothing but a peculiar logical network constructed out of gates AND and OR. The author invented the «long flip-flop» (see chapter II) in 1980 when he investigated possibilities to make logical networks more suitable for being tested for defects. The «long flip-flop» is able, in particular, to implement the $N$-argument AND or the $N$-argument OR. But it is surprising that for being tested for multiple stuck-at-faults, the «long flip-flop» needs only two test vectors for informational inputs, whereas an ordinary analogue which implements $N$-argument OR or AND needs $N+1$. This fact forced the author to surmise that exceptionally easily testable logical networks can be constructed by using the «long flip-flop».

The surmise was successfully confirmed. To achieve this, the author invented the «long flip-flop» which has width (see chapter III), proved two theorems concerning testing of the «widened long flip-flop» (see chapter III and appendix) and created two methods to implement a finite-state automaton within the «widened long flip-flop»: the first method yields a network 1 which can be easily tested for unfitness (see chapter IV) whereas the second method yields a network 2 which can be easily tested for intactness (see chapter V).

Next properties define high suitability of the network 2 for being tested for multiple stuck-at-faults:

- Here principle of «the back-doors testing», when the main testing process is being performed through lateral inputs and outputs while neutral input vectors are being applied to non-lateral inputs, was successfully implemented. In our case such testing uses 22 beats, 18 lateral inputs and 10 lateral outputs. So size of the test is no only small but independent of the finite-state automaton complexity. It is amazing because principle «divide et impera» for the first time is denied: to simplify the task of testing of several logical devices a joining up of these devices (instead of dividing a device into several devices) becomes advantageous.
- There are only 2 neutral vectors to be applied to non-lateral inputs: 00...0 and 11...1. At least two vectors are necessary to examine such inputs for stuck-at-faults, so the minimal size of test as applied to non-lateral inputs is achieved.
- There is no need to observe values of non-lateral outputs.

«Back-doors testing» for multiple stuck-at-faults in case of network 1 is even simpler: 6 beats, 2 lateral outputs, 4 lateral inputs and just one neutral vector are used. But some stuck-at-faults are not being detected. On the other hand, every case, when bits of an output vector of an automaton implemented by network 1 are containing errors forced by these undetected faults, can be immediately discovered as network 1 is outputting cautionary one-bit signal. Several very important real cases when this is adoptable were investigated.

It is easy to see that both methods the author suggested provide successful implementing of principle of «back-doors testing».

It is very important from the point of view of improving the reliability of computer-aided control so widely used in mission-critical systems. As is well known, these systems always are characterised by too high price of misoperation.

By using of any of two mentioned method, the digital equipment, whatever its complexness, can be easily, quickly and amazingly exhaustively tested if necessary. This is what is important from the practical point of view.

As to the scientifical importance, the author hopes that he offered the non-bad decision which for the first time eliminates at once 3 main problems in the sphere of testing logic networks for stuck-at-faults:

- 1) «An exponential increasing of the complexity of the testing with the complexity of the equipment». To solve this problem, a testing in parts was always used, begetting the non-trivial problem of a reliability of the means which divide the equipment into parts during the test run;
- 2) «Who will test the watchman of watchmen?», i.e. the problem of the testing of the tester. This problem arises if the tester is complex. Till now, a tester for a complex equipment was complex always;
- 3) «Mutual masking of faults». This problem is usually ignored on the grounds that mutual masking of faults is unlikely. In reality, this problem is being ignored due to difficulties in decision.

One can easily see that the said 3 problems are absent for both network 1 and network 2.

# Chapter I. Model of multiple stuck-at-faults

To describe failures, a model denoted by $C^{\mathrm{mpl}}$ will be used. This model seems to be the strictest of all those models of multiple stuck-at-faults which the author knows.

In accordance with model $C^{\mathrm{mpl}}$, any failure of a logic network comes down to a multiple stuck-at-fault representing *an arbitrary nonempty subset of all possible elementary stuck-at-faults*.

Here an elementary stuck-at-fault represents a stuck-at-fault of a pole of a logic network.

As poles of a logic network, we consider logic inputs and logic outputs of the network, as well as logic inputs and logic outputs of logic gates of the network.

Each logic gate represents or gate AND, or gate OR, or gate NOT.

(As a rule, further we´ll omit the adjective «logic» when saying about networks, gates, inputs, outputs, poles, and other logic things.)

We have two definitions for elementary stuck-at-fault:

- For $x \in \{0, 1\}$ the fault « $\equiv x$» (alias «stuck-at-$x$») of a network´s pole representing an output of a network or an input of a gate is equivalent to tearing off the pole from the logic link leading towards the pole together with connecting the link´s start to a source of constant «$x$».
- For $x \in \{0, 1\}$ the fault « $\equiv x$» (alias «stuck-at-$x$») of a network´s pole representing an input of a network or the output of a gate is equivalent to tearing off the pole from each logical link leading from the pole together with connecting the link´s start to a source of constant «$x$».

One can easily see, that all cases when the value of the pole becomes unchangeable because of a damage of this pole or a damage of the rest of the network, are covered with these 2 definitions of elementary stuck-at-fault.

The network´s failure covered with model $C^{\mathrm{mpl}}$ should be considered as $C^{\mathrm{mpl}}$-fault.

If the network is intact or has $C^{\mathrm{mpl}}$-fault, then a network´s pole having no elementary stuck-at-fault should be considered as $C^{\mathrm{mpl}}$-intact.

**Remarks** relating to model $C^{\mathrm{mpl}}$:

- 1) each pole of the network may have only 3 alternative technical states:
  - having fault « $\equiv 0$»,
  - having fault « $\equiv 1$»,
  - being $C^{\mathrm{mpl}}$-intact;
- 2) some part of the network may be intact and at the same time not be $C^{\mathrm{mpl}}$-intact. *For example*, some gate $x$, that represents gate AND, is intact, however the value «0» is forever fixed on the output of gate $x$, because the output of some gate $z$ has fault « $\equiv 0$» and the output of gate $z$ is connected to some input $y$ of gate $x$; such case *should mean* that and the output of gate $z$, and input $y$ of gate $x$, and the output of gate $x$ have fault « $\equiv 0$»;
- 3) $C^{\mathrm{mpl}}$-intactness of some input $x$ of some gate $y$, that represents either gate AND or gate OR, means that there is such combination of logic constants on the rest of $C^{\mathrm{mpl}}$-intact inputs of gate $y$, that any logic step (that represents transition of a logic value into the opposite logic value) is able to go from input $x$ of gate $y$ towards the output of gate $y$;
- 4) if the output of some gate $y$ is $C^{\mathrm{mpl}}$-intact, then gate $y$ has at least one $C^{\mathrm{mpl}}$-intact input;
- 5) the fact that the output of some gate $y$ has a stuck-at-fault, doesn´t mean, that those outputs of gates of the network or those inputs of the network, which are connected with inputs of gate $y$, have a stuck-at-faults;

- 6) if the output of some gate $x$ is connected with some input $y$ of some gate $z$, then there are only 7 alternatives presented in Table I;
- 7) the amount of elementary stuck-at-faults which form a multiple stuck-at-fault is not limited, i.e. it is allowed even that *all* network poles have stuck-at-faults;
- 8) a break of any logic link is allowed only if it leads to a stuck-at-fault;
- 9) short-circuits between poles or links are not allowed.

An attribute (component, part, property, ...) belonging to some network $S$ should be referred to as $S$-attribute (-component, -part, -property, ...). For example: $S$-input, $S$-output, $S$-gate, $S$-link.

A time moment, at which there are no transients in some network $S$, will be referred to as the moment of $S$-halt.

It is evident that such a moment, which is known as a moment of $S$-halt when network $S$ is intact, remains the moment of $S$-halt when any $C^{mpl}$-fault appears in network $S$.

**Table I. Possible cases when the output of gate $x$ is connected with input $y$ of gate $z$**

| The state of the output of gate $x$ | The state of input $y$ of gate $z$ |
|---|---|
| $C^{mpl}$-intactness | Fault « $\equiv 1$» |
| | Fault « $\equiv 0$» |
| | $C^{mpl}$-intactness |
| Fault « $\equiv 0$» | Fault « $\equiv 0$» |
| | Fault « $\equiv 1$» |
| Fault « $\equiv 1$» | Fault « $\equiv 0$» |
| | Fault « $\equiv 1$» |

# Chapter II. «Long flip-flop»

A «canonical long flip-flop» of length $N$ is a logical network (Figure II-1) which is constructed out of two-input gates AND $1_1 \dots 1_{N+1}$ and two-input gates OR $2_1 \dots 2_N$, $3_1 \dots 3_N$, moreover for $i = \overline{1,\ N}$ the input 1 of gate $2_i$ is connected with the output of gate $1_i$, the output of gate $2_i$ is connected with input 1 of gate $1_{i+1}$, the input 1 of gate $3_i$ is connected with the output of gate $1_{i+1}$, the output of gate $1_{i+1}$ is connected with input 2 of gate $1_i$.

A «minimal long flip-flop» of length $N$ is a logical network constructed out of gates $1_1 \dots 1_{N+1}$, $2_1 \dots 2_N$, $3_1 \dots 3_N$, every of which is a gate AND or differs from a gate AND by that that at least one input is made an inverting input and/or by that that the output is made an inverting output, in addition the following conditions must be comlied with for $i = \overline{1,\ N}$:

- — the output of gate $1_i$ is made an inverting output, otherwise the input 1 of gate $2_i$ is made an inverting input;
- — the output of gate $2_i$ is made an inverting output, otherwise the input 2 of gate $1_{i+1}$ is made an inverting input;
- — the output of gate $1_{i+1}$ is made an inverting output, otherwise the input 1 of gate $3_i$ is made an inverting input;
- — the output of gate $3_i$ is made an inverting output, otherwise the input 2 of gate $1_i$ is made an inverting input;
- — the input 1 of gate $2_i$ is connected with the output of gate $1_i$, whereas the output of gate $2_i$ is connected with the input 1 of gate $1_{i+1}$;
- — the input 1 of gate $3_i$ is connected with the output of gate $1_{i+1}$, whereas the output of gate $3_i$ is connected with the input 2 of gate $1_i$.

If the «minimal long flip-flop» is to be constructed out of gates AND, OR and NOT, then a gate OR may replace any gate NOT-AND-NOT, each inverting input of any gate may be implemented by inserting an invertor ahead of this input, and the inverting output of any gate may be implemented by inserting an invertor after this output.

(We mean that a gate NOT-AND-NOT differs from a gate AND by that that each input is made an inverting input and by that that the output is made an inverting output. We also mean that an invertor represents a gate NOT.)

In this connexion, the «canonical long flip-flop» corresponds to the variant (Figure II-2) of the «minimal long flip-flop» after replacing gates NOT-AND-NOT by gates OR.

In general case the «long flip-flop» differs from the «minimal long flip-flop» by that that any gate may have more than two inputs every of which may be made an inverting input.

**THEOREM 1.** Let a «canonical long flip-flop» stand the test (Table II-1). Then this «canonical long flip-flop» doesn´t have stuck-at-faults, in addition $u_{1,1} = u_{1,2} = \dots = u_{1,N} = u_{4,1} = u_{4,2} = \dots = u_{4,N} = 0$ and $u_{2,1} = u_{2,2} = \dots = u_{2,N} = u_{3,1} = u_{3,2} = \dots = u_{3,N} = 1$.

The proof.

1. During interval $[t_1,\ t_2]$ logical step «$0 \rightarrow 1$» goes by the single possible path which passes through the input 1 and the output of each of gates $1_1 \dots 1_{N+1}$ and $2_1 \dots 2_N$. It means the following:

- a) the input 1 and the output of each of gates $1_1 \dots 1_{N+1}$ and $2_1 \dots 2_N$ don´t have any stuck-at-faults,
- b) the outputs 2 of gates $2_1 \dots 2_N$ don´t have faults «$\equiv 1$»,
- c) those inputs 2 of gates $2_1 \dots 2_N$ which don´t have faults «$\equiv 0$» have the value «0» during whole interval $[t_1,\ t_2]$.

2. During interval $[t_3,\ t_4]$ logical step «$0 \to 1$» goes by the single possible path which passes through the input 2 and the output of each of gates $\mathit{1}_1 \ldots \mathit{1}_{N+1}$ and through the input 1 and the output of each of gates $\mathit{3}_1 \ldots \mathit{3}_N$. It means the following:

- a) the input 2 and the output of each of gates $\mathit{1}_1 \ldots \mathit{1}_{N+1}$, as well as the input 1 and the output of each of gates $\mathit{3}_1 \ldots \mathit{3}_N$, don´t have any stuck-at-faults,
- b) the inputs 2 of gates $\mathit{3}_1 \ldots \mathit{3}_N$ don´t have faults « $\equiv 1$»,
- c) those inputs 2 of gates $\mathit{3}_1 \ldots \mathit{3}_N$ which don´t have faults « $\equiv 0$» have the value «0» during whole interval $[t_3,\ t_4]$.

3. Due to items 1 and 2, the «canonical long flip-flop» don´t have faults except, maybe, faults « $\equiv 0$» of inputs 2 of gates $\mathit{2}_1 \ldots \mathit{2}_N$ and $\mathit{3}_1 \ldots \mathit{3}_N$. But we see the following:

- a) let the input 2 of some gate $\mathit{2}_x$ have fault « $\equiv 0$» or the value $u_{2,\,x} = 0$. But then the loop, composed of gates $\mathit{1}_x$, $\mathit{2}_x$, $\mathit{3}_x$, and $\mathit{1}_{x+1}$, to moment $t_1$ must enter such an internally-stable-state 0 that outputs of the gates have the value 0. And this state 0 remains during whole interval $[t_1,\ t_2]$ in contradiction (!) to item 1;
- b) let the input 2 of some gate $\mathit{3}_y$ have fault « $\equiv 0$» or the value $u_{3,\,y} = 0$. But then the loop, composed of gates $\mathit{1}_y$, $\mathit{2}_y$, $\mathit{3}_y$ and $\mathit{1}_{y+1}$, to moment $t_3$ must enter such an internally-stable-state 0 that outputs of the gates have the value 0. And this state 0 remains during whole interval $[t_3,\ t_4]$ in contradiction (!) to item 2.

4. The «canonical long flip-flop» is intact owing to items 1, 2, 3*a*, and 3*b*. Therefore items 1*c* and 2*c* entail that $u_{1,\,1} = u_{1,\,2} = \ \ldots = u_{1,\,N} = u_{4,\,1} = u_{4,\,2} = \ \ldots = u_{4,\,N} = 0$, whereas items 3*a* and 3*b* entail that $u_{2,\,1} = u_{2,\,2} = \ \ldots = u_{2,\,N} = u_{3,\,1} = u_{3,\,2} = \ \ldots = u_{3,\,N} = 1$.

The proof is over.

*A remark to **Theorem 1***. If $\exists\ x \in \overline{1,\ N} \mid (u_{1,\,x} \neq 0) \vee (u_{2,\,x} \neq 1) \vee (u_{3,\,x} \neq 1) \vee (u_{4,\,x} \neq 0)$, then the test (Table II-1) would fail, no matter whether the «canonical long flip-flop» has stuck-at-faults. Thus if the «canonical long flip-flop» has stood the test (Table II-1), then as a matter of fact the test (Table II-2) was successfully executed. Or in other words, $u_{1,\,1} = u_{1,\,2} = \ \ldots = u_{1,\,N} = u_{4,\,1} = u_{4,\,2} = \ \ldots = u_{4,\,N} = 0$ and $u_{2,\,1} = u_{2,\,2} = \ \ldots = u_{2,\,N} = u_{3,\,1} = u_{3,\,2} = \ \ldots = u_{3,\,N} = 1$ are right test stimuli, whose deviations (i.e. errors) are being fully detected during performing the test.

The «long flip-flop» has many features which are used below, but the following two ones are most amazing:

- 1) it is possible to use the «long flip-flop» (Figure II-3) for comparing $2N$ logical variables with the pattern in accordance with Theorem 1. I.e., to compare variables $x_1 \ldots x_{2N}$ with «0» during some time interval we should regard this interval as interval $[t_1,\ t_2]$ of Tables II-1 and II-2, whereas for comparing them with «1» we should regard this interval as interval $[t_3,\ t_4]$ of Tables II-1 and II-2. If the pattern alternates its values, then the «long flip-flop» executes useful function and in the same time is being tested for stuck-at-faults;
- 2) it is possible to use «long flip-flop» (Figure II-4) for implementing $N$–argument AND, and «long flip-flop» (Figure II-5) — for implementing $N$–argument OR.

It is important that traditional analogues in all three mentioned cases of using the «long flip-flop» (Figures II-3, II-4, II-5) are being characterized by that that complexity of testing for stuck-at-faults fast increases with $m$ (where $m$ — amount of arguments of the implemented logical function):

- — the test of the $m$-argument AND includes $m+1$ input vectors because consists of subtests «all bits of the input vector are equal to 1» and «all bits of the input vector by turns are equal to 0»,
- — the test of the $m$-argument OR includes $m+1$ input vectors because consists of subtests «all bits of the input vector are equal to 0» and «all bits of the input vector by turns are equal to 1»,
- — the test of the $m$-argument comparator (Figure II-6) includes $2m+2$ input vectors because consists of

subtests «all bits of the input vector are equal to 1», «all bits of the input vector by turns are equal to 0», «all bits of the input vector are equal to 0», and «all bits of the input vector by turns are equal to 1».

Furthermore, traditional analogues (unlike the «long flip-flop») require absence of errors in input vectors of test because any such error leads to erroneous testing results.

The fact that testing of any «long flip-flop» of those presented on Figures II-3, II-4, II-5 requires only two input vectors to be applied towards informational inputs, and in addition all errors of these vectors are being caught, once has very surprised the author and caused him to suppose that the «long flip-flop» can be used for constructing extremely easily testable logical devices.

## Table II-1. The conditional test of the «canonical long flip-flop»

| The inputs and the outputs of the «canonical long flip-flop» | Moments $t_1 \ldots t_4$ of halts of the «canonical long flip-flop», where $t_1 < t_2$ and $t_3 < t_4$. The time intervals which are constrained by the said moments | | | | | |
|---|---|---|---|---|---|---|
| | $t_1$ | $[t_1, t_2]$ | $t_2$ | $t_3$ | $[t_3, t_4]$ | $t_4$ |
| The input 2 of gate $2_i$ for $i = \overline{1, N}$ | $u_{1,i}$ | | | $u_{2,i}$ | | |
| The input 2 of gate $3_i$ for $i = \overline{1, N}$ | $u_{3,i}$ | | | $u_{4,i}$ | | |
| The input 1 of gate $1_1$ | 0 | | 1 | 1 | | |
| The output of gate $1_{N+1}$ | 0 | | 1 | | | |
| The input 2 of gate $1_{N+1}$ | 1 | | | 0 | | 1 |
| The output of gate $1_1$ | | | | 0 | | 1 |

## Table II-2. The test of the «canonical long flip-flop»

| The inputs and the outputs of the «canonical long flip-flop» | Moments $t_1 \ldots t_4$ of halts of the «canonical long flip-flop», where $t_1 < t_2$ and $t_3 < t_4$. The time intervals which are constrained by the said moments | | | | | |
|---|---|---|---|---|---|---|
| | $t_1$ | $[t_1, t_2]$ | $t_2$ | $t_3$ | $[t_3, t_4]$ | $t_4$ |
| The inputs 2 of gates $2_1 \ldots 2_N$ | 0 | | | 1 | | |
| The inputs 2 of gates $3_1 \ldots 3_N$ | 1 | | | 0 | | |
| The input 1 of gate $1_1$ | 0 | | 1 | 1 | | |
| The output of gate $1_{N+1}$ | 0 | | 1 | | | |
| The input 2 of gate $1_{N+1}$ | 1 | | | 0 | | 1 |
| The output of gate $1_1$ | | | | 0 | | 1 |

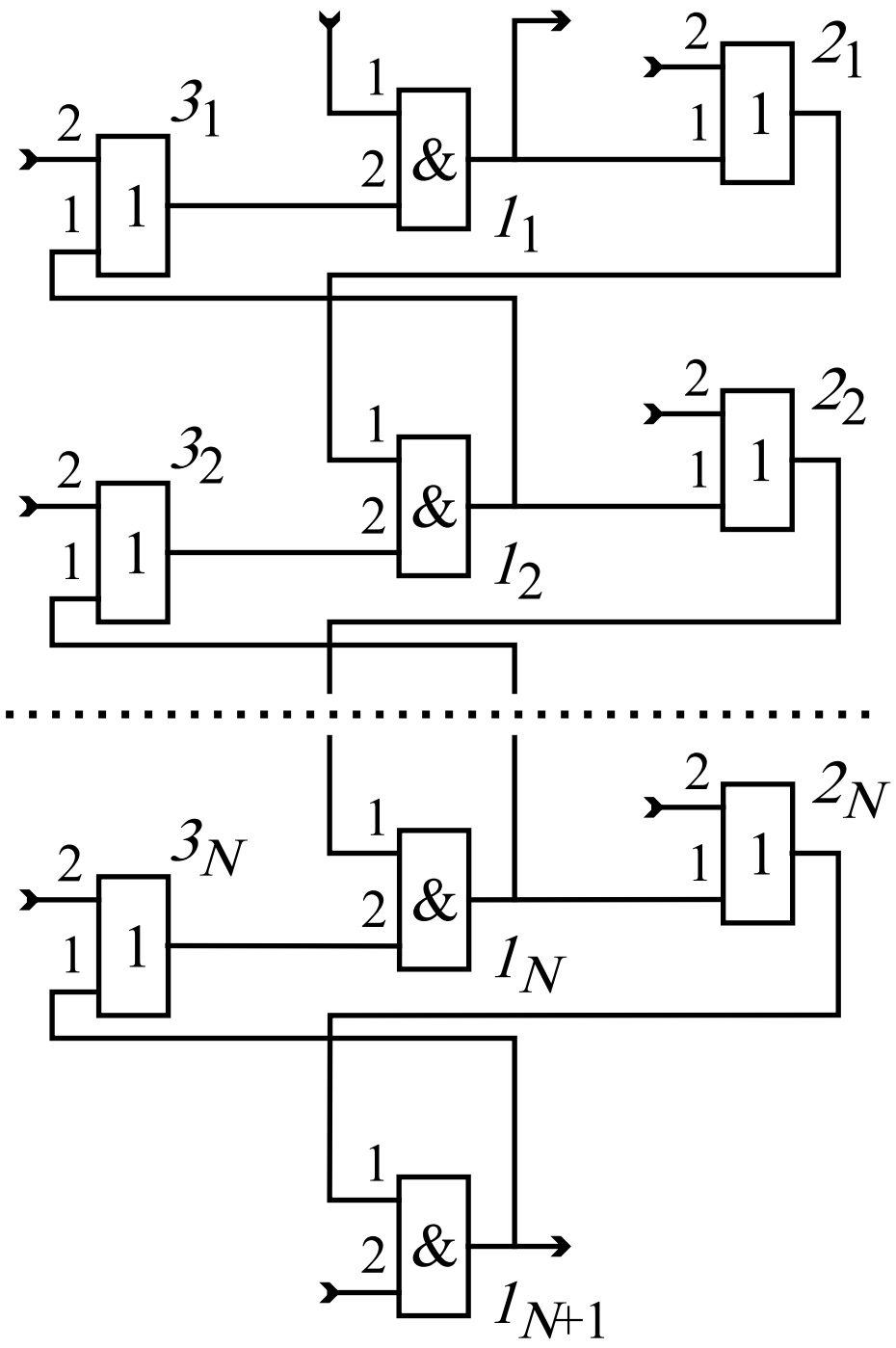

Figure II-1. The «canonical long flip-flop» of length $N$.

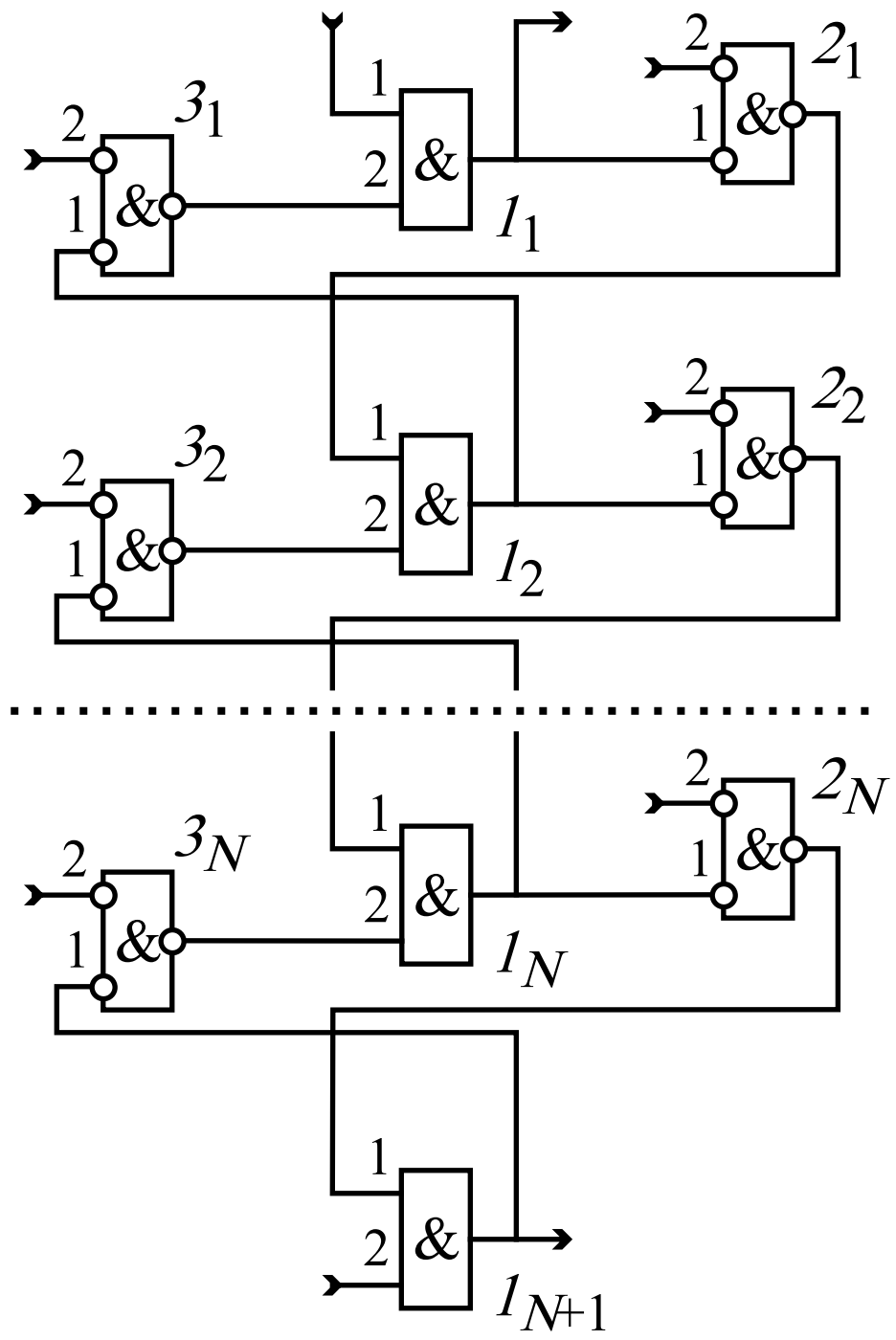

Figure II-2. The variant of the «minimal long flip-flop» which conforms to the «canonical long flip-flop» presented on Figure II-1.

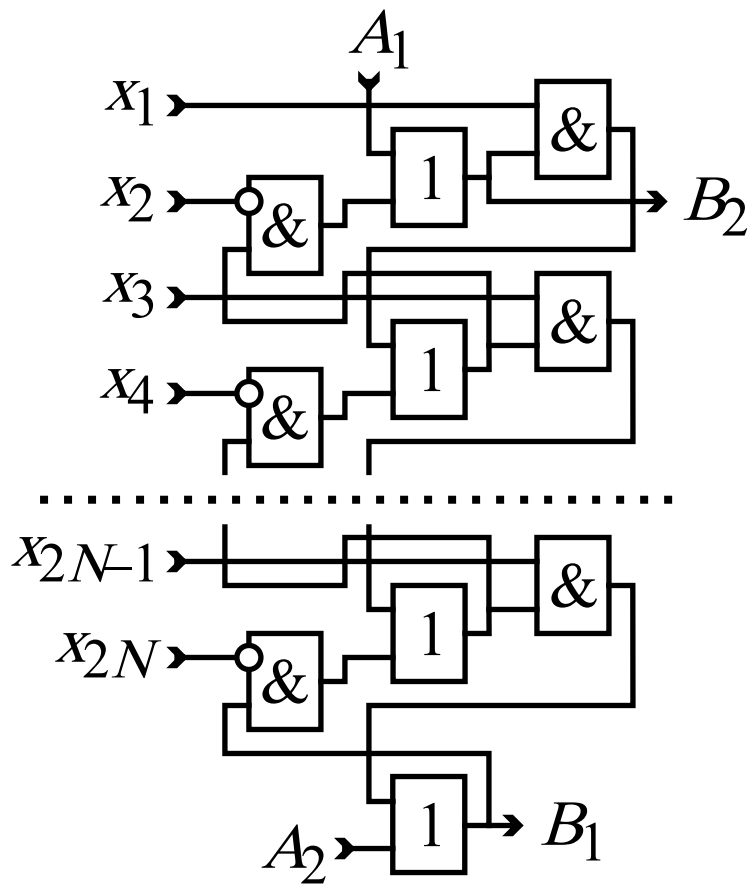

Figure II-3. The «long flip-flop» which implements the function of comparing logical variables $x_1, x_2, \ldots, x_{2N}$ with the pattern.

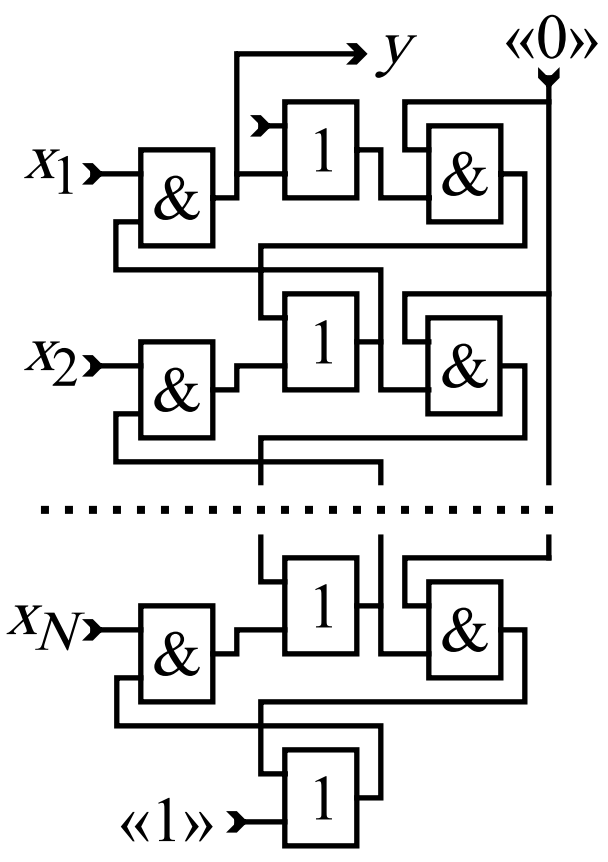

Figure II-4. The «long flip-flop» which implements the function $y = x_1 \ \& \ x_2 \ \& \ \ldots \ \& \ x_N$.

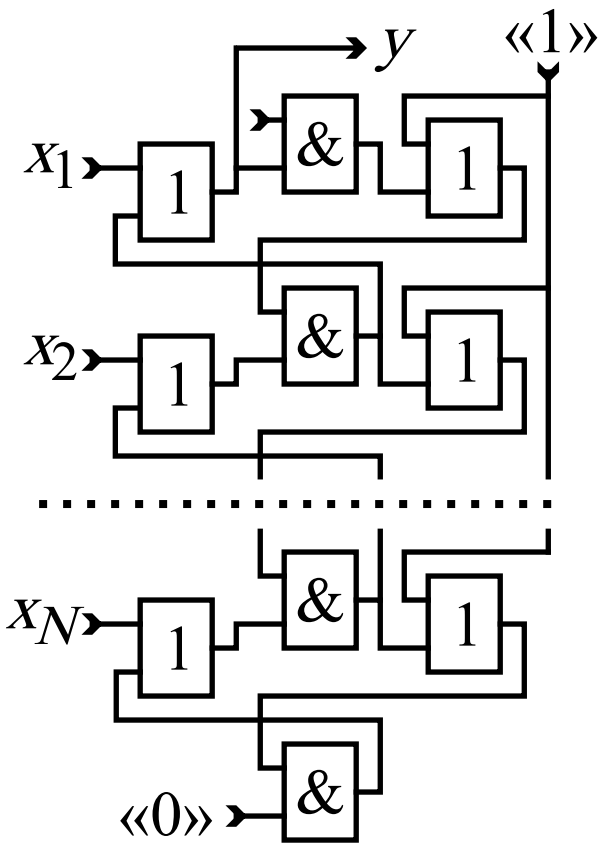

Figure II-5. The «long flip-flop» which implements the function $y = x_1 \vee x_2 \vee \ldots \vee x_N$.

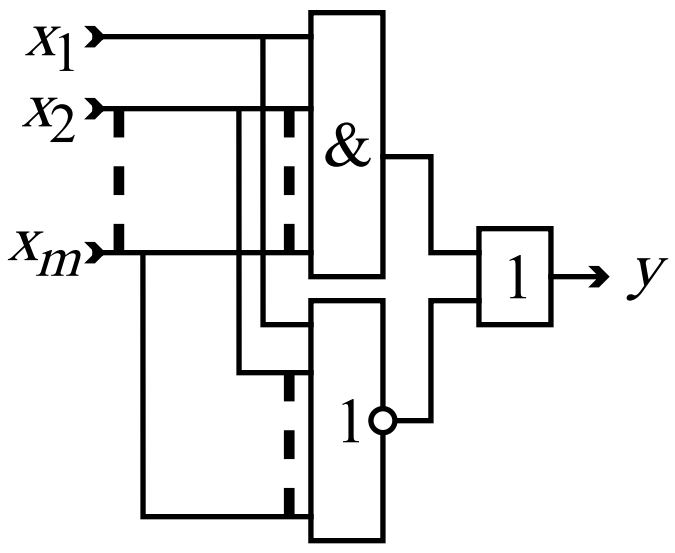

Figure II-6. The ordinary logic network which implements the $m$-argument function of comparing logical variables $x_1, x_2, \ldots, x_{m-1}$ with the pattern which is represented by variable $x_m$.

# Chapter III. «Widened long flip-flop»

Let´s view a logic network $\Delta$ presented on Figure III-1.

To avoid possible mistakes, we give exhaustive textual description of network $\Delta$:

«Network $\Delta$ consists of $\Delta$-gates, $\Delta$-inputs, $\Delta$-outputs and $\Delta$-links, in addition:

- 1. $\Delta$-gates divide into 5 groups: $\Delta$-gates $1_1 \ldots 1_{N+1}$ (the group $1$), $\Delta$-gates $2_1 \ldots 2_N$ (the group $2$), $\Delta$-gates $3_1 \ldots 3_N$ (the group $3$), $\Delta$-gates $4_1 \ldots 4_N$ (the group $4$) and $\Delta$-gates $5_1 \ldots 5_N$ (the group $5$). $\Delta$-gates $1_1 \ldots 1_{N+1}$, $4_1 \ldots 4_N$ and $5_1 \ldots 5_N$ are direct-current logic gates AND, and $\Delta$-gates $2_1 \ldots 2_N$ and $3_1 \ldots 3_N$ are direct-current logic gates OR. Each of $\Delta$-gates $1_1 \ldots 1_{N+1}$ has at least 2 inputs. $\Delta$-gates $2_1 \ldots 2_N$ and $3_1 \ldots 3_N$ are two-input. Each of $\Delta$-gates $4_1 \ldots 4_N$ and $5_1 \ldots 5_N$ has at least 1 input.
- 2. A set of $\Delta$-inputs consists of non-lateral $\Delta$-inputs $6_1 \ldots 6_K$ and of lateral $\Delta$-inputs *9*, *10*, *11*, and *12*.
- 3. A set of $\Delta$-inputs consists of non-lateral $\Delta$-inputs $7_1 \ldots 7_N$ and $8_1 \ldots 8_N$, as well as of lateral $\Delta$-inputs *13* and *14*.
- 4. Only the following $\Delta$-links (for $i = \overline{1,\ N}$) exist:
  - — $\Delta$-gate $2_i$ is connected by own input 1 with the output of $\Delta$-gate $1_i$, by own input 2 — with the output of $\Delta$-gate $4_i$, and by own output — with $\Delta$-output $7_i$, which in turn is connected with input 1 of $\Delta$-gate $1_{i+1}$,
  - — $\Delta$-gate $3_i$ is connected by own input 1 with the output $\Delta$-gate $1_{i+1}$, by own input 2 — with the output of $\Delta$-gate $5_i$, and by own output — with $\Delta$-output $8_i$, which in turn is connected with input 2 of $\Delta$-gate $1_i$,
  - — each input of $\Delta$-gates $1_1 \ldots 1_{N+1}$, except inputs 1 and 2, is connected with one of $\Delta$-inputs $6_1 \ldots 6_K$,
  - — the input 1 of each of $\Delta$-gates $4_1 \ldots 4_N$ is connected with $\Delta$-input *11*,
  - — the input 1 of each of $\Delta$-gates $5_1 \ldots 5_N$ is connected with $\Delta$-input *12*,
  - — each input of $\Delta$-gate $4_i$, except input 1, is connected or with one of $\Delta$-inputs $6_1 \ldots 6_K$, or with the output of one of $\Delta$-gates $3_{i+1} \ldots 3_N$,
  - — each input of $\Delta$-gate $5_i$, except input 1, is connected or with one of $\Delta$-inputs $6_1 \ldots 6_K$, or with the output of one of $\Delta$-gates $2_1 \ldots 2_{i-1}$,
  - — $\Delta$-gate $1_N$ is connected by own input 1 with $\Delta$-input *9*, as well as by own output — with $\Delta$-output *13*,
  - — $\Delta$-gate $1_{N+1}$ is connected by own input 2 with $\Delta$-input *10*, as well as by the output — with $\Delta$-output *14*.
- 5. We mean that the $\Delta$-outputs $7_1 \ldots 7_N$ and $8_1 \ldots 8_N$ represent only points on logical links. It means, that for each output $x \in \{7_1,\ 7_2,\ \ldots,\ 7_N,\ 8_1,\ 8_2,\ \ldots,\ 8_N\}$ we suppose only 2 alternatives:
  - — when testing network $\Delta$, we don´t need to test a path of delivery of signals from output $x$ towards a consumer, as we regard testing this path as a problem of the consumer;
  - — we must test a path of delivery of signals from output $x$ towards a consumer, in this connexion, we regard output $x$ as a gap and mean that a signal goes from the gap´s start towards the consumer and during known time comes back from the consumer, not being subjected to any logic transformation (except a case when the consumer has a failure and therefore always returns the same logic value).

- 6. There is some set $U_{\Delta}$ consisting of:
  - — $\Delta$-links connecting inputs of $\Delta$-gates $1_1 \ldots 1_{N+1}$ to $\Delta$-inputs $6_1 \ldots 6_K$,
  - — $\Delta$-links connecting inputs of $\Delta$-gates $4_1 \ldots 4_N$ and $5_1 \ldots 5_N$ to $\Delta$-inputs $6_1 \ldots 6_K$,
  - — $\Delta$-links connecting inputs of $\Delta$-gate $4_1 \ldots 4_N$ and $5_1 \ldots 5_N$ to outputs of $\Delta$-gates $2_1 \ldots 2_N$ and $3_1 \ldots 3_N$.»

We´ll name described network $\Delta$ the **«widened long flip-flop»**, characterized by length $N$, by width $K$ and by allowable internal links´ set $U_{\Delta}$.

* * *

One can easily see that the gates $1_1 \ldots 1_{N+1}$, $2_1 \ldots 2_N$ and $3_1 \ldots 3_N$ along with their interconnections form the «long flip-flop», which is transformed into the «widened long flip-flop» at the expense of adding of gates $4_1 \ldots 4_N$, $5_1 \ldots 5_N$ and of some interconnections.

The generalized chart of such a «widened long flip-flop» which is characterized by the length 7 and by the width 11 is presented on Figure III-2. The transition from the generalized chart towards the chart of a real network requires, that instead of all allowable links only real allowable links be presented.

* * *

**THEOREM 2.** Let following conditions be met:

- 1°) network $\Delta$ either has no failures or has a failure which is covered with model $C^{mpl}$,
- 2°) network $\Delta$ has no transients at moments $t_1 \ldots t_6$, where $t_1 < t_2 < t_3$ and $t_4 < t_5 < t_6$;
- 3°) $\Delta$-inputs $6_1 \ldots 6_K$ keep constant values during intervals $[t_1,\ t_3]$ and $[t_4,\ t_6]$,
- 4°) $\Delta$-inputs $11$ and $12$ keep constant values during intervals $[t_1,\ t_3]$ and $[t_4,\ t_6]$,
- 5°) the value of $\Delta$-input $9$ is constant during interval $[t_4,\ t_6]$ and can change just once within each of intervals $[t_1,\ t_2]$ and $[t_2,\ t_3]$,
- 6°) the value of $\Delta$-input $10$ is constant during interval $[t_1,\ t_3]$ and can change just once within each of intervals $[t_4,\ t_5]$ and $[t_5,\ t_6]$,
- 7°) the value of $\Delta$-output $14$ during interval $[t_1,\ t_3]$ conforms with Table III-1,
- 8°) the value of $\Delta$-output $13$ during interval $[t_4,\ t_6]$ conforms with Table III-1.

Then

- 1°°) network $\Delta$ either has no failure or has a failure representing an arbitrary subset of following elementary stuck-at-faults:
  - — the fault «$\equiv 1$» of some input $x \in \{6_1,\ 6_2,\ \ldots,\ 6_K\}$,
  - — the fault «$\equiv 0$» of some input $x \in \{6_1,\ 6_2,\ \ldots,\ 6_K\}$, if there is no such $y \in \{1_1,\ 1_2,\ \ldots,\ N+1,\ 4_1,\ 4_2,\ \ldots,\ 4_N,\ 5_1,\ 5_2,\ \ldots,\ 5_N\}$ that the input $x$ is connected with $C^{mpl}$-intact input of gate $y$,
  - — the fault «$\equiv 1$» of some input, except input 1, of some gate $x \in \{4_1,\ 4_2,\ \ldots,\ 4_N,\ 5_1,\ 5_2,\ \ldots,\ 5_N\}$,
  - — the fault «$\equiv 1$» of some input, except inputs 1 and 2, of some gate $x \in \{1_1,\ 1_2,\ \ldots,\ 1_{N+1}\}$,
- 2°°) the values of $\Delta$-inputs $9$, $10$, $11$, and $12$ conform with Table III-1,

- 3°°) the value of each $C^{mpl}$-intact input $x \in \{6_1, 6_2, ..., 6_K\}$ conforms with Table III-1 in case the input $x$ is connected with some $C^{mpl}$-intact input of some gate $x \in \{1_1, 1_2, ..., 1_{N+1}, 4_1, 4_2, ..., 4_N, 5_1, 5_2, ..., 5_N\}$.

The proof of Theorem 2 is presented in the Appendix.

Theorem 2 recognize that network $\Delta$ has the certain level of intactness, in case certain responses appear on $\Delta$-outputs *13* and *14*. But what if these responses appear never? In this case Theorem 2 would be useless, as its hypothesis are never carried out. As an insurance against this trouble, there is

**THEOREM 3.** If items 1°°—3°° of Theorem 2 are true, then items 7° and 8° of Theorem 2 are also true.

The proof of Theorem 3 is presented in the Appendix.

## An interpretation of theorems 2 and 3:

- 1. If network $\Delta$ stands test $T_\Delta$ (Table III-1), then only 3 alternatives are possible:
  - a) network $\Delta$ is intact,
  - b) network $\Delta$ has a failure not covered with model $C^{mpl}$,
  - c) network $\Delta$ has a fault effecting only asymmetric errors «1 instead of 0» on $\Delta$-outputs $7_1 ... 7_N$ and $8_1 ... 8_N$.
- 2. If network $\Delta$ stands the test $T_\Delta^{corr}$ (Table III-2), where $u_{1,1} ... u_{1,K}$ and $u_2 ... u_{13}$ are logic constants, then only 2 alternatives are possible:
  - a) network $\Delta$ has a failure not covered with model $C^{mpl}$,
  - b) the test $T_\Delta^{corr}$ (Table III-2) coincides with test $T_\Delta$ (Table III-1), because $u_{1,1} = u_{1,2} = ... = u_{1,K} = u_3 = u_4 = u_6 = u_8 = u_9 = u_{10} = u_{11} = u_{13} = 1$ and $u_2 = u_5 = u_7 = u_{12} = 0$.

Item 2*b* has the easy proof that follows from process of proving Theorem 2.

As a matter of fact, the test $T_\Delta^{corr}$ (Table III-2) serves as an attempt of performing of test $T_\Delta$ (Table III-1) in case the guarantee, that stimuli of test $T_\Delta$ (Table III-1) are free of errors, is absent. The success of the attempt (i.e. observing proper responses) means errors are absent.

## Table III-1. Test $T_{\Delta}$

| $\Delta$-inputs and $\Delta$-outputs | Moments $t_1 \ldots t_6$ of $\Delta$-halts, where $t_1 < t_2 < t_3$ and $t_4 < t_5 < t_6$. The time intervals which are constrained by the said moments | | | | | | | | | |
|---|---|---|---|---|---|---|---|---|---|---|
| | $t_1$ | $[t_1, t_2]$ | $t_2$ | $[t_2, t_3]$ | $t_3$ | $t_4$ | $[t_4, t_5]$ | $t_5$ | $[t_5, t_6]$ | $t_6$ |
| $б_1 \ldots б_K$ | 1 | | | | | | | | | |
| *11* | 0 | | | | | 1 | | | | |
| *12* | 1 | | | | | 0 | | | | |
| *9* | 1 | | 0 | | 1 | 1 | | | | |
| *10* | 1 | | | | | 1 | | 0 | | 1 |
| *13* | | | | | | 1 | | 0 | | 1 |
| *14* | 1 | | 0 | | 1 | | | | | |

## Table III-2. Test $T_{\Delta}^{\mathbf{corr}}$

| $\Delta$-inputs and $\Delta$-outputs | Moments $t_1 \ldots t_6$ of $\Delta$-halts, where $t_1 < t_2 < t_3$ and $t_4 < t_5 < t_6$. The time intervals which are constrained by the said moments | | | | | | | | | |
|---|---|---|---|---|---|---|---|---|---|---|
| | $t_1$ | $[t_1, t_2]$ | $t_2$ | $[t_2, t_3]$ | $t_3$ | $t_4$ | $[t_4, t_5]$ | $t_5$ | $[t_5, t_6]$ | $t_6$ |
| $б_i$ for $i = \overline{1, K}$ | $u_{1,i}$ | | | | | | | | | |
| *11* | $u_2$ | | | | | $u_3$ | | | | |
| *12* | $u_4$ | | | | | $u_5$ | | | | |
| *9* | $u_6$ | | $u_7$ | | $u_8$ | $u_9$ | | | | |
| *10* | $u_{10}$ | | | | | $u_{11}$ | | $u_{12}$ | | $u_{13}$ |
| *13* | | | | | | 1 | | 0 | | 1 |
| *14* | 1 | | 0 | | 1 | | | | | |

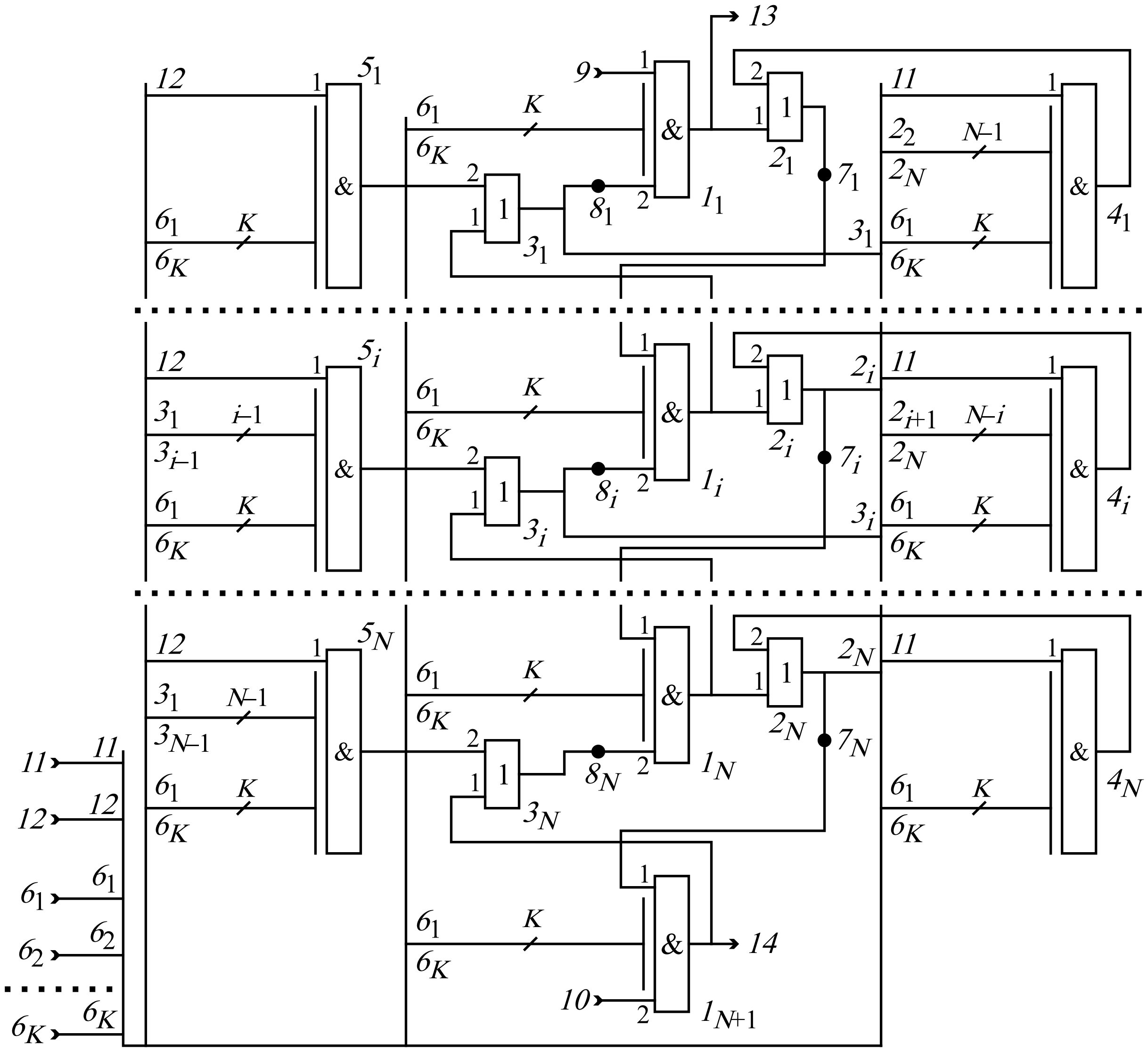

Figure III-1. The generalized structure of the «widened long flip-flop» of length $N$ and of width $K$. Allowability for some links to lead towards inputs of some gate AND we showed in that way that these links lead towards a shelf added to the said gate.

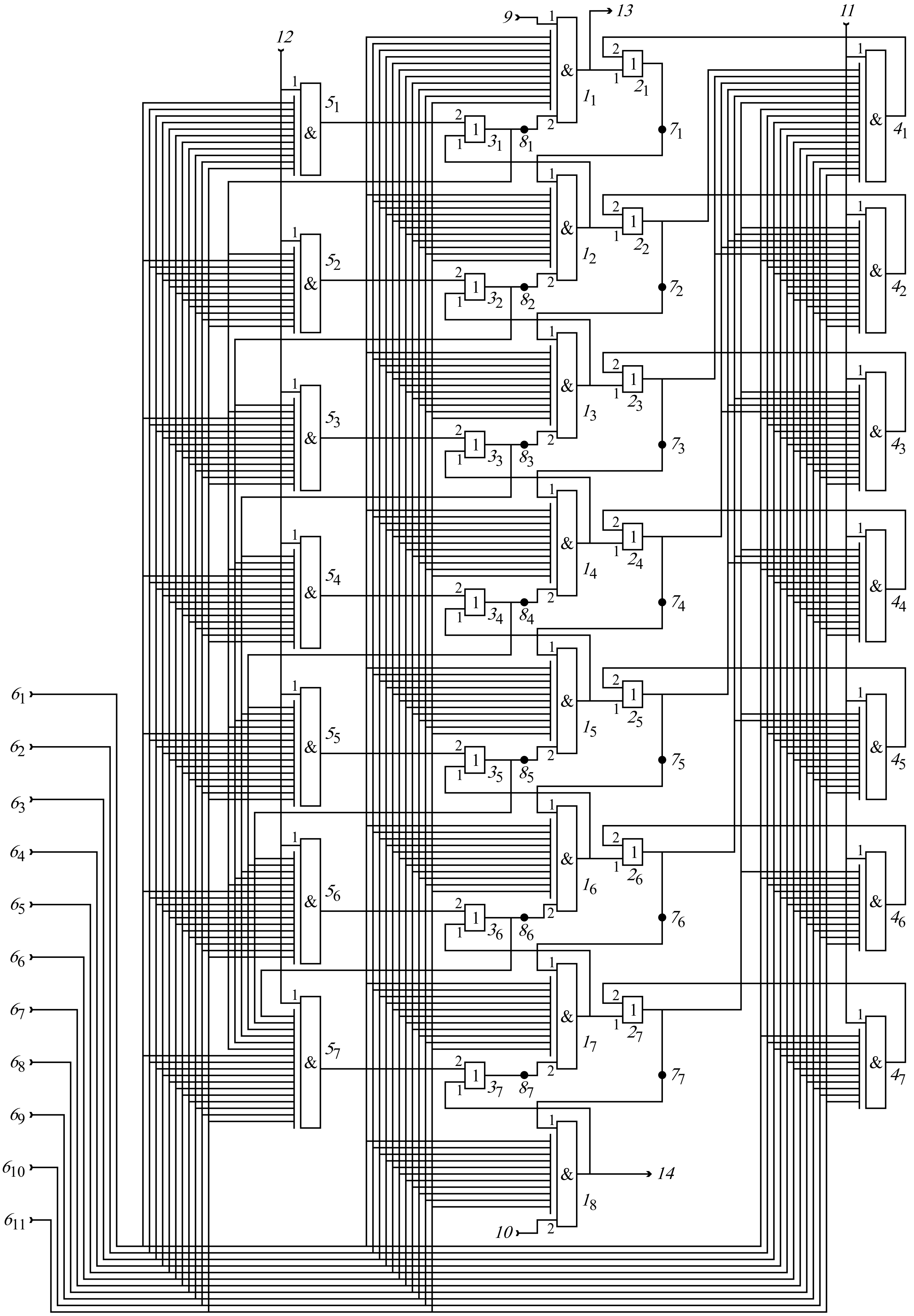

Figure III-2. The generalized structure of the «widened long flip-flop» of length 7 and of wideness 11. Allowability for some links to lead towards inputs of some gate AND we showed in that way that these links lead towards a shelf added to the said gate.

# Chapter IV. Finite-state automata being easily tested for unfitness

In this chapter we´ll offer Method IV for synthesis of an arbitrary abstract finite-state automaton $A$ within the «widened long flip-flop» $\Delta$.

As is well known, an abstract finite-state automaton is an device having input and output channels and acting in discrete time moments which are referred to as beats.

During current beat, the automaton is staying in some internal state $q \in Q$, some symbol $x$ of an input alphabet $X$ is being received through the input channel, and some symbol $y$ of an output alphabet $Y$ is being sent into the output channel.

Symbol $y$ according to law $y = \lambda(q, x)$ depends on symbol $x$ and on state $q \in Q$. An internal state $q' \in Q$ of the automaton for the next beat is being defined by state $q$ and by symbol $x$ according to law $q' = \delta(q, x)$. Herein $\delta : Q \times X \to Q$ — a transition function, and $\lambda : Q \times X \to Y$ — an output function. Alphabets $X$, $\Delta$ and $Q$ ought to be finite. At the primary moment the automaton ought to be in known state $q_0 \in Q$.

Thus, automaton $A$ is defined by the five-element collection $< X, Y, Q, \lambda, \delta >$.

According to Method IV, automaton $A$ must be presented as the combination of automata $A'$ and $A''$ (Figure IV-1). Automaton $A'$ is being defined by the same five-element collection $< X, Y, Q, \lambda, \delta >$, like automaton $A$ is. Logical vectors, which are forming a code $\left\{ < \vec{\alpha}, \vec{\beta} >_i \right\}_i$ able to detect asymmetric errors, are being used as symbols of input alphabet of automaton $A'$. A vector $< \vec{\alpha}, \vec{\beta} >$ consists of informational bits forming subvector $\vec{\alpha}$ and of check bits forming subvector $\vec{\beta}$. An automaton $A''$ reduces vector $< \vec{\alpha}, \vec{\beta} >$ up to bit $\gamma$ that equals 1 as soon as there is at least one error in the vector. A vector $< \vec{\alpha}, \gamma >$ codes a symbol of the output alphabet of automaton $A$. As is easy to see, the code $\left\{ < \vec{\alpha}, \gamma >_i \right\}_i$ is characterised by minimal redundancy as each vector of the code has just one check bit.

As symbols of the input alphabet of automaton $A$ (like of automaton $A'$), logical vectors forming a two-rail code are used. As is well known, there is a check bit for each informational bit in a vector of *a two-rail code* and both bits have contrary values, which are either 0 and 1, or 1 and 0.

As a code detecting asymmetric errors, a two-rail code mentioned, Berger code or an equilibrium code can be used.

*An equilibrium code* is also referred to as code «$m$ of $n$». A vector of this code has $n$ bits of which $m$ bits are equal to 1 whereas the others are equal to 0. Choozing different combinations of $n$ and $m$ gives different variants of an equilibrium code.

In *Berger code*, check bits taken together contain the amount of informational bits equal to 1.

According to Method IV, in the beginning a functional prototype of network $\Delta$ (that will be referred to as *$\Delta$-prototype*) should be created. Any known manner should be used for constructing the $\Delta$-prototype out of elementary units $e_1 \ldots e_D$, which have no gates in joint posession, by means of creating links between these units along with creating links of these units with inputs and outputs of the $\Delta$-prototype. The amount $D$ is arbitrary.

Each of elementary units $e_1 \ldots e_D$ is specified as a «black box» which implements some finite-state automaton. But constructing this «black box» out of gates AND and OR should be possible.

Each input of the constructed $\Delta$-prototype apparently is a point of applying of some inputting logical variable which is either the bit of a vector coding a symbol of the input alphabet of automaton $A$ or a signal of a clock, of a reset etc.

Moreover each output of the $\mathit{\Delta}$-prototype is a point of observing of some outputting logical variable which is the bit of a vector coding a symbol of the output alphabet of automaton $A$.

The $\mathit{\Delta}$-prototype should be transformed into the network $\mathit{\Delta}$ in the following manner:

- a) the $\mathit{\Delta}$-inputs $6_2 \dots 6_K$ should be used as points of applying of the inputting logical variables;
- b) $\mathit{\Delta}$-fragment $F_{\varepsilon(i)}$ (where $\varepsilon : \overline{1,\, D} \to \overline{1,\, D}$ — some one-to-one function) should be used in the quality of the elementary unit $e_i \in \{e_1,\, e_2,\; \dots,\, e_D\}$.

This $\mathit{\Delta}$-fragment $F_{\varepsilon(i)}$ is formed by gates $1_{m(i)} \dots 1_{k(i)+1}$, $2_{m(i)} \dots 2_{k(i)}$, $3_{m(i)} \dots 3_{k(i)}$, $4_{m(i)} \dots 4_{k(i)}$, $5_{m(i)} \dots 5_{k(i)}$ and by connections between these gates, where

$$m(i) = \begin{cases} 1, \text{ if } \varepsilon(i) = 1, \\ k\left(\varepsilon^{-1}(\varepsilon(i) - 1)\right), \text{ if } \varepsilon(i) > 1; \end{cases} \qquad \text{(IV)}$$

$\varepsilon^{-1}(\,\cdot\,)$ is a function which is inverse to the function $\varepsilon(\,\cdot\,)$; $N = k(D)$.

One can easily see that in accordance with expression (IV) gate $1_{k\left(\varepsilon^{-1}(d-1)\right)+1}$ of $\mathit{\Delta}$-fragment $F_{d-1}$ is used in the quality of gate $1_{m\left(\varepsilon^{-1}(d)\right)}$ of $\mathit{\Delta}$-fragment $F_d$ for $d = \overline{2,\, D}$.

Besides, there are the following restrictions for $\mathit{\Delta}$-fragment $F_{\varepsilon(i)}$:

- 1) only inputs (except inputs 1 and 2) of gates $1_{m(i)} \dots 1_{k(i)+1}$ and inputs (except inputs 1) of gates $4_{m(i)} \dots 4_{k(i)}$, $5_{m(i)} \dots 5_{k(i)}$ may be used as those inputs of elementary unit $e_i$ which are connected with inputs of $\mathit{\Delta}$-prototype,
- 2) only inputs (except inputs 1) of gates $4_{m(i)} \dots 4_{k(i)}$, $5_{m(i)} \dots 5_{k(i)}$ can be used as those inputs of elementary unit $e_i$ which are connected with outputs of elementary units,
- 3) only outputs of gates $2_{m(i)} \dots 2_{k(i)}$, $3_{m(i)} \dots 3_{k(i)}$ may be used as those outputs of elementary unit $e_i$ which are connected with inputs of elementary units,
- 4) only points $7_{m(i)} \dots 7_{k(i)}$ and $8_{m(i)} \dots 8_{k(i)}$ may be used as outputs of the $\mathit{\Delta}$-prototype.

$\mathit{\Delta}$-fragment $F_{\varepsilon(i)}$ must model the finite-state automaton, implemented by elementary unit $e_i$ in the $\mathit{\Delta}$-prototype, only *in operating condition* of network $\mathit{\Delta}$, when constant «0» remains on input $6_1$ and constant «1» remains on inputs *9...12*.

Allowed in network $\mathit{\Delta}$ (according to the definition of the «widened long flip-flop») connections of outputs of gates $2_1 \dots 2_N$ with inputs of gates $4_1 \dots 4_N$ will be referred to as links $2 \to 4$, whereas connections of outputs of gates $3_1 \dots 3_N$ with inputs of gates $5_1 \dots 5_N$ — as links $3 \to 5$.

The transformation of the $\mathit{\Delta}$-prototype into network $\mathit{\Delta}$ may be confronted only by following difficulties:

- 5) each elementary unit must allow an implementation by some $\mathit{\Delta}$-fragment,
- 6) each such link between elementary units that is inherited from $\mathit{\Delta}$-prototype must be one of links $2 \to 4$ or $3 \to 5$; in this connexion the restrictions for structure of $\mathit{\Delta}$-fragments and for function $\varepsilon^{-1}(\,\cdot\,)$ appear.

For overcoming these difficulties, in Method IV $\mathit{\Delta}$-prototype must have a structure in accordance with Figure IV-2 and the clock cycle in accordance with Table IV. Here each elementary unit $e_i \in \{e_{2\nu+1},\, e_{2\nu+2},\; \dots,\, e_D\}$ shown on Figure IV-2 must conform with Figures IV-3*a* and IV-3*b*, whereas each elementary unit $e_i \in \{e_1,\, e_2,\; \dots,\, e_{2\nu}\}$ shown on Figure IV-2 must conform with Figures IV-4*a* and IV-4*b*.

In order that each such link between elementary units of network $\mathit{\Delta}$ which is inherited from the $\mathit{\Delta}$-prototype were one of links $2 \to 4$ or $3 \to 5$, the elementary units of the $\mathit{\Delta}$-prototype first must be ranked by means of following Algorithm IV:

- *Step 1.* Flip-flops $e_{\nu+1}$ ...$e_{2\nu}$ and all those elementary units AND-OR, whose inputs are connected only with inputs of the $\Delta$-prototype, should be refered to the rank 1. Value $R$ should be set to 1.
- *Step 2.* Each such elementary unit AND-OR whose one input is connected with an output of an elementary unit of rank $R$,
  whereas every another input is connected with either an output of an elementary unit of a rank not greater than $R$ or with an input of the $\Delta$-prototype, should be refered to the rank $R+1$.
- *Step 3.* If rank $R+1$ is not empty, then we should increase $R$ by 1 and go to Step 2.
- *Step 4.* Flip-flops $e_1$ ...$e_\nu$ should be refered to the rank $R+1$.
- *Step 5.* If rank $R+1$ is not empty, then we should increase $R$ by 1.
- *Step 6.* We should set $\rho$ equal to achieved value of $R$.

Next for $R = \overline{1,\ \rho}$ we should use $\Delta$-fragments $F_{a(R)+1}$ ...$F_{a(R)+D(R)}$ in the quality of elementary units of rank $R$, where $D(R)$ is the amount of elementary units of rank $R$;

$$a(R) = \begin{cases} 0, \text{ if } R = 1, \\ \sum_{i=1}^{R-1} D(i), \text{ if } R > 1. \end{cases}$$

Besides, when $i = \overline{1,\ D}$, $\Delta$-fragment $F_{\varepsilon(i)}$, presented on Figures IV-5*a* and IV-5*b* (if $i \in \overline{1,\ \nu}$), or on Figures IV-6*a* and IV-6*b* (if $i \in \overline{\nu+1,\ 2\nu}$), or on Figures IV-7*a* and IV-7*b* (if $i \in \overline{2\nu+1,\ D}$), should be used in the quality of elementary unit $e_i$.

Such is Method IV.

## An example of applying of Method IV

Let the $\Delta$-prototype used in Method IV be presented on Figure IV-8*a*.

Obviously in this case the $\Delta$-prototype implements the automaton $A$ whose symbol of the input alphabet presents the 4-bit vector $< x_1, \overline{x}_1, x_2, \overline{x}_2 >$ of a two-rail code, whose internal state presents the 2-bit vector $< q_1, \overline{q}_1 >$ of a two-rail code, and whose symbol of the output alphabet presents the 3-bit vector $< y_1, y_2, y_3 >$.

The structure of this $\Delta$-prototype conforms with Figure IV-1, i.e. $\gamma = < y_3 >$, $\vec{\alpha} = < y_1, y_2 >$ and $\vec{\beta} = < q_2, f_4 >$. Here elementary units $e_1$ ...$e_8$ along with their interconnections implement the automaton $A'$, whereas elementary unit $e_9$ implements the automaton $A''$.

In accordance to Algorithm IV, elementary units $e_3$ and $e_4$ should be refered to the rank 1, elementary units $e_5$ and $e_6$ — to the rank 2, elementary units $e_7$ and $e_8$ — to the rank 3, elementary unit $e_9$ — to the rank 4, and elementary units $e_1$ and $e_2$ — to the rank 5.

The network $\Delta$, constructed from this $\Delta$-prototype by Method IV, is presented on Figures IV-8*b* and IV-8*c*.

Here the elementary unit $e_3$ is formed by gates $1_1$ ...$1_4$, $2_1$ ...$2_3$, $3_1$ ...$3_3$, $4_1$ ...$4_3$, $5_1$ ...$5_3$ and by connections between these gates.

The elementary unit $e_4$ is formed by gates $1_4$ ...$1_7$, $2_4$ ...$2_6$, $3_4$ ...$3_6$, $4_4$ ...$4_6$, $5_4$ ...$5_6$ and by connections between these gates, moreover gate $1_4$ belongs both to elementary unit $e_3$ and to elementary unit $e_4$.

The elementary unit $e_5$ is formed by gates $1_7$, $1_8$, $2_7$, $3_7$, $4_7$, $5_7$ and by connections between these gates, moreover gate $1_7$ belongs both to elementary unit $e_5$ and to elementary unit $e_4$.

The elementary unit $e_6$ is formed by gates $1_8$ ...$1_{11}$, $2_8$ ...$2_{10}$, $3_8$ ...$3_{10}$, $4_8$ ...$4_{10}$, $5_8$ ...$5_{10}$ and by connections between these gates, moreover gate $1_8$ belongs both to elementary unit $e_6$ and to elementary unit $e_5$.

The elementary unit $e_7$ is formed by gates $1_{11}$ ...$1_{13}$, $2_{11}$, $2_{12}$, $3_{11}$, $3_{12}$, $4_{11}$, $4_{12}$, $5_{11}$, $5_{12}$ and by connections between these gates, moreover gate $1_{11}$ belongs both to elementary unit $e_7$ and to elementary unit $e_6$.

The elementary unit $e_8$ is formed by gates $1_{13}$ ...$1_{15}$, $2_{13}$, $2_{14}$, $3_{13}$, $3_{14}$, $4_{13}$, $4_{14}$, $5_{13}$, $5_{14}$ and by connections between these gates, moreover gate $1_{13}$ belongs both to elementary unit $e_8$ and to elementary unit $e_7$.

The elementary unit $e_9$ is formed by gates $1_{15}$ ...$1_{17}$, $2_{15}$, $2_{16}$, $3_{15}$, $3_{16}$, $4_{15}$, $4_{16}$, $5_{15}$, $5_{16}$ and by connections between these gates, moreover gate $1_{15}$ belongs both to elementary unit $e_9$ and to elementary unit $e_9$.

The elementary unit $e_1$ is formed by gates $1_{17}$ ...$1_{20}$, $2_{17}$ ...$2_{19}$, $3_{17}$ ...$3_{19}$, $4_{17}$ ...$4_{19}$, $5_{17}$ ...$5_{19}$ and by connections between these gates, moreover gate $1_{17}$ belongs both to elementary unit $e_1$ and to elementary unit $e_9$.

The elementary unit $e_2$ is formed by gates $1_{20}$ ...$1_{23}$, $2_{20}$ ...$2_{22}$, $3_{20}$ ...$3_{22}$, $4_{20}$ ...$4_{22}$, $5_{20}$ ...$5_{22}$ and by connections between these gates, moreover gate $1_{20}$ belongs both to elementary unit $e_2$ and to elementary unit $e_1$.

$\Delta$-fragments conforming with Figures IV-5*a* and IV-5*b* are used in quality of elementary units $e_1$ and $e_2$, respectively. $\Delta$-fragments conforming with Figures IV-6*a* and IV-6*b* are used in quality of elementary units $e_3$ and $e_4$, respectively. $\Delta$-fragments conforming with Figures IV-7*a* and IV-7*b* are used in quality of elementary units $e_5$ ...$e_9$, respectively.

*In operating condition*, constant «0» remains on the input $6_1$ and constant «1» remains on inputs *9...12*.

The inputs $6_2$ ...$6_9$ are used for applying the inputting logical variables $\overline{c}_1$, $c_1$, $\overline{x}_2$, $x_2$, $\overline{x}_1$, $x_1$, $\overline{c}_2$, and $c_2$, respectively.

The poles $8_2$, $8_{11}$, and $8_{15}$ are used for observing the outputting logical variables $y_1$, $y_2$, and $y_3$, respectively.

The outputting logical variable $y_3$ presents the check bit of the output vector of the implemented automaton $A$.

## Properties of network $\Delta$ constructed by Method IV

Theorems 2 and 3 assert that testing of network $\Delta$ in accordance with Table III-1 detects all stuck-at-faults covered by model $C^{\text{mpl}}$, excepting stuck-at-1 faults of some inputs of gates AND.

(More precisely, the full set of poles whose stuck-at-1 faults are undetected consists of inputs (except inputs 1 and 2) of gates $1_1$ ...$1_{N+1}$ and of inputs (except inputs 1) of gates $4_1$ ...$4_N$, $5_1$ ...$5_N$.)

In automaton $A$ these undetected faults are able to cause only asymmetric errors «1 instead of 0» in bits of an output vector.

But if and only if there is at least one such an error in bits of the output vector, then the value «1» appears in the check bit of the output vector.

(To be easily mentioned, the $\Delta$-output generating the check bit of the output vector will be referred to as *«the ban output»*.)

If to bear with such behaviour of the network $\Delta$ concerning undetected faults is possible, then:

- — network $\Delta$, which stood the test described by Table III-1, is **fit to working** despite faults undetected,
- — the test being described by Table III-1 fully examines network $\Delta$ for **unfitness to working**.

Let´s point out important practical cases when network $\Delta$ is what is necessary:

- a) when faults «stuck-at-0» of inputs of gates AND have much more intensity of appearance than faults «stuck-at-1» have. This is true for some sorts of integrated logic. An additional condition is the manufacturing of a network $\Delta$ as an unmendable component (for example, as an integrated circuit). Then probability of the network $\Delta$ being on strike (with warning about it by value «1» on the ban output) is tiny. In troth, inasmuch as the probability of appearance of faults «stuck-at-0» during time between neighbouring runnins of the test significantly exceeds the probability of appearance of faults

«stuck-at-1», then the unmendable component according to results of testing will be as a rule replaced before fault «stuck-at-1» appeared;

- b) when network *Δ* is used to oversee intactness of some equipment and the check bit of the output vector of network *Δ* is to give warning about the damage of the equipment (the rest bits of the output vector of the network *Δ* may qualify the damage of the equipment). If mentioned warning appeared, acting of the personnel (i.e. full repair of both the equipment and network *Δ*) is needed to garantee against using the damaged equipment;
- c) when network *Δ* performs transactions changing a data base. There are two alternatives to act:
  - — perform the transaction, but if during this the value «1» appears on the ban output of the network *Δ*, do the rollback;
  - — perform the transaction in dummy mode, and if during this the value «1» doesn´t appear on the ban output of the network *Δ*, do the transaction indeed;
- d) when network *Δ* is added to some network *X* to store states of flip-flops of network *X* during testing of network *X*. Before testing network *X*, the states of flip-flops of network *X* should be copied into network *Δ*. If value «1» does not appear on the ban output of network *Δ* during this copying, then it is safe to test network *X* and next restore states of flip-flops of network *X* from network *Δ*.

(In cases «*b*»—«*d*» we mean that network *Δ* stands the test described in Table III-1. To insure this, the test must be run often enough and all repairs needed must be made in time.)

**Table IV. The clock cycle for the network being presented on Figure IV-2**

| The inputs and the outputs of the network | Moments $t'$ and $t''$ of the halts of the network, where $t' < t''$ | |
|---|---|---|
| | $t'$ | $t''$ |
| $c_1, \bar{c}_2$ | 1 | 0 |
| $\bar{c}_1, c_2$ | 0 | 1 |
| $\vec{x}, \vec{y}$ | acting values of bits | indifferent values of bits |

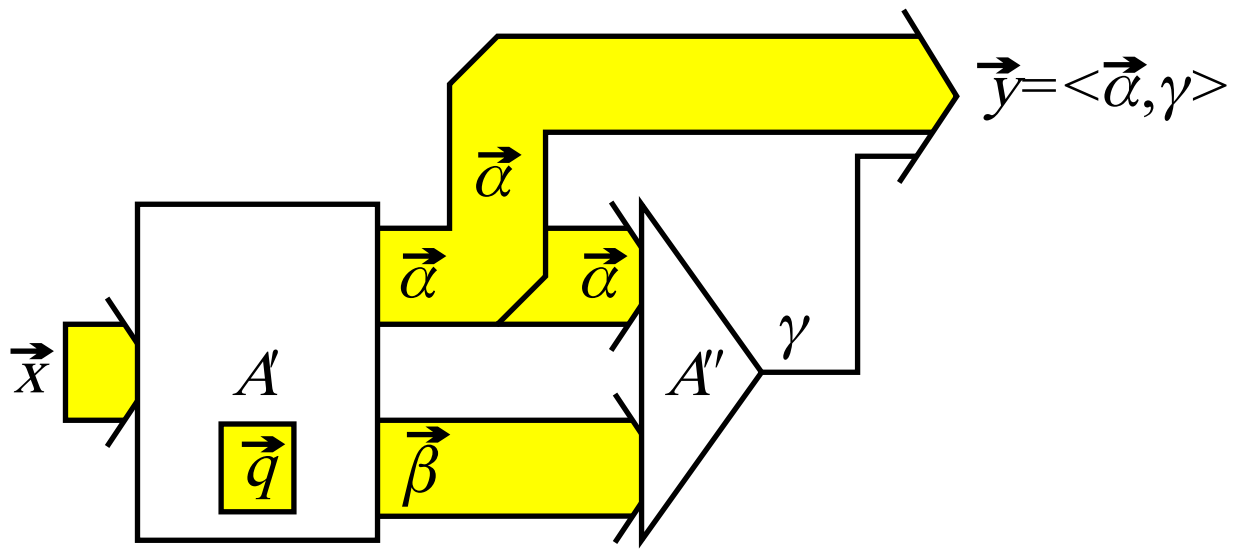

Figure IV-1. The structure of automaton *A* in Method IV.

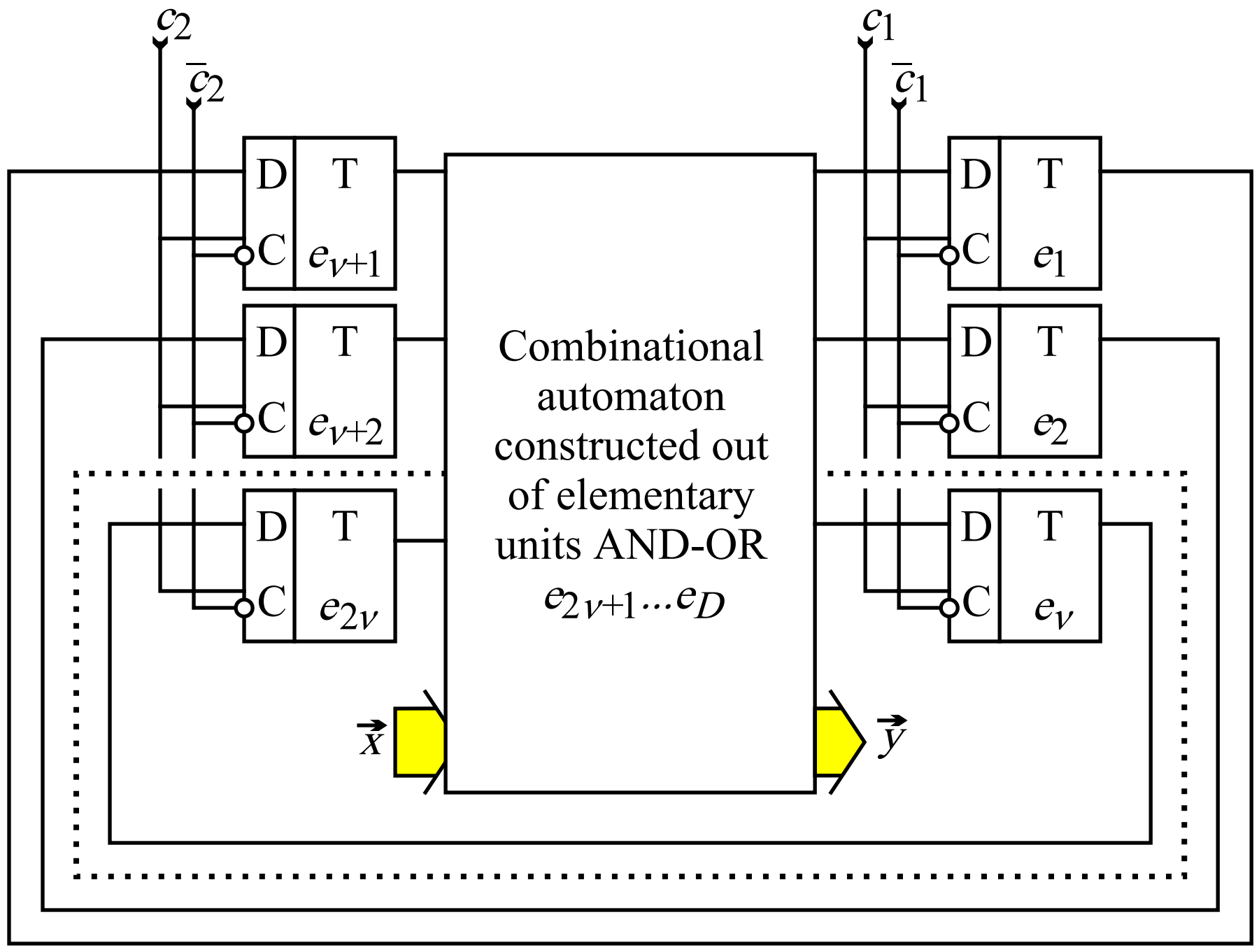

Figure IV-2. The structure of the $\varDelta$-prototype in accordance with Method IV. Here $\nu$ — the amount of bits in the logical vector which codes the internal state of the automaton $A$; $\vec{x} = \langle x_1, x_2, \ldots, x_{n_{\mathrm{IN}}} \rangle$ — the logical vector which codes the symbol of the input alphabet of automaton $A$; $\vec{y} = \langle y_1, y_2, \ldots, y_{n_{\mathrm{OUT}}} \rangle$ — the logical vector which codes the symbol of the output alphabet of automaton $A$. Here each elementary unit $e_i \in \{e_{2\nu+1}, e_{2\nu+2}, \ldots, e_D\}$ must conform with Figures IV-3*a* and IV-3*b*, whereas each elementary unit $e_i \in \{e_1, e_2, \ldots, e_{2\nu}\}$ must conform with Figures IV-4*a* and IV-4*b*.

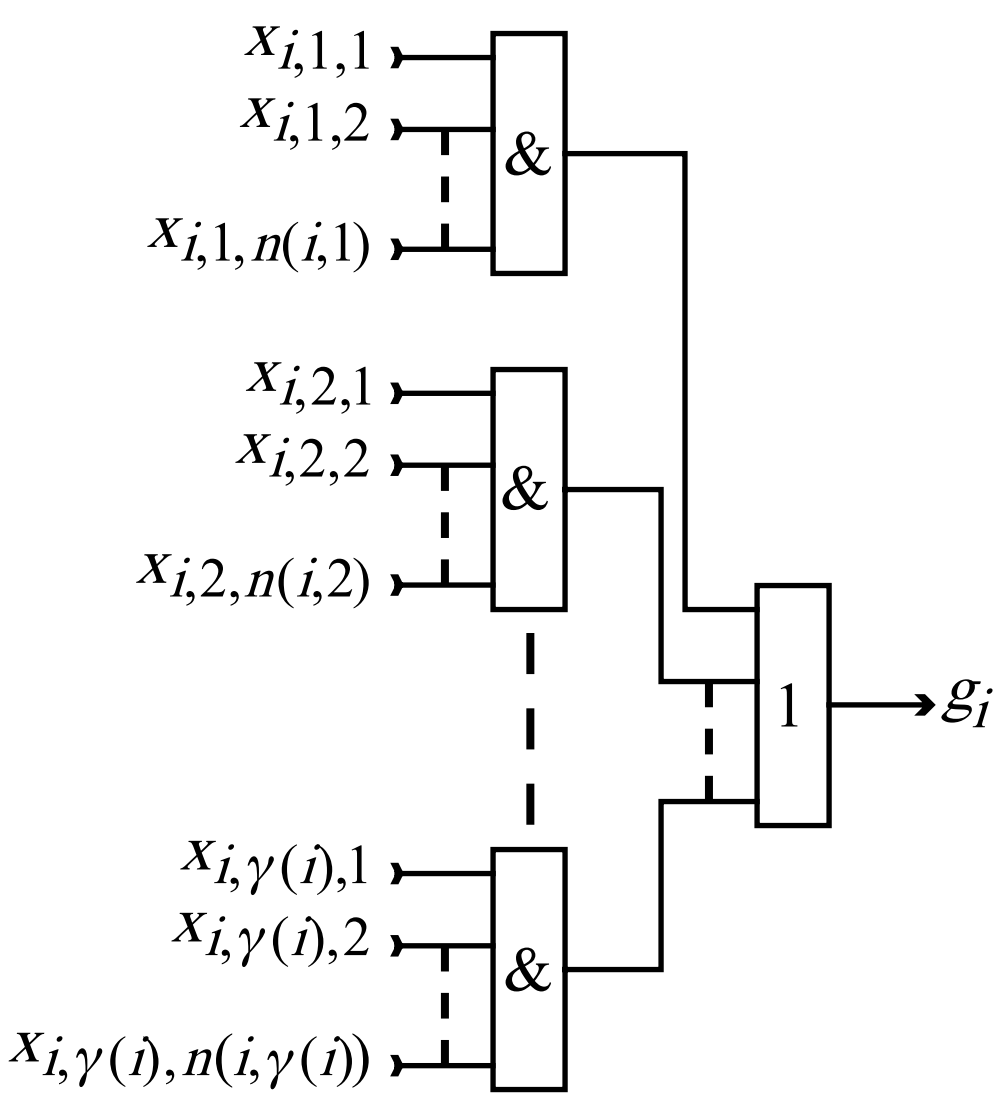

Figure IV-3*a*. The structure of elementary unit $e_i$ which represents an element AND-OR in the $\varDelta$-prototype.

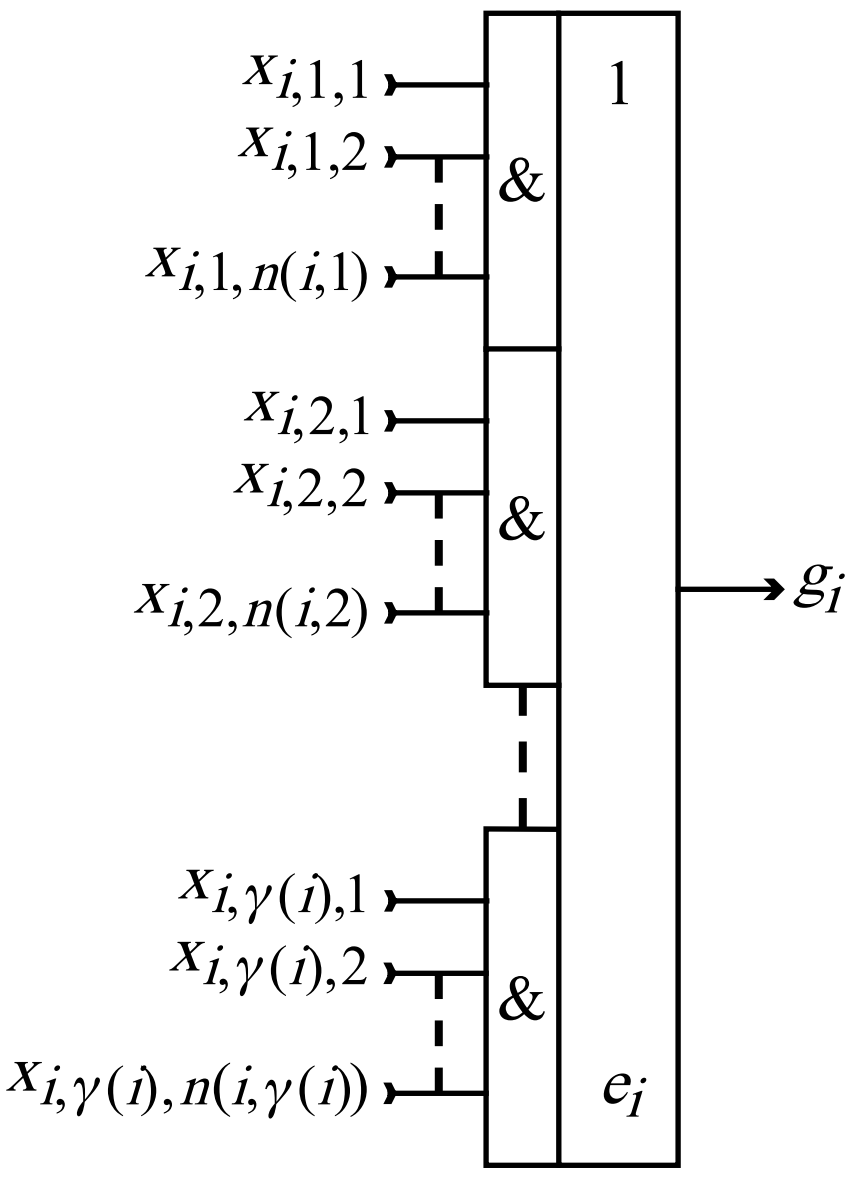

Figure IV-3*b*. The special graphical symbol (SGS) for elementary unit $e_i$ which is presented on Figure IV-3*a*.

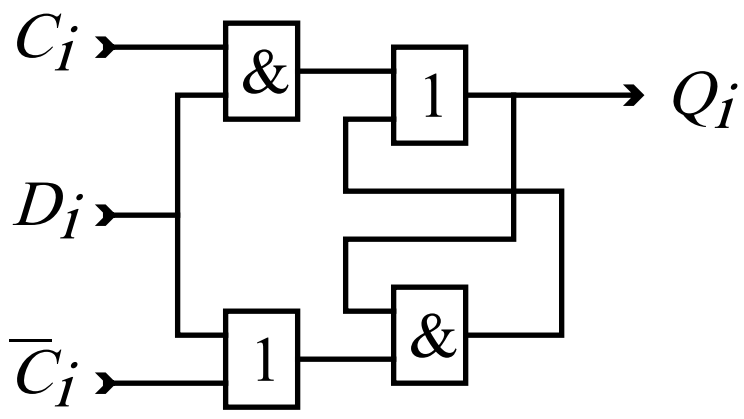

Figure IV-4*a*. The structure of elementary unit $e_i$ which represents a flip-flop in the $\varDelta$-prototype.

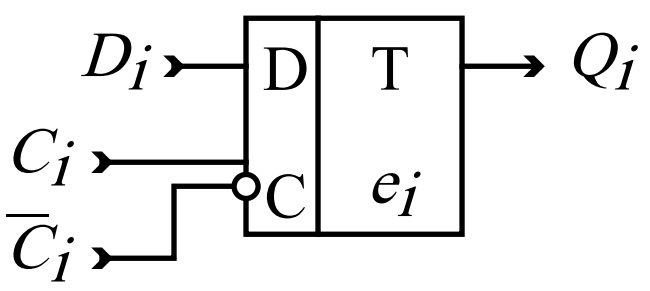

Figure IV-4*b*. The special graphical symbol (SGS) for elementary unit $e_i$ which is presented on Figure IV-4*a*.

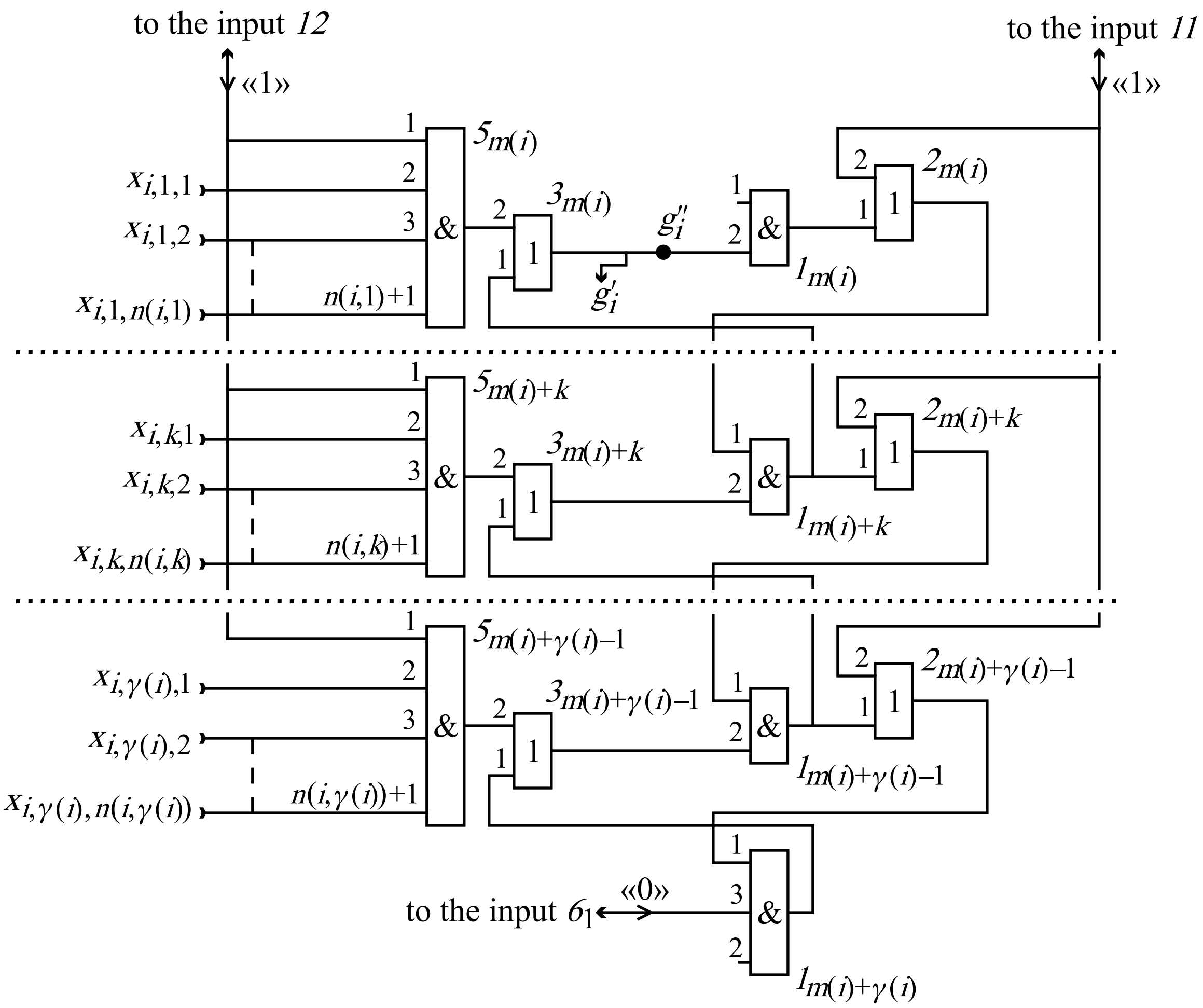

Figure IV-5*a*. The $\Delta$-fragment $F_{\varepsilon(i)}$ to be used in the quality of elementary unit $e_i$ which is nothing else than element AND-OR, where $i = \overline{1, D}$. Logical values shown on the drawing characterize the operating condition of network $\Delta$.

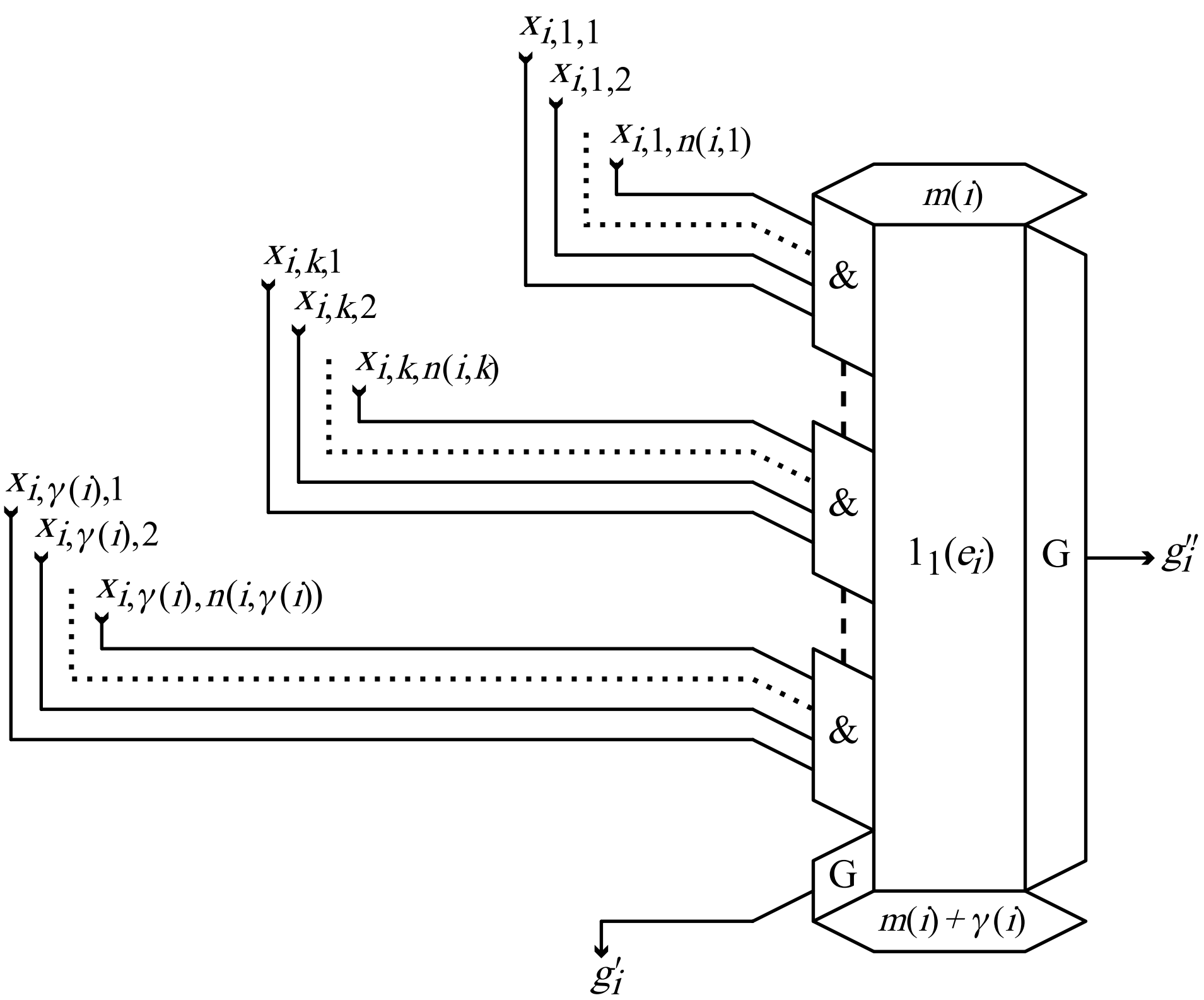

Figure IV-5*b*. The special graphical symbol (SGS) for $\Delta$-fragment $F_{\varepsilon(i)}$ which is presented on Figure IV-5*a*.

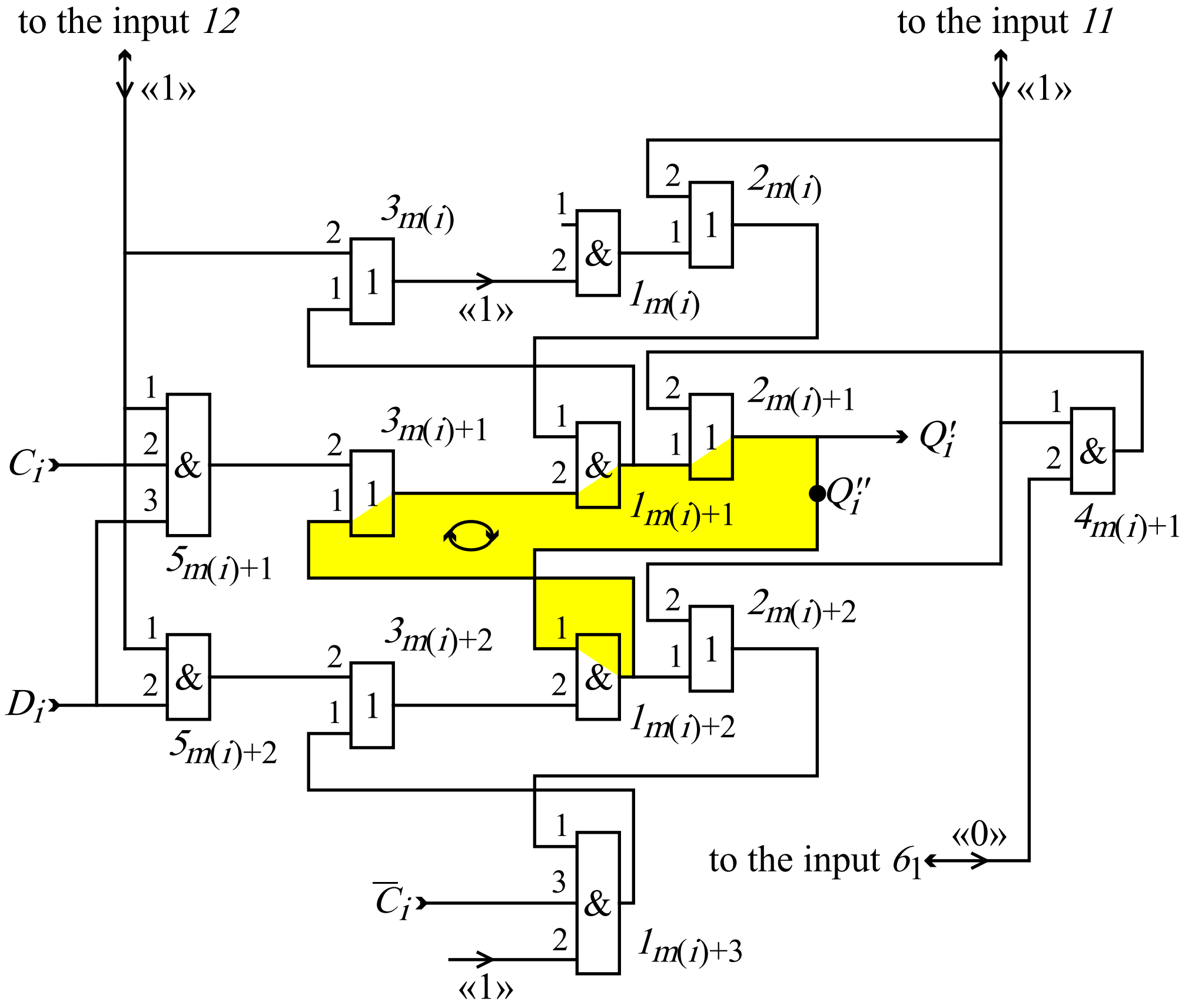

Figure IV-6*a*. The $\Delta$-fragment $F_{\varepsilon(i)}$ to be used in the quality of elementary unit $e_i$ which is nothing else than a flip-flop, where $i \in \overline{\nu+1,\, 2\nu}$. Logical values shown on the drawing characterize the operating condition of network $\Delta$. The loop formed by gates $1_{m(i)+1}$, $2_{m(i)+1}$, $3_{m(i)+1}$, $1_{m(i)+2}$ is used to store an internally-stable state of flip-flop $e_i$.

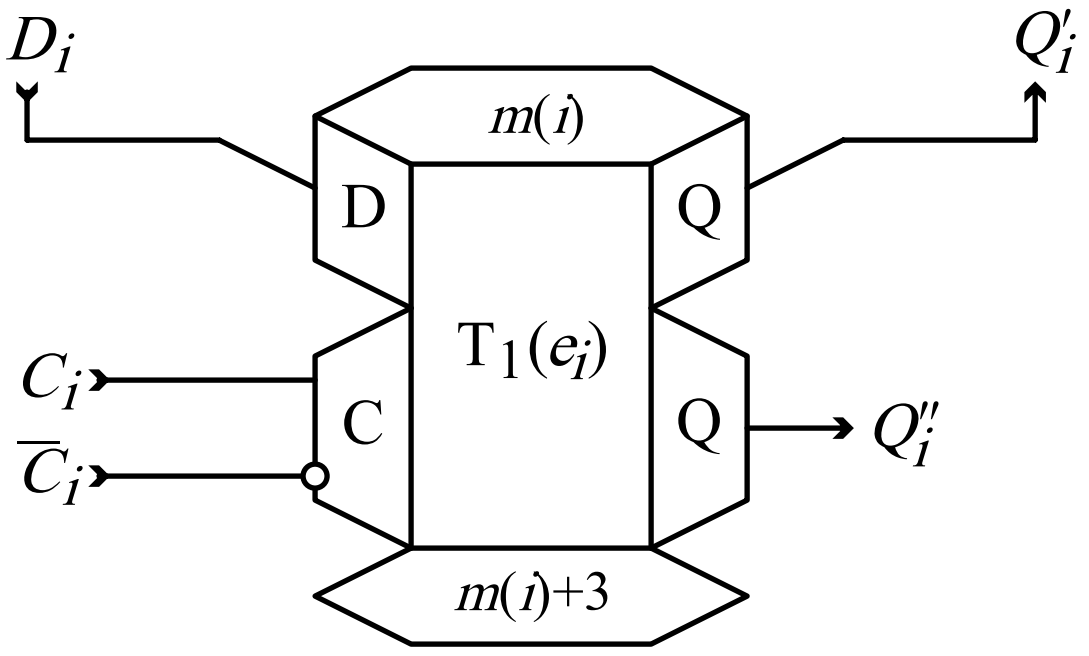

Figure IV-6*b*. The special graphical symbol (SGS) for $\Delta$-fragment $F_{\varepsilon(i)}$ which is presented on Figure IV-6*a*.

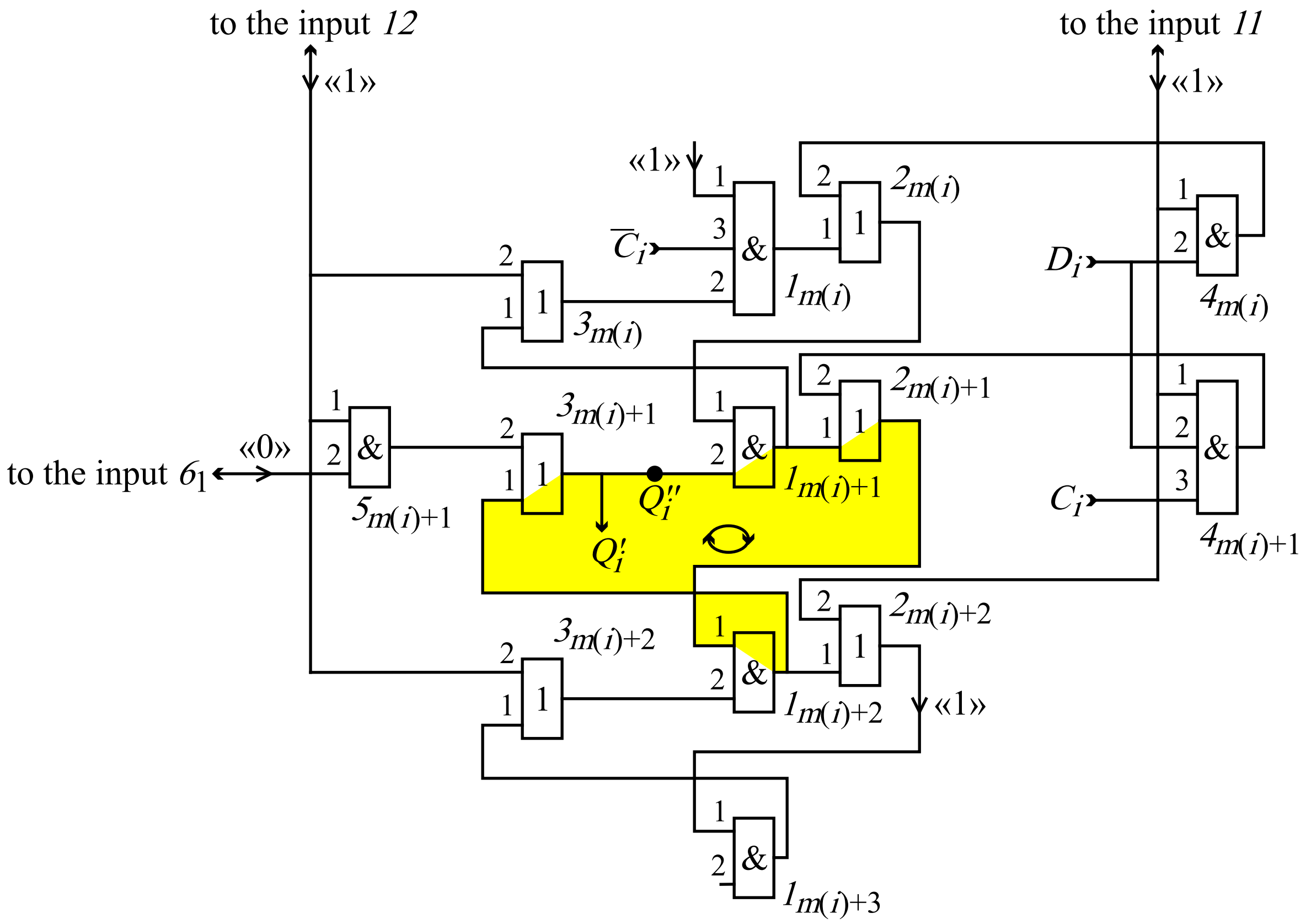

Figure IV-7*a*. The *Δ*-fragment $F_{\varepsilon(i)}$ to be used in the quality of elementary unit $e_i$ which is nothing else than a flip-flop, where $i \in \overline{2\nu + 1,\ D}$. Logical values shown on the drawing characterize the operating condition of network *Δ*. The loop formed by gates $1_{m(i)+1}$, $2_{m(i)+1}$, $3_{m(i)+1}$, $1_{m(i)+2}$ is used to store an internally-stable state of flip-flop $e_i$.

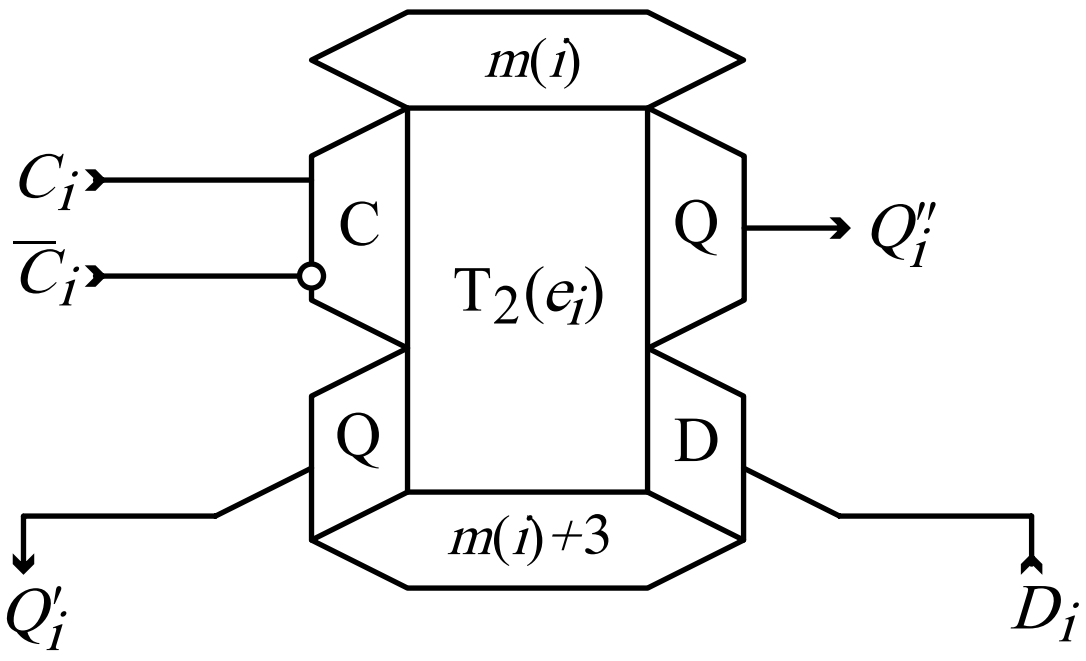

Figure IV-7*b*. The special graphical symbol (SGS) for *Δ*-fragment $F_{\varepsilon(i)}$ which is presented on Figure IV-7*a*.

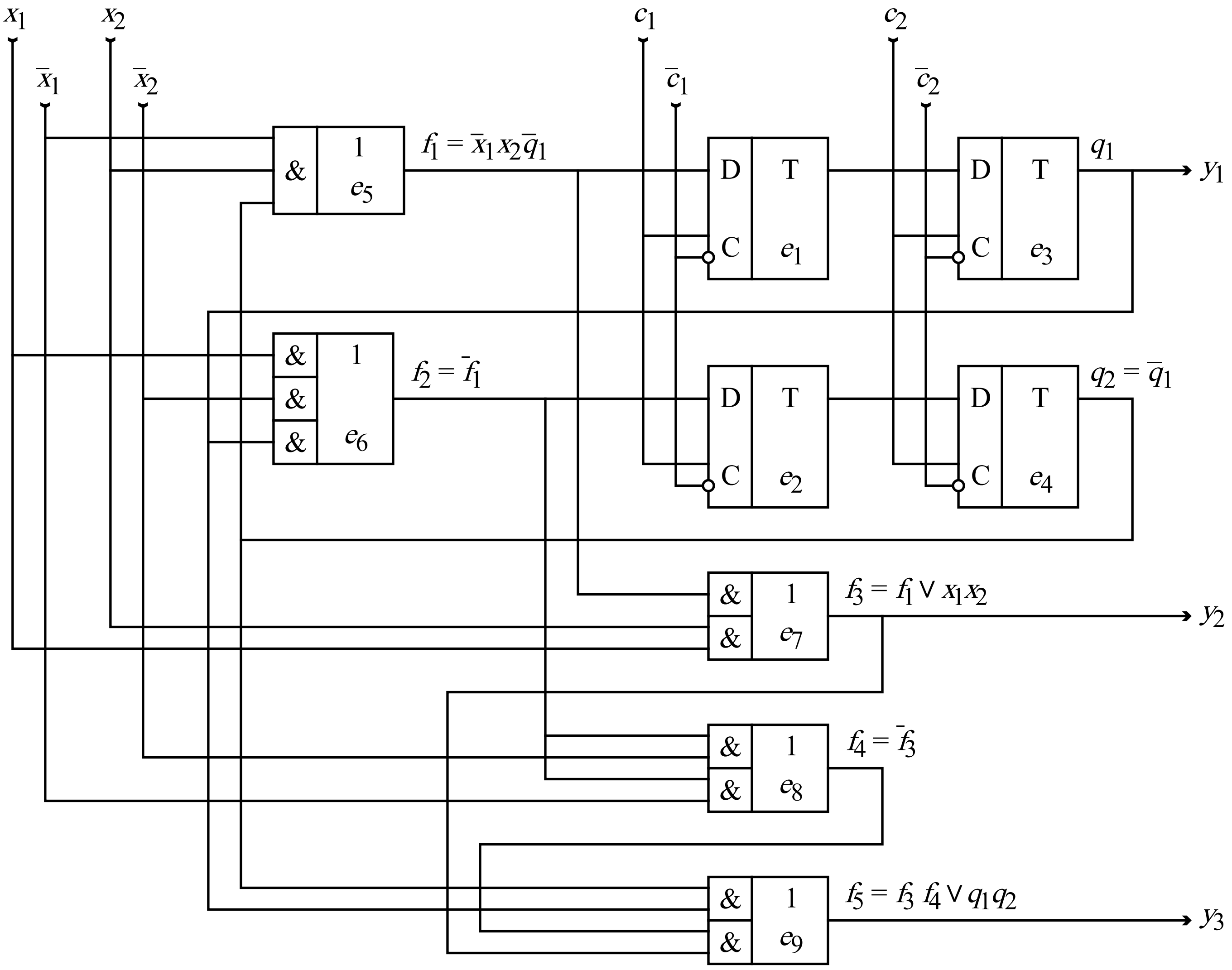

Figure IV-8*a*. The example of the *Δ*-prototype for Method IV.

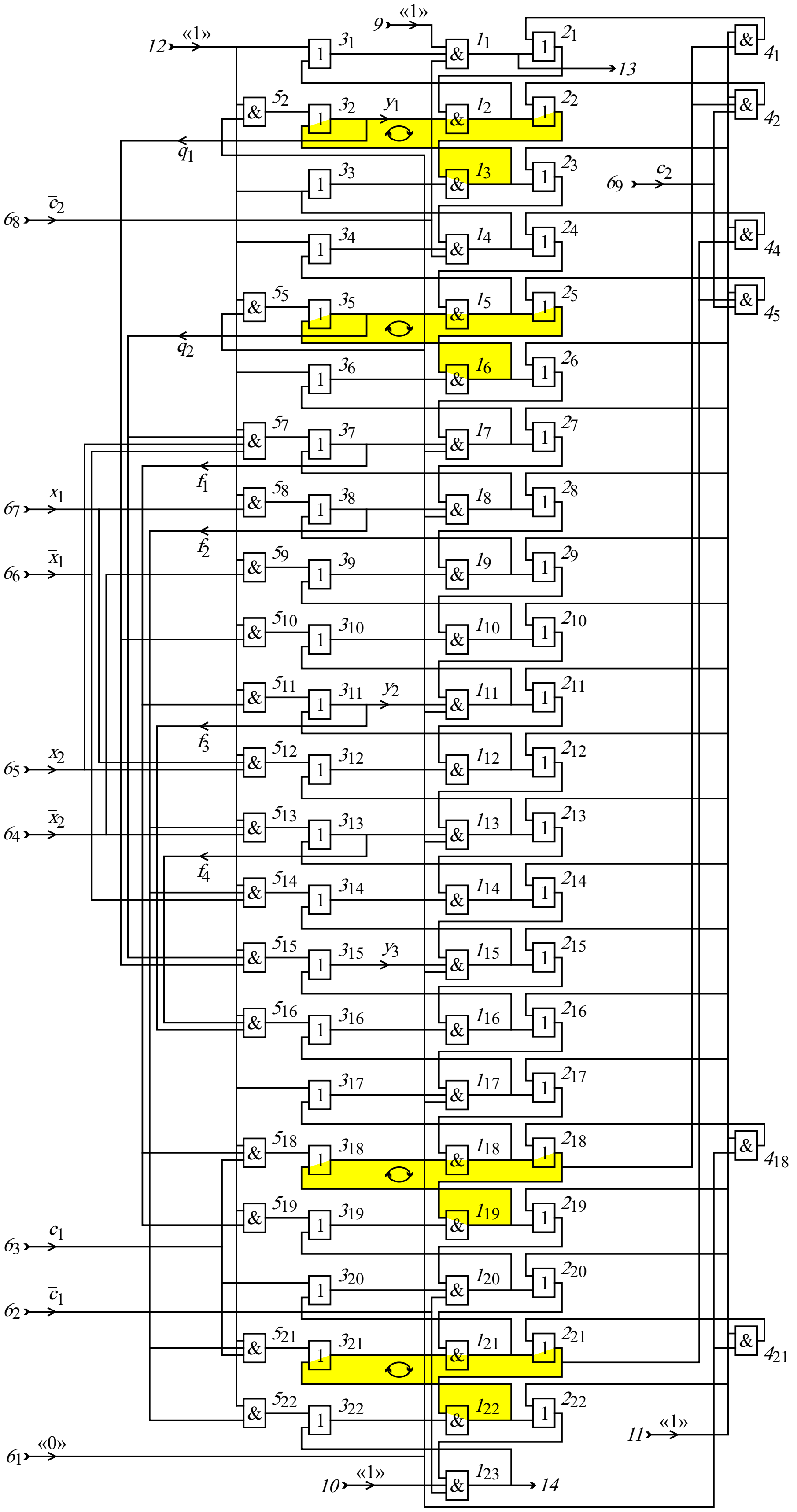

Figure IV-8*b*. The network *Δ* which was obtained from the *Δ*-prototype presented on Figure IV-8*a* by Method IV. Logical values shown on the drawing characterize the operating condition of network *Δ*.

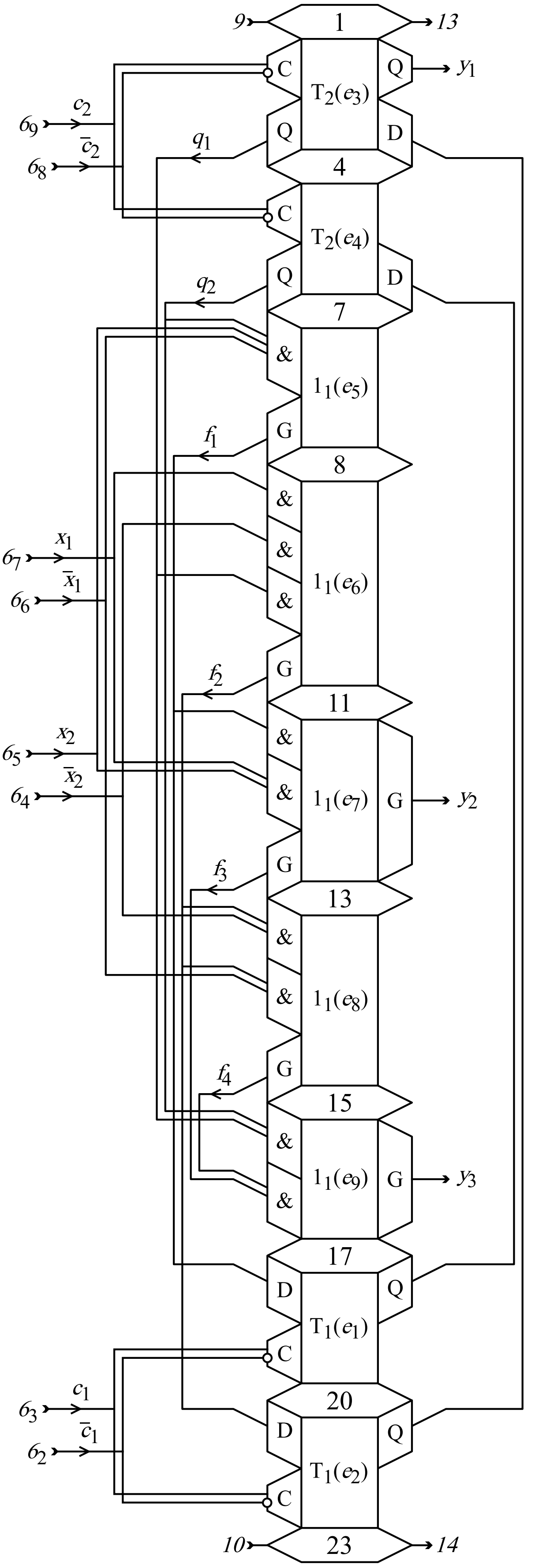

Figure IV-8*c*. The network *Δ*, which coincides with the network presented on Figure IV-8*a*, but is sketched out with using special graphical symbols (SGS) for *Δ*-fragments $F_{\varepsilon(1)} \ldots F_{\varepsilon(9)}$ (see Figures IV-5*b*, IV-6*b*, and IV-7*b*). Logical values shown on the drawing characterize the operating condition of network *Δ*.

# Chapter V. Finite-state automata being easily tested for intactness

Let´s consider Method V which has following 3 distinctions from Method IV:

- 1) $\Delta$-prototype is to implement not the union of automata $A'$ and $A''$ (Figure IV-1), but the primary finite-state automaton $A$ (Figure V-1). It allows to use an arbitrary irredundant code for symbols of alphabet $Y$;
- 2) $\Delta$-gates $1_1 \dots 1_{N+1}$ are to be strictly two-input, in this connexion now an elementary unit $e_i$ (where $i = \overline{1,\ D}$) is to be implemented by $\Delta$-fragment $F_{\varepsilon(i)}$, presented on Figures V-2*a* and V-2*b* (if $i \in \overline{1,\ \nu}$), or on Figures V-3*a* and V-3*b* (if $i \in \overline{\nu+1,\ 2\nu}$), or on Figures V-4*a* and V-4*b* (if $i \in \overline{2\nu+1,\ D}$); *the operating condition* now is to be specified by Table V-1;
- 3) gates $4_1 \dots 4_N$ are to be substituted by the system of «long flip-flops» ($4_1$)...($4_N$), ($4'$), ($4''$), as well as gates $5_1 \dots 5_N$ — by the system of «long flip-flops» ($5_1$)...($5_N$), ($5'$), ($5''$), in accordance with Figure V-5 (we advise you to compare Figures V-5 and III-1).

## An example of applying Method V

Let´s take previous example of $\Delta$-prototype (Figure IV-8*a*). In accordance with item «*1)*», we simplify this example of prototype up to that presented on Figure V-6*a*.

In accordance with item «*2)*», we transform $\Delta$-prototype (Figure V-6*a*) into network $\Delta$ which is presented on Figures V-6*b* and V-6*c*.

In accordance with item «*3)*», further we transform gates $4_1 \dots 4_{25}$, which are presented on Figure V-6*b*, into the system of «long flip-flops» ($4_1$)...($4_N$), ($4'$), ($4''$), and we transform gates $5_1 \dots 5_{25}$, which are presented on Figure V-6*b*, into the system of «long flip-flops» ($5_1$)...($5_N$), ($5'$), ($5''$). As a result we obtain network $\Delta$ which is presented on Figure V-6*d*.

## Properties of the network $\Delta$ constructed by Method V

The network $\Delta$ is being tested for all stuck-at-faults, covered by model $C^{\mathrm{mpl}}$, in accordance with Table V-2 (here both Theorems 2 and 3 should be used) and in accordance with Tables V-3 and V-4 (here Theorem 1 should be used).

One can easily see that in this case the principle of «back-doors testing» is implemented. Here «back-doors test» of network $\Delta$ uses 22 beats, 18 lateral inputs and 10 lateral outputs.

Under conditions $t_6 < t_7$, $t_{10} < t_{15}$, $t_{18} < t_{11}$ and $t_{14} < t_{19}$, during test run first neutral vector 11...1 (during interval $[t_1,\ t_{18}]$) and next neutral vector 00...0 (during interval $[t_{11},\ t_{22}]$) are being applied towards $\Delta$-inputs $6_2 \dots 6_N$ (we remind they are the only non-lateral inputs).

## Table V-1. Controlling the operating condition of network $\Delta$ obtained by Method V

| The inputs and the outputs of network $\Delta$ | The moment* $t_0$, which is a moment of $\Delta$-halt preceding the start of the operating condition of network $\Delta$ | The time interval during which network $\Delta$ is in the operating condition |
|---|---|---|
| $б_1$ | 0** | |
| $б_2 \ldots б_K$ | indifferent values | acting values |
| *9...12* | 0 | 1 |
| $C_1$, $C_2$ | indifferent values | 0 |
| $A_{1,1}$, $A_{2,1}$ | indifferent values | 1 |

* The moment is used for installing loops which must supply zeros, necessary for $\Delta$-fragments $F_1 \ldots F_D$, into zero internally-stable states.

** This zero must remains from moment $t_0$ till the end of the operating condition.

## Table V-2. The test of «long flip-flop» (*0*)

| The inputs and the outputs of network $\Delta$ | Moments $t_1 \ldots t_6$ of $\Delta$-halts, where $t_1 < t_2 < t_3$ and $t_4 < t_5 < t_6$. The time intervals which are constrained by the said moments | | | | | | | | | |
|---|---|---|---|---|---|---|---|---|---|---|
| | $t_1$ | $[t_1, t_2]$ | $t_2$ | $[t_2, t_3]$ | $t_3$ | $t_4$ | $[t_4, t_5]$ | $t_5$ | $[t_5, t_6]$ | $t_6$ |
| $б_1 \ldots б_K$ | 1 | | | | | | | | | |
| *11* | 0 | | | | | 1 | | | | |
| *12* | 1 | | | | | 0 | | | | |
| *9* | 1 | | 0 | | 1 | 1 | | | | |
| *10* | 1 | | | | | 1 | | 0 | | 1 |
| *13* | | | | | | 1 | | 0 | | 1 |
| *14* | 1 | | 0 | | 1 | | | | | |
| $A_{1,1}$ | 1 | | | | | const | | | | |
| $A_{2,1}$ | const | | | | | 1 | | | | |
| $A_{1,2}$, $A_{2,2}$, $C_1$, $C_2$ | const | | | | | const | | | | |

## Table V-3. The test of «long flip-flops» ($5_1$)...($5_N$), ($5'$), ($5''$)

<table>
<tr><td rowspan="2">The inputs and the outputs of network $\Delta$</td><td colspan="14">Moments $t_7 \ldots t_{14}$ of $\Delta$-halts, where $t_7 < t_8 < t_9 < t_{10}$ and $t_{11} < t_{12} < t_{13} < t_{14}$.<br>The time intervals which are constrained by the said moments</td></tr>
<tr><td>$t_7$</td><td>$[t_7, t_8]$</td><td>$t_8$</td><td>$[t_8, t_9]$</td><td>$t_9$</td><td>$[t_9, t_{10}]$</td><td>$t_{10}$</td><td>$t_{11}$</td><td>$[t_{11}, t_{12}]$</td><td>$t_{12}$</td><td>$[t_{12}, t_{13}]$</td><td>$t_{13}$</td><td>$[t_{13}, t_{14}]$</td><td>$t_{14}$</td></tr>
<tr><td>$6_1 \ldots 6_K$, 10, 12</td><td colspan="7">1</td><td colspan="7">0</td></tr>
<tr><td>11, $A_{2,2}$, $C_2$</td><td colspan="7">1</td><td colspan="7">const</td></tr>
<tr><td>$C_1$</td><td colspan="7">0</td><td colspan="7">1</td></tr>
<tr><td>$A_{1,1}$</td><td colspan="3">1</td><td></td><td colspan="3">0</td><td colspan="7">0</td></tr>
<tr><td>$A_{1,2}$</td><td colspan="7">0</td><td colspan="3">1</td><td></td><td colspan="3">0</td></tr>
<tr><td>$A_{3,1}$</td><td>1</td><td></td><td>0</td><td></td><td colspan="3">0</td><td colspan="7"></td></tr>
<tr><td>$B_{3,1}$</td><td>1</td><td></td><td>0</td><td colspan="4"></td><td colspan="7"></td></tr>
<tr><td>$A_{3,2}$</td><td colspan="3">0</td><td></td><td>1</td><td></td><td>0</td><td colspan="7"></td></tr>
<tr><td>$B_{3,2}$</td><td colspan="4"></td><td>1</td><td></td><td>0</td><td colspan="7"></td></tr>
<tr><td>$A_{4,1}$</td><td colspan="7"></td><td>1</td><td></td><td>0</td><td></td><td colspan="3">0</td></tr>
<tr><td>$B_{4,1}$</td><td colspan="7"></td><td>1</td><td></td><td>0</td><td colspan="4"></td></tr>
<tr><td>$A_{4,2}$</td><td colspan="7"></td><td colspan="3">0</td><td></td><td>1</td><td></td><td>0</td></tr>
<tr><td>$B_{4,2}$</td><td colspan="7"></td><td colspan="4"></td><td>1</td><td></td><td>0</td></tr>
</table>

## Table V-4. The test of «long flip-flops» ($4_1$)...($4_N$), ($4'$), ($4''$)

<table>
<tr><td rowspan="2">The inputs and the outputs of network $\Delta$</td><td colspan="14">Moments $t_{15} \ldots t_{22}$ of $\Delta$-halts, where $t_{15} < t_{16} < t_{17} < t_{18}$ and $t_{19} < t_{20} < t_{21} < t_{22}$.<br>The time intervals which are constrained by the said moments</td></tr>
<tr><td>$t_{15}$</td><td>$[t_{15}, t_{16}]$</td><td>$t_{16}$</td><td>$[t_{16}, t_{17}]$</td><td>$t_{17}$</td><td>$[t_{17}, t_{18}]$</td><td>$t_{18}$</td><td>$t_{19}$</td><td>$[t_{19}, t_{20}]$</td><td>$t_{20}$</td><td>$[t_{20}, t_{21}]$</td><td>$t_{21}$</td><td>$[t_{21}, t_{22}]$</td><td>$t_{22}$</td></tr>
<tr><td>$6_1 \ldots 6_K$, 9, 11</td><td colspan="7">1</td><td colspan="7">0</td></tr>
<tr><td>12, $A_{1,2}$, $C_1$</td><td colspan="7">1</td><td colspan="7">const</td></tr>
<tr><td>$C_2$</td><td colspan="7">0</td><td colspan="7">1</td></tr>
<tr><td>$A_{2,1}$</td><td colspan="3">1</td><td></td><td colspan="3">0</td><td colspan="7">0</td></tr>
<tr><td>$A_{2,2}$</td><td colspan="7">0</td><td colspan="3">1</td><td></td><td colspan="3">0</td></tr>
<tr><td>$A_{5,1}$</td><td>1</td><td></td><td>0</td><td></td><td colspan="3">0</td><td colspan="7"></td></tr>
<tr><td>$B_{5,1}$</td><td>1</td><td></td><td>0</td><td colspan="4"></td><td colspan="7"></td></tr>
<tr><td>$A_{5,2}$</td><td colspan="3">0</td><td></td><td>1</td><td></td><td>0</td><td colspan="7"></td></tr>
<tr><td>$B_{5,2}$</td><td colspan="4"></td><td>1</td><td></td><td>0</td><td colspan="7"></td></tr>
<tr><td>$A_{6,1}$</td><td colspan="7"></td><td>1</td><td></td><td>0</td><td></td><td colspan="3">0</td></tr>
<tr><td>$B_{6,1}$</td><td colspan="7"></td><td>1</td><td></td><td>0</td><td colspan="4"></td></tr>
<tr><td>$A_{6,2}$</td><td colspan="7"></td><td colspan="3">0</td><td></td><td>1</td><td></td><td>0</td></tr>
<tr><td>$B_{6,2}$</td><td colspan="7"></td><td colspan="4"></td><td>1</td><td></td><td>0</td></tr>
</table>

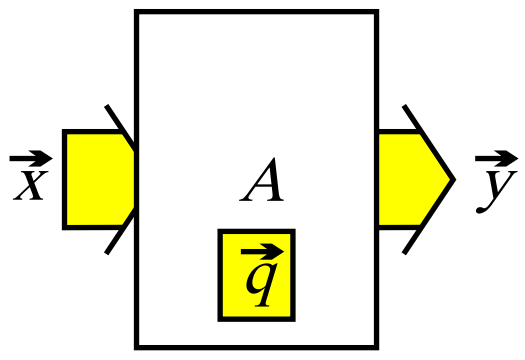

Figure V-1. Unlike Method IV, the automaton *A* isn´t divided into two automata in Method V.

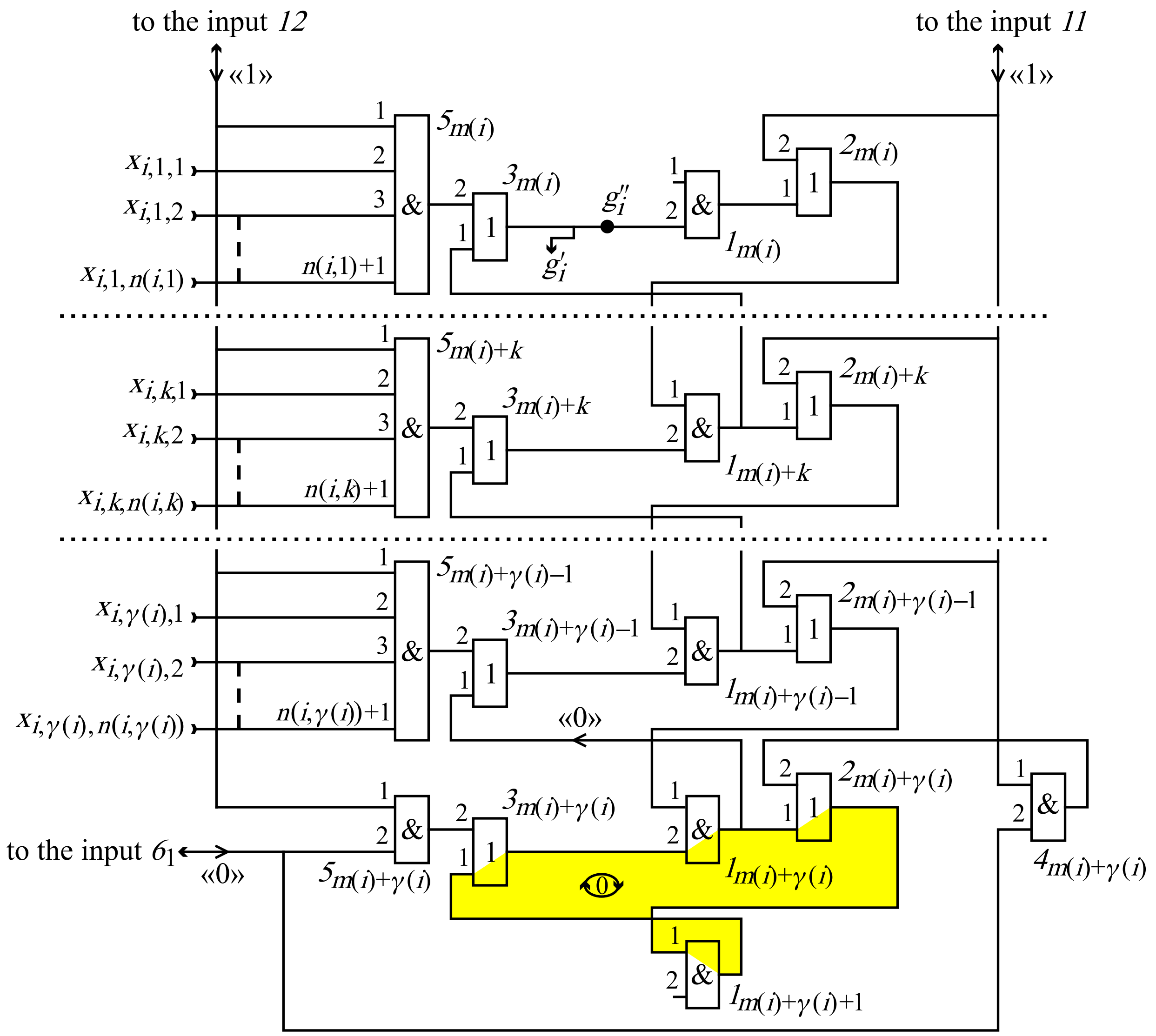

Figure V-2*a*. The $\Delta$-fragment $F_{\varepsilon(i)}$ to be used in the quality of elementary unit $e_i$ which is nothing else than element AND-OR, where $i = \overline{1,\ D}$. Logical values shown on the drawing characterize the operating condition of network $\Delta$. Gates $1_{m(i)+\gamma(i)}$, $2_{m(i)+\gamma(i)}$, $3_{m(i)+\gamma(i)}$, $1_{m(i)+\gamma(i)+1}$ form the loop staying in the zero internally-stable state during the whole operating condition of network $\Delta$ to supply the said fragment with the constant of 0.

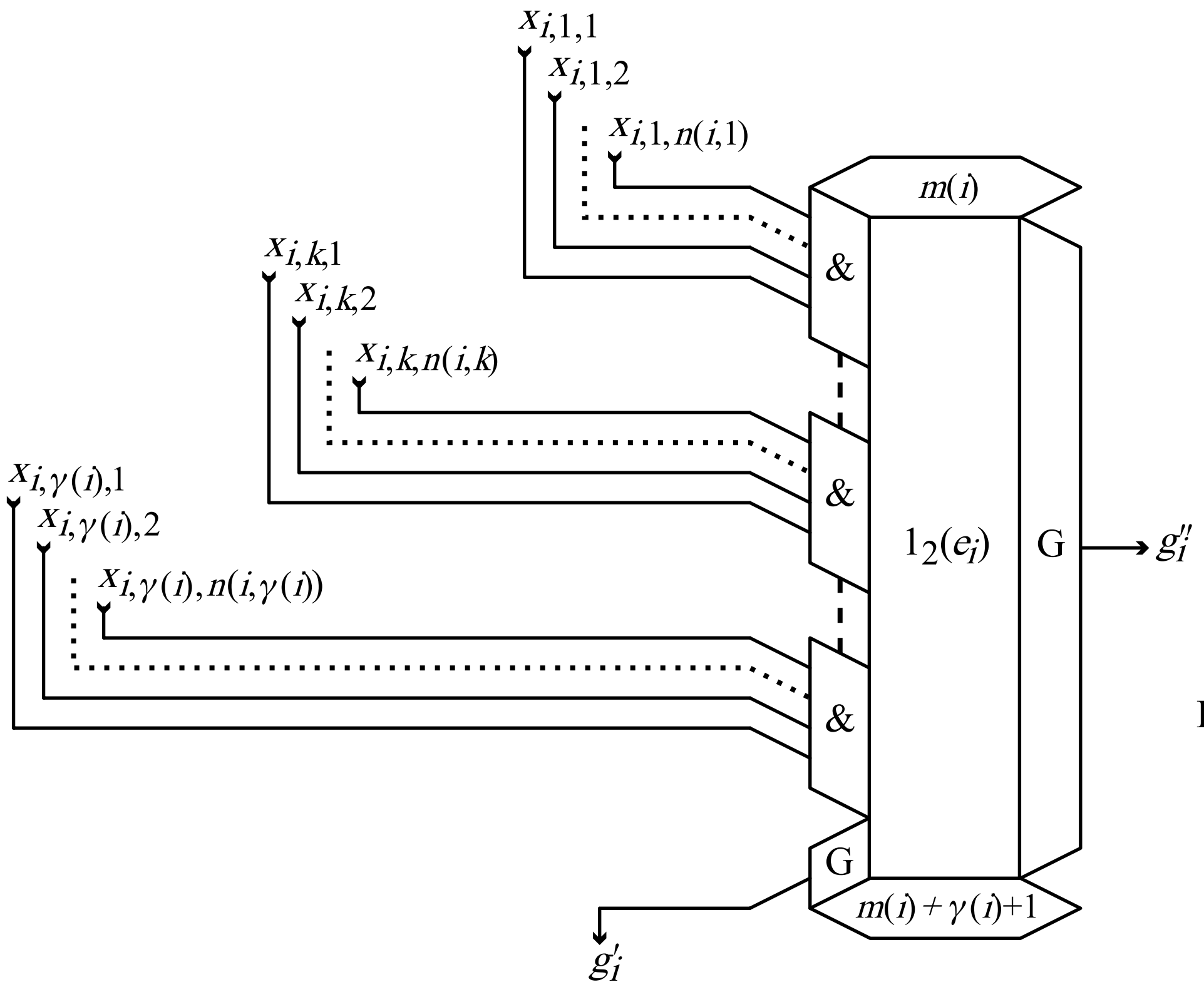

Figure V-2*b*. The special graphical symbol (SGS) for $\Delta$-fragment $F_{\varepsilon(i)}$ which is presented on Figure V-2*a*.

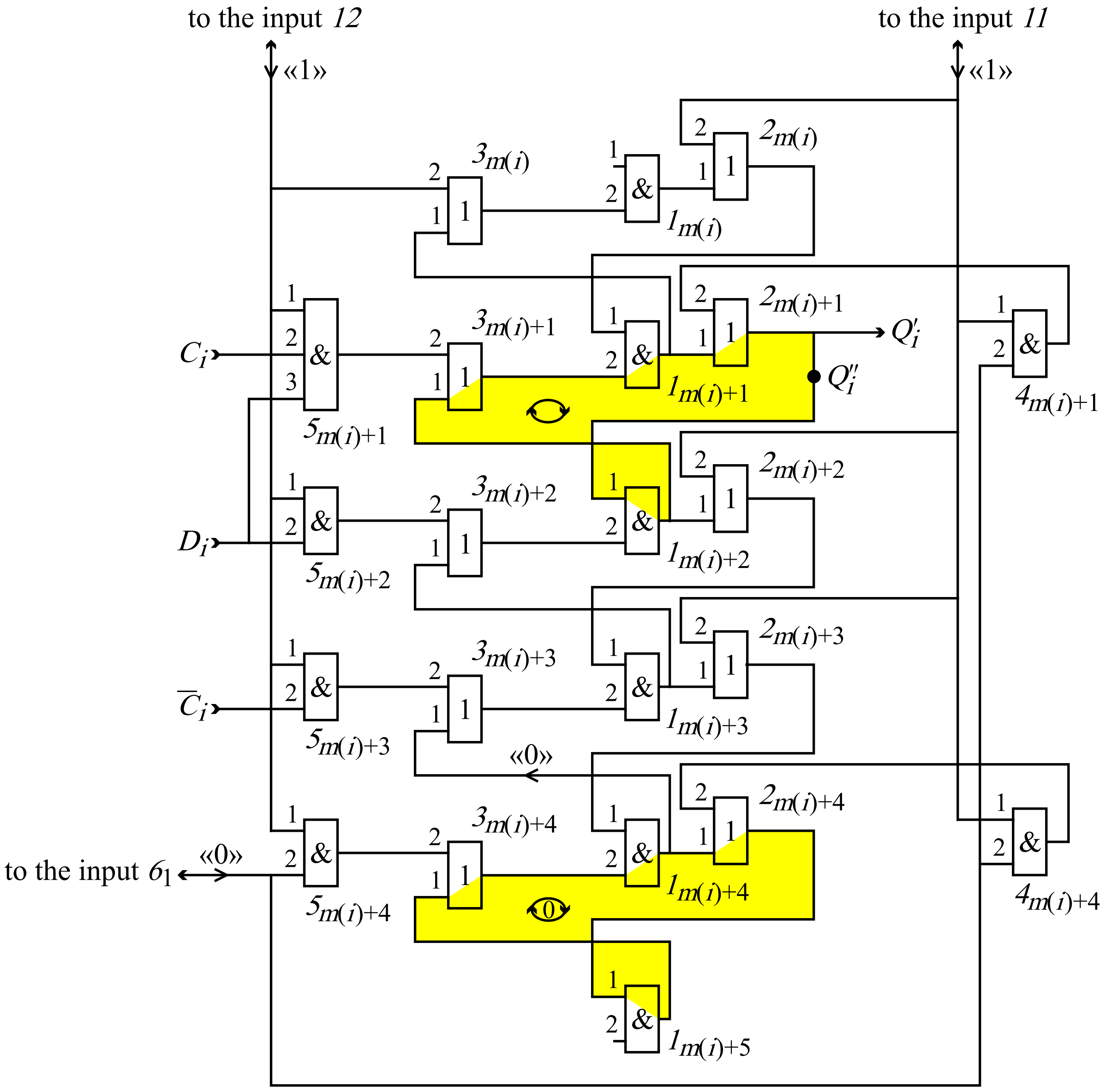

Figure V-3*a*. The *Δ*-fragment $F_{\varepsilon(i)}$ to be used in the quality of elementary unit $e_i$ which is nothing else than a flip-flop, where $i \in \overline{\nu+1,\, 2\nu}$. Logical values shown on the drawing characterize the operating condition of network *Δ*. The loop formed by gates $1_{m(i)+1}$, $2_{m(i)+1}$, $3_{m(i)+1}$, $1_{m(i)+2}$ is used to store an internally-stable state of flip-flop $e_i$. Gates $1_{m(i)+4}$, $2_{m(i)+4}$, $3_{m(i)+4}$, $1_{m(i)+5}$ form the loop staying in the zero internally-stable state during the whole operating condition of network *Δ* to supply the said fragment with the constant of 0.

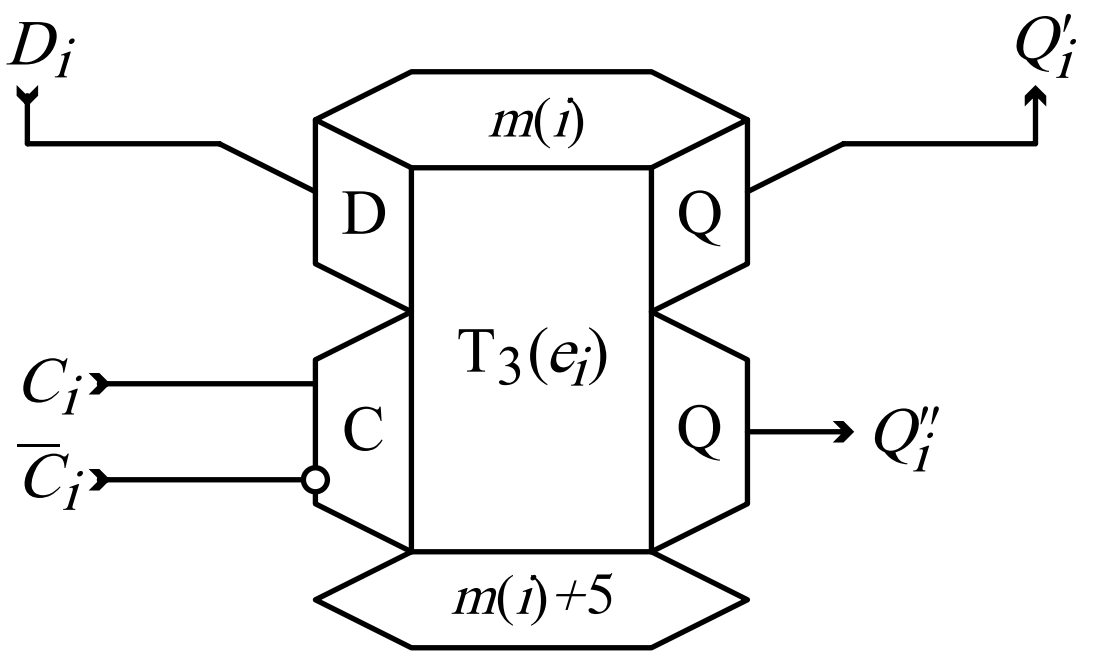

Figure V-3*b*. The special graphical symbol (SGS) for *Δ*-fragment $F_{\varepsilon(i)}$ which is presented on Figure V-3*a*.

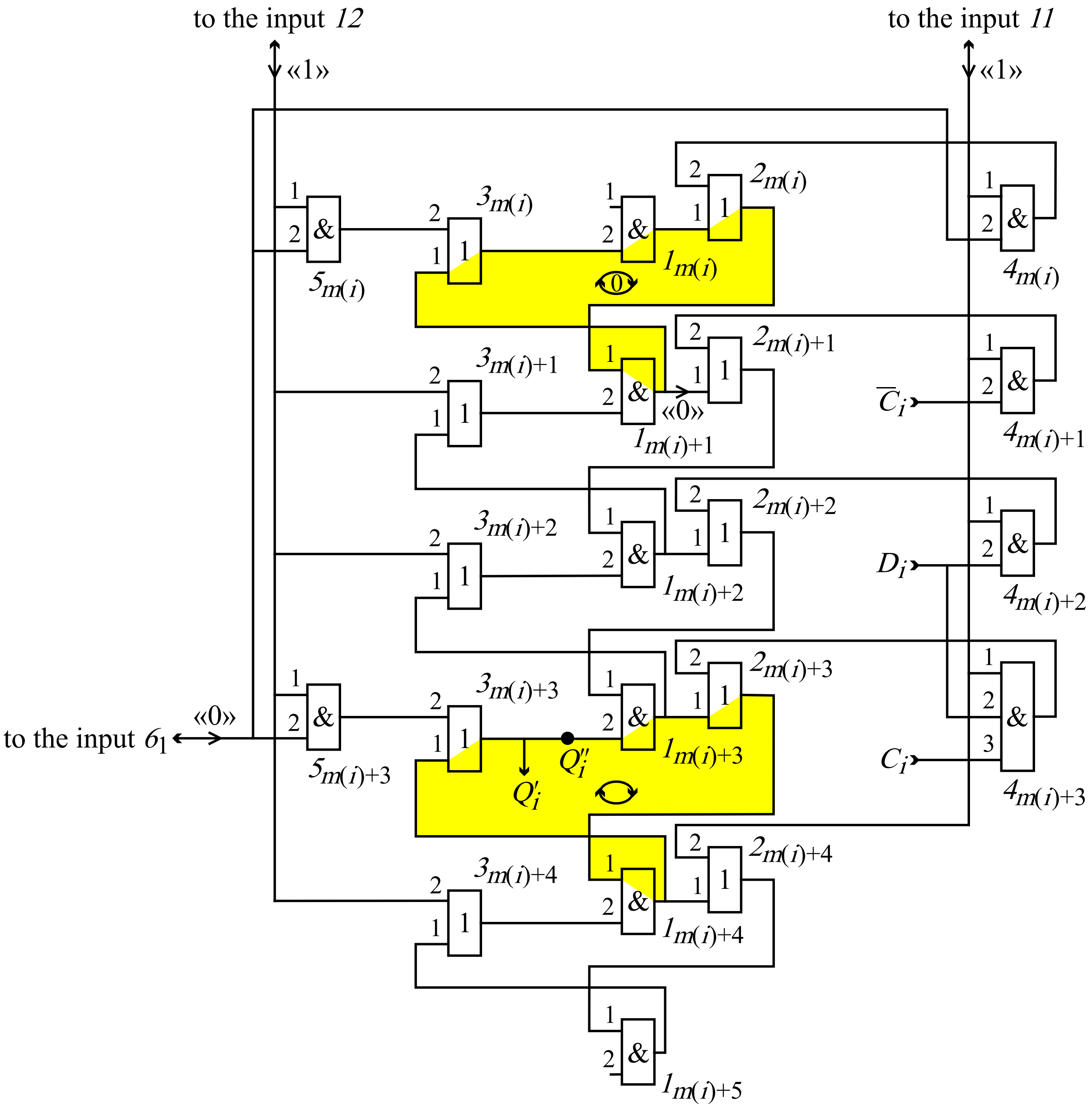

Figure V-4*a*. The $\Delta$-fragment $F_{\varepsilon(i)}$ to be used in the quality of elementary unit $e_i$ which is nothing else than a flip-flop, where $i \in \overline{2\nu+1,\ D}$. Logical values shown on the drawing characterize the operating condition of network $\Delta$. The loop formed by gates $1_{m(i)+3}$, $2_{m(i)+3}$, $3_{m(i)+3}$, $1_{m(i)+4}$ is used to store an internally-stable state of flip-flop $e_i$. Gates $1_{m(i)}$, $2_{m(i)}$, $3_{m(i)}$, $1_{m(i)+1}$ form the loop staying in the zero internally-stable state during the whole operating condition of network $\Delta$ to supply the said fragment with the constant of 0.

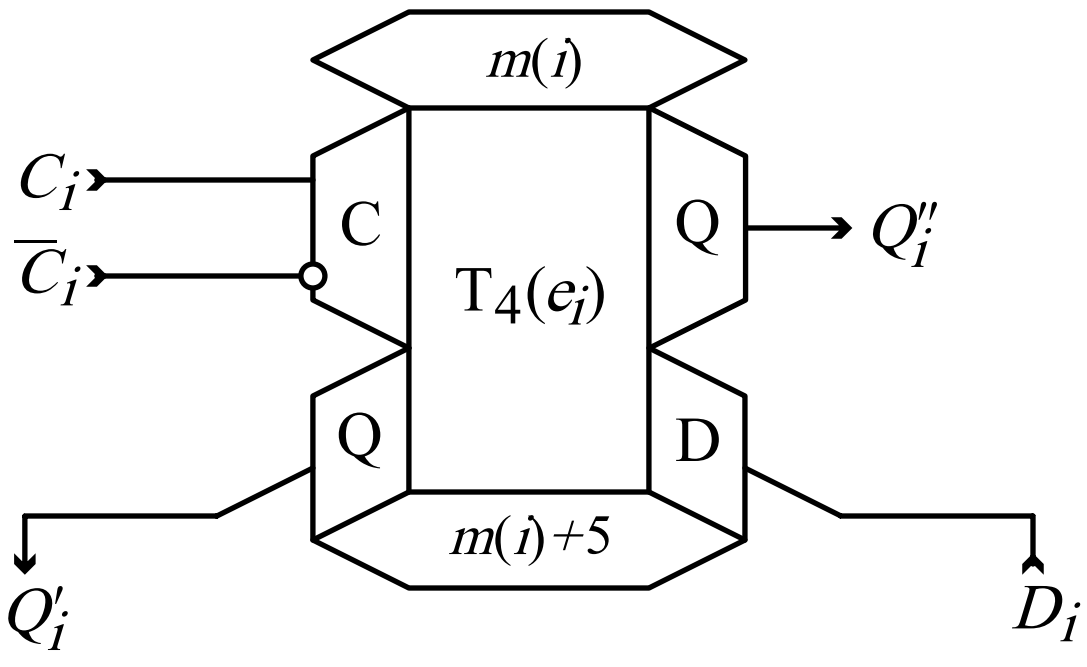

Figure V-4*b*. The special graphical symbol (SGS) for $\Delta$-fragment $F_{\varepsilon(i)}$ which is presented on Figure V-4*a*.

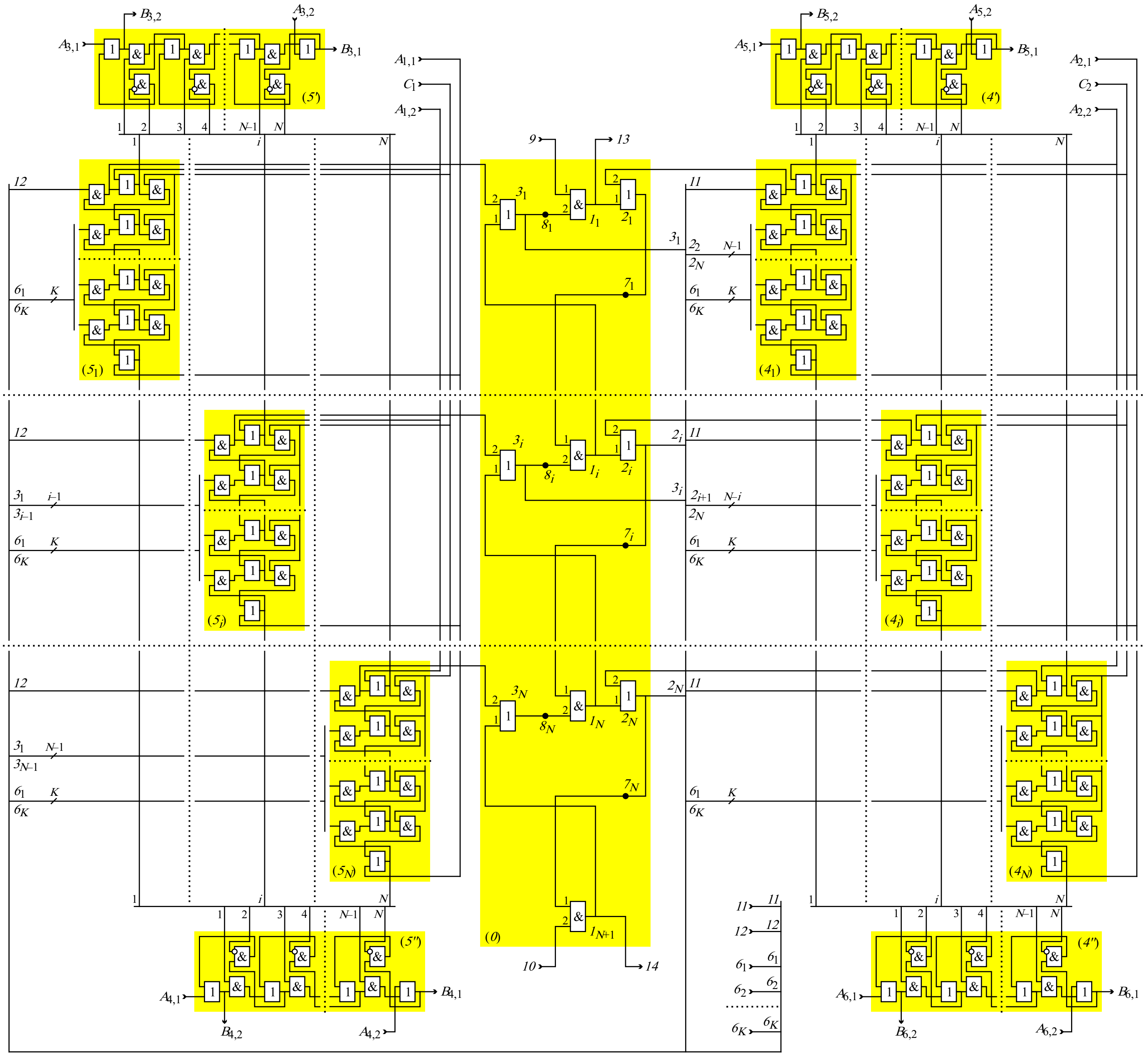

Figure V-5. The network *Δ* constructed by Method V. This network has only the following two distinctions from the network presented on Figure III-1: gates $4_1$ ...$4_N$ are replaced by the system of «long flip-flops» $(4_1)$...$(4_N)$, $(4')$, $(4'')$, and gates $5_1$ ...$5_N$ are replaced by the system of «long flip-flops» $(5_1)$...$(5_N)$, $(5')$, $(5'')$.

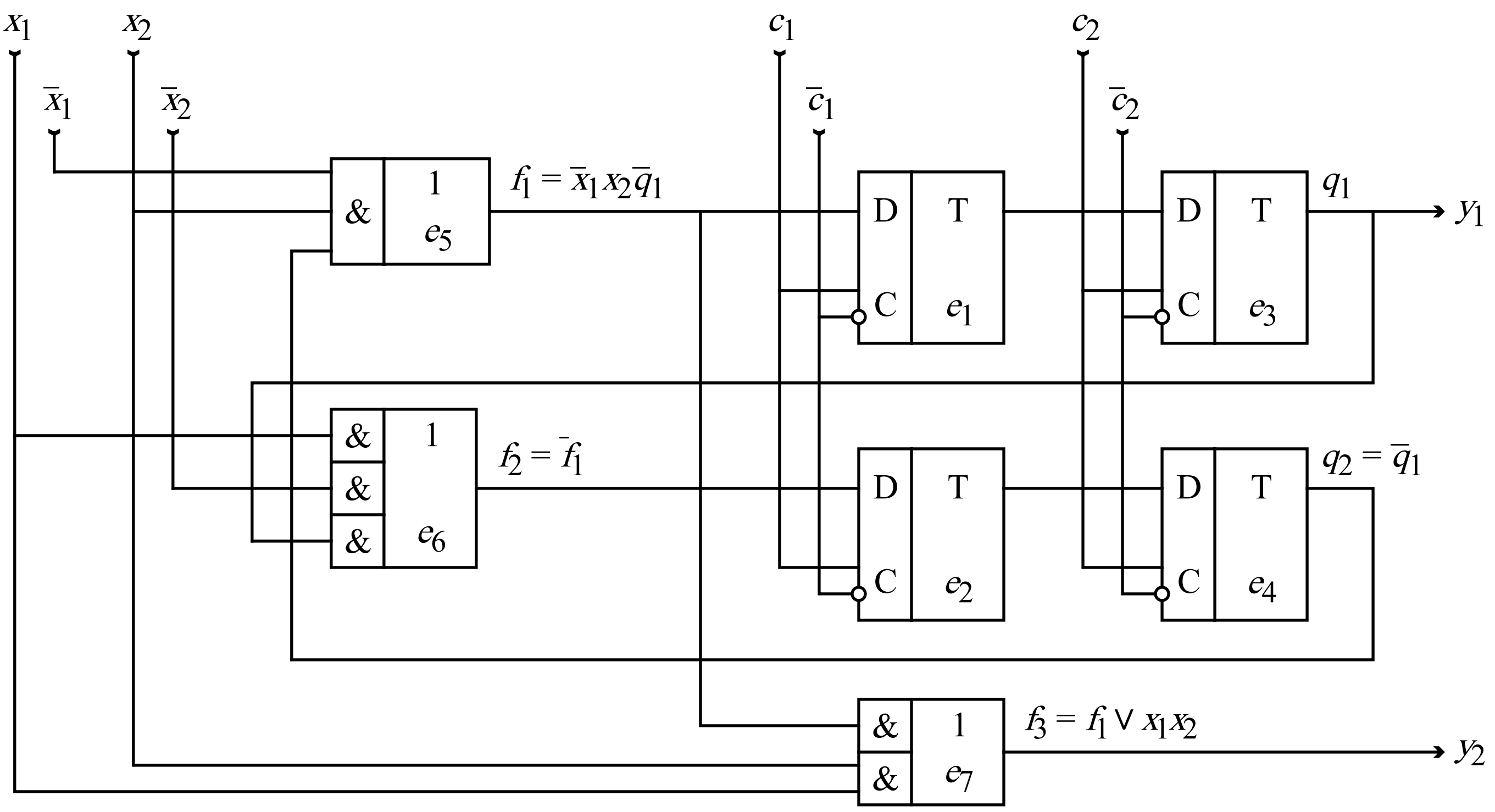

Figure V-6*a*. An example of the $\Delta$-prototype for Method V. The example is obtained from the example presented on Figure IV-8*a* by means of deleting of elements $e_8$ and $e_9$ and of output $y_3$ as a need to form the check bit $\gamma$ disappeared.

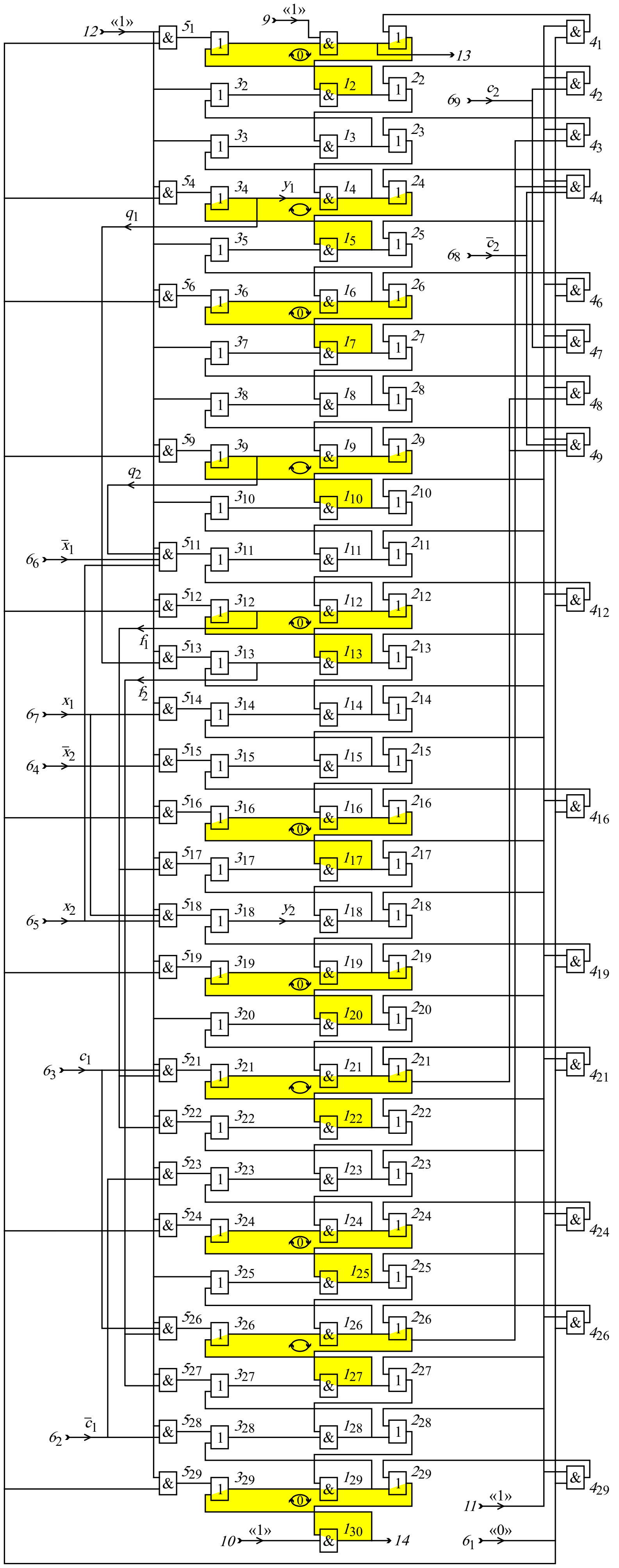

Figure V-6*b*. The network *Δ* which was obtained from the *Δ*-prototype presented on Figure V-6*a*, by Method V. Logical values shown on the drawing characterize the operating condition of the network.

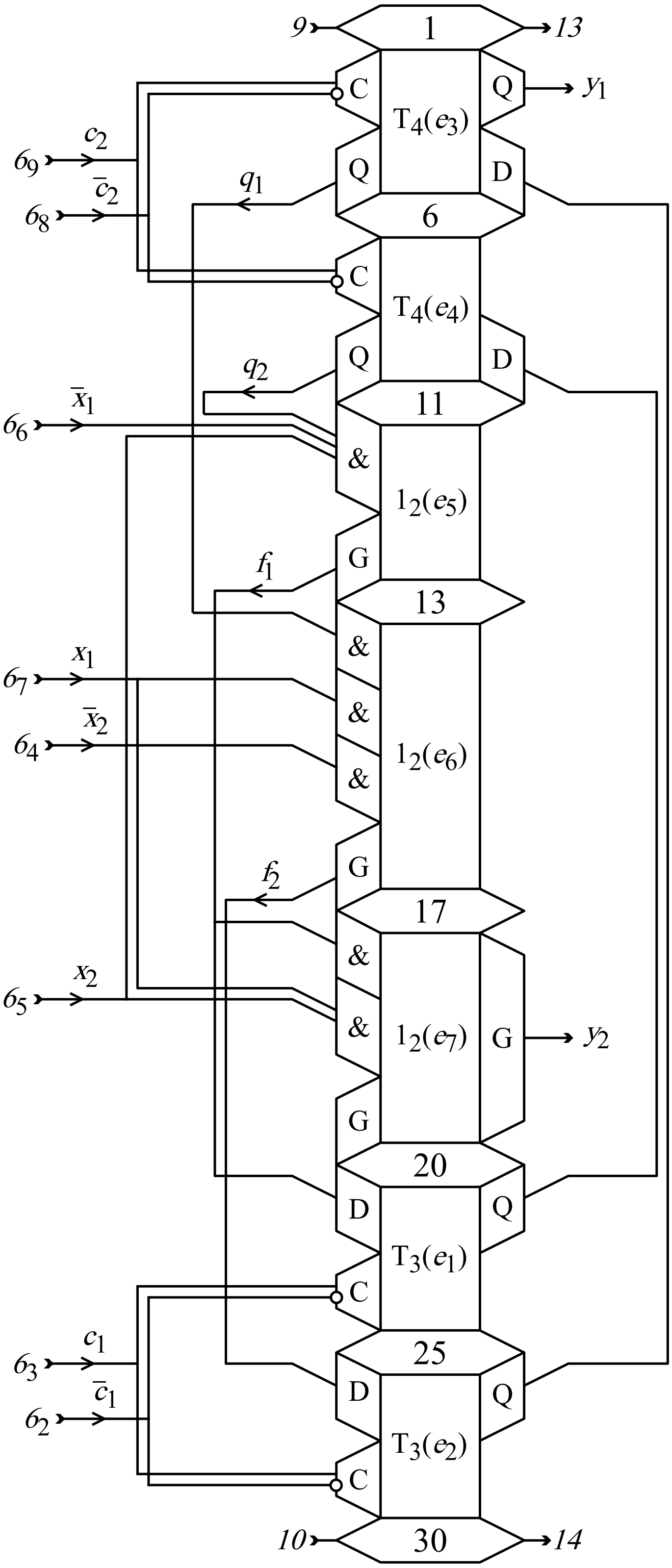

Figure V-6*c*. The network *Δ*, which coincides with the network presented on Figure V-6*b*, but is sketched out with using special graphical symbols (SGS) for *Δ*-fragments $F_{\varepsilon(1)}$ ...$F_{\varepsilon(9)}$ (see Figures V-2*b*, V-3*b*, and V-4*b*). Logical values shown on the drawing characterize the operating condition of network *Δ*.

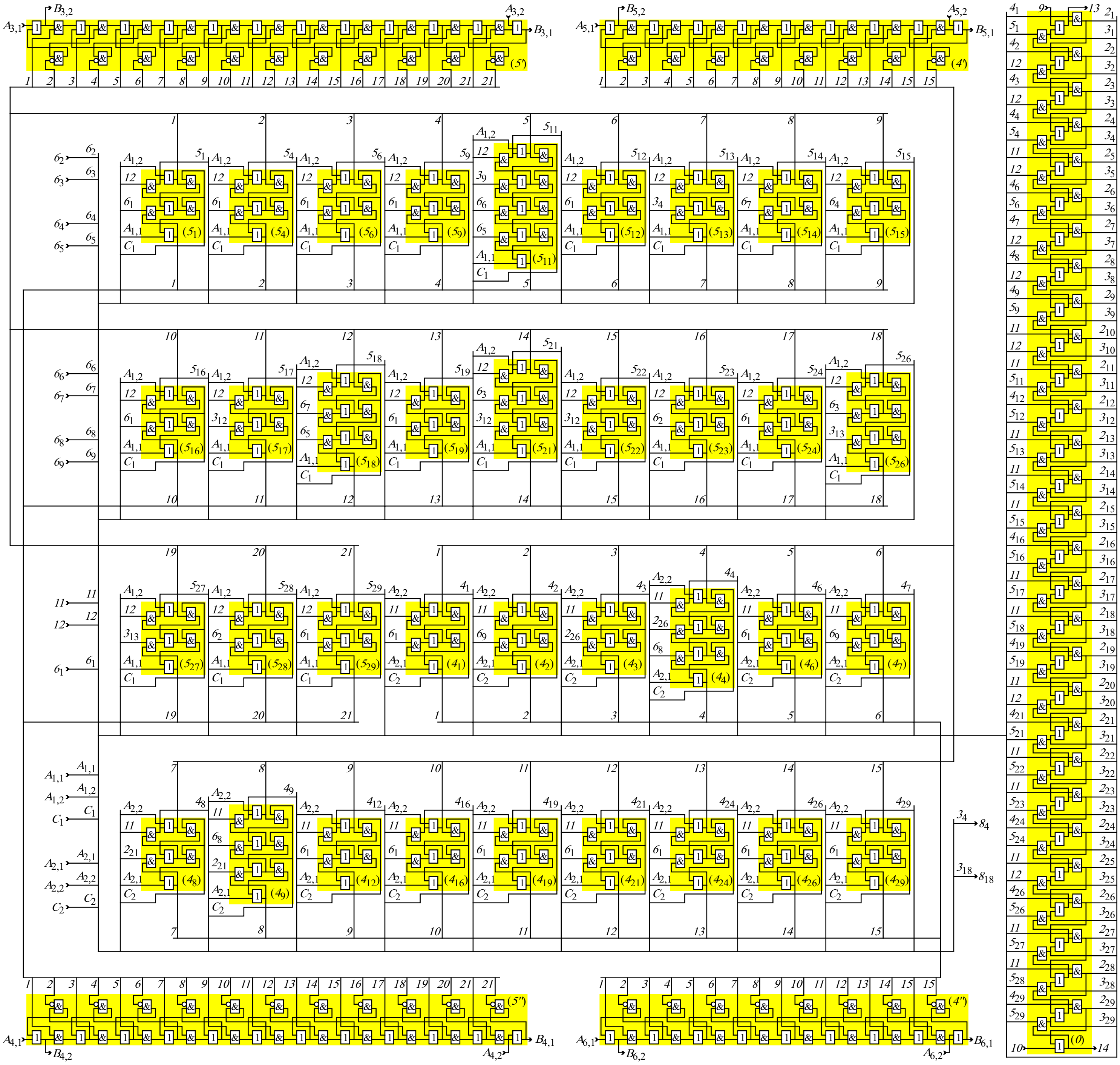

Figure V-6*d*. The network *Δ* which is obtained from *Δ*-prototype presented on Figure V-6*a*, by Method V. This network has the following distinctions from the network presented on Figure V-6*b*: gates $4_1$ ...$4_{29}$ are replaced by the system of «long flip-flops» ($4_1$)...($4_{29}$), ($4'$), ($4''$), and gates $5_1$ ...$5_{29}$ are replaced by the system of «long flip-flops» ($5_1$)...($5_{29}$), ($5'$), ($5''$).

# Conclusion

For the sake of simplification, by *network 1* we´ll mean network *Δ* constructed by Method IV, whereas by *network 2* — network *Δ* constructed by Method V.

## Comparing networks 1 and 2

- 1. When comparing complexity of networks 1 and 2, we see 2 factors which counteract:
    - — during transition from network 1 into network 2 multiinput gates AND of groups *4* and *5* are being replaced by more complex functional analogues, furthermore, in view of impossibility of using inputs 1 of gates AND of group *1*
      for feeding constant «0» in operating condition, elementary units which represent AND-OR and flip-flops become more complex;
    - — network 2 ceases to need redundant coding of output vectors which is required in network 1 for protecting against faults undetected by the test.
- 2. For testing network 1, unlike network 2, there is a need to apply just one neutral vector, namely 11...1, to informational inputs. It is a considerable advantage of network 1 over network 2, because vector 11...1 can be applied by means of switching off the power supply of followers inserted in front of informational inputs of network 1 (see [4, 6, 8]). However, faults «stuck-at-0» of the followers must be detectable while this power supply is switched off. An ability to meet this condition depends on a type of logic used during constructing network 1 (by types of logic, we mean a transistor-transistor logic, an emitter-coupled logic, an n-channel metal-oxide-semiconductor logic, a complementary metal-oxide-semiconductor logic, and so on). For example, using an integrated injection logic ($I^2L$) meets the condition mentioned. Advantages are following:
    - — a network being tested is not to be extracted from the equipment which uses it;
    - — size of the test ceases to depend on complexity of a network being tested, because instead of applying vector 11...1 to informational inputs a logical signal which switches off the power supply of input followers is used.

## Redundancy

Networks 1 and 2 differ from traditional analogues by noticeably greater complexity estimated by the amount of gates: a prototype which is presented on Figure V-6*a* is constructed out of 25 gates, whereas networks 1 and 2 which conform with this prototype have 124 and 272 gates, respectively. One can think that networks 1 and 2 have an unacceptable (giant, incredible) redundancy.

But during an epoch of very-large-scale integration circuits (VLSI), there is a need to estimate redundancy in another manner. Indeed, a time delay, a die size and a power consumption of a typical VLSI chip are being defined mainly by long inner links of the chip.

Short links and gates, even if their amount fundamentally increases, have slight influence in the general case.

Of course all depends on the structure of the chip: if the amount of long internal links is small or they are not long enough, then it be not so.

But usually an influence of long internal links is determinative. In this connexion we can expect that the deciding factor causing redundancy of our decisions with respect to traditional ones represents using of two-rail coding for logical signals. This coding is needed because only gates AND and OR may be used whereas invertors may not. The rest distinctions of our decisions from traditional ones are local (i.e. they cause restricted complication of the average elementary unit) or add structures passing serially through all elementary units (this gives tiny complication for the average elementary unit).

One can see that using a two-rail coding for logical signals on the average increases a die size in 4 times, a time delay in 2 times and a power consumption in 2 times.

In any case we need increasing of chip complexity in a few times (say in 10 times) whatever complexity (including memory size) of automaton being implemented. This complication due to Moor law will be adoptable as early as in a few years.

This circumstance along with the fact that already now industry doesn´t know what to do with great amount of elements that can be placed into a chip (the increasing the amount of cores, of a cache size, of a clock frequency or of a scale of processing according to principle «a single-instruction, multiple-data» gives decreasing effect) cause us to ask ourself: isn´t it time that we have to go in for providing logical devices by new useful properties — by an ability of being fully and simply tested for unfitness or intactness, for example?

But the discussed redundancy is of interest only if we desire the wide use of networks 1 and 2. If we mean using networks 1 and 2 only in the mission-critical equipment, then their redundancy is not important, only achieved level of protecting the equipment against refusals is of real importance.

# A1. Some important properties of direct-current logic networks constructed out of gates AND and OR

The arbitrary element $r$ of arbitrary logic network $q$ will be referred to as $q$-element $r$. For example, an arbitrary $q$-input $x$ means the input $x$ of logic network $q$, an arbitrary $q$-gate $y$ — the logic gate $y$ of logic network $q$, an arbitrary $q$-pole $z$ — the pole $z$ of logic network $q$, and so on.

Let $S$ denotes *the arbitrary correct logical network*. We mean that such network consists of arbitrary $S$-gates, $S$-inputs, $S$-outputs and $S$-links, where

- each $S$-link represents an unidirectional logic connection (in wide sense — a channel for transmitting a logic signal) leading from the source of a logic signal towards the consumer of a logic signal;
- each $S$-link leads or from some $S$-input towards some input of some $S$-gate, or from the output of some $S$-gate towards some input of some $S$-gate, or from the output of some $S$-gate towards some $S$-output;
- there are no two $S$-links leading towards the same $S$-pole;
- a number of $S$-links may lead from the same $S$-pole;
- each $S$-pole represents either any $S$-input, or any $S$-output, or any input of any $S$-gate, or the output of any $S$-gate;
- there are no connections between $S$-poles except those implemented by $S$-links.

Furthermore we´ll regard an arbitrary correct logical network $S$ as *correct direct-current AND-OR-network* $S$, if each $S$-gate is either a direct-current gate AND or a direct-current gate OR, besides

- no $S$-link leads towards any input of any $S$-gate AND from the output of any $S$-gate AND,
- no $S$-link leads towards any input of any $S$-gate OR from the output of any $S$-gate OR,
- no $S$-link leads towards any input of any $S$-gate AND from any $S$-input connected to any input of any $S$-gate OR,
- no $S$-link leads towards any input of any $S$-gate OR from any $S$-input connected to any input of any $S$-gate AND,
- no $S$-link leads towards any input of any $S$-gate AND from any $S$-output connected to the output of any $S$-gate AND,
- no $S$-link leads towards any input of any $S$-gate OR from any $S$-output connected to the output of any $S$-gate OR.

We´ll regard an arbitrary $S$-pole $a'$ as $S$-ancestor of arbitrary $S$-pole $a''$, if either there is $S$-link leading from pole $a''$ to pole $a'$, or there is some such $S$-gate $x$ that pole $a'$ is an input of gate $x$ while pole $a''$ is the output of gate $x$.

By an arbitrary $S$-*path* $L$, we´ll mean such sequence

$$L = a_1 \to a_2 \to \ \dots \to a_n \tag{1}$$

of $S$-poles, that $S$-pole $a_j$ is $S$-ancestor of $S$-pole $a_{j+1}$ for $i = \overline{1,\ n-1}$. Let´s agree that any subsequence $\lambda = a_{j'} \to a_{j'+1} \to \ \dots \to a_{j''}$ in (1) (where $j' \in \overline{1,\ n-1}$, $j'' \in \overline{j',\ n}$) may be replaced by $|\lambda|$. (Obviously, this implies that there is $S$-path $\lambda$ and that $S$-path $\lambda$ is a subpath of $S$-path $L$.) Therefore there can be some $S$-path

$$L = A_1 \to A_2 \to \ \dots \to A_m, \tag{2}$$

where $A_i$ for $i = \overline{1,\ m}$ is either a denotation of a pole or a denotation of a subpath. The right part of equation (1) or (2) will be referred to as *trace* of $S$-path $L$. Let´s agree to put this trace in quotes when using within text (not within formulas).

An arbitrary $S$-gate, whose both some input and the output are poles of some $S$-path $L$, will be regarded as $L$-gate.

All $S$-poles, belonging to some $S$-path $L$, will be regarded as $L$-poles.

We´ll mean that arbitrary $S$-path $L$ *goes within* network $S$.

We´ll mean that arbitrary $S$-path $L$ *leads* from the first $L$-pole towards the last $L$-pole.

We´ll mean that arbitrary $S$-path $L$ *goes* through each $L$-pole, each $L$-gate, and each own subpath.

In case both the first pole and the last pole of some $S$-path $L$ coincide, this path will be regarded as $S$-*loop* $L$. (In fact, each pole of $S$-loop $L$ can be regarded both as first and as last.) We´ll allow an arbitrary $S$-loop $L$ to go through any $L$-pole more than ones.

We´ll regard arbitrary $S$-path $L$ as $C^{\text{mpl}}$-*intact*, if all $L$-poles are $C^{\text{mpl}}$-intact.

As *moment of* $S$-*halt*, we´ll regard such time moment, at which network $S$ has no transients, both in case network $S$ is intact and in case network $S$ has an arbitrary $C^{\text{mpl}}$-fault.

Obviously, if during some time interval $\left[t - \tau_S^{\max},\ t\right]$ (where $\tau_S^{\max}$ is duration of assured settling time of network $S$) all $S$-inputs have constant values, then $t$ is moment of $S$-halt.

Existence of time $\tau_S^{\max}$ is easy to prove, as $S$ is direct-current logic network constructed from gates AND and OR.

We´ll mean that an arbitrary value $x \in \{0,\ 1\}$ goes by some $S$-path $L$ at some moment $t$ of $S$-halt, if all $L$-poles have value $x$ at moment $t$.

We´ll mean that some $S$-pole $x$ is being passed through by some $[t',\ t'']$-step «$x \to y$» (where $t'$ and $t''$ are moments of $S$-halt, $t' < t''$, $x,\ y \in \{0,\ 1\}$ and $x \neq y$), if pole $a$ has the value $x$ at moment $t'$ and the value $y$ at some moment of interval $(t',\ t'']$.

We´ll say that some path $L$, which goes within some correct direct-current AND-OR-network $S$, is being passed through by some $[t',\ t'']$-step «$x \to y$» (where $t'$ and $t''$ are moments of $S$-halt, $t' < t''$, $x,\ y \in \{0,\ 1\}$ and $x \neq y$), if all $L$-poles have the value $x$ at moment $t'$, whereas for each pair of $L$-poles $a'$ and $a''$ (where pole $a'$ is $S$-ancestor of pole $a''$) there is $\exists\, t_{a''} \exists\, t_{a'} \mid (t_{a'} < t_{a''})$, where $t_a$ for $a \in \{a',\ a''\}$ is the earliest of those moments of interval $[t',\ t'']$, at which pole $a$ has the value $y$.

By *a test* of network $S$, we´ll mean a system of rules which define both applying test stimuli towards $S$-inputs and interpreting those responses of network $S$ which appear on $S$-outputs in answer to the test stimuli. As *static*, we´ll regard such test of network $S$, which is defined only at those time moments which represent moments of $S$-halt, and only during those time intervals which are bounded by moments of $S$-halt.

* * *

**LEMMA 1.** Let some loop $L$ go within some correct direct-current AND-OR-network $S$. Let $t'$ and $t''$ (where $t' < t''$) be moments of $S$-halt. Let following conditions be met:

- **i**) each $L$-pole is $C^{\text{mpl}}$-intact,
- **ii**) the output of each $L$-gate AND has value «1» at moment $t'$,
- **iii**) for each such input of each $L$-gate AND, which is not $L$-pole, there are only 2 alternatives:
  - **iii$_1$**) this input keeps constant value during whole interval $[t',\ t'']$,
  - **iii$_2$**) each such $S$-path, which both leads towards this input and can be passed through by

$[t', t'']$-step «$1 \to 0$», goes through $L$-gate.

Then there is true

**Statement** $A_1(L, [t', t''])$: outputs of all $L$-gates keep value «1» during whole interval $[t', t'']$.

A proof.

1. If at moment $t'$ an arbitrary $S$-gate $q$ is in state «1» (state of any gate is defined as value of the output of this gate), then a reason for this $S$-gate to switch into state «0» at some moment $t_q^+ \in (t', t'']$ appears, obviously, during interval $(t', t_q^-]$, where $t_q^- < t_q^+$.

2. All $L$-gates OR at moment $t'$ have the state «1», because

- — the output of each $L$-gate OR is connected to an input of at least one $L$-gate AND,
- — in accordance with item «**ii**», $L$-gates AND have state «1» at moment $t'$,
- — in accordance with item «**i**», all those inputs of $L$-gates AND, which belong to path $L$, are $C^{\text{mpl}}$-intact,
- — $t'$ is a moment of $S$-halt.

Owing to items 1 and 2, for proving lemma 1 we must only show, that no reason for any $L$-gate to switch into state «0» appears during whole interval $(t', t'')$.

Let us assume the opposite. Then some such $L$-gate $x$ exists, for which a reason to switch into state «0» appears during interval $(t', t_x^-] \subseteq (t', t'')$, whereas for no $L$-gate a reason to switch into state «0» appears during interval $(t', t_x^-)$. There are 2 alternatives:

- — gate $x$ is gate OR. Let pole $y$ be such input of gate $x$ which is $L$-pole. Loop $L$ goes within a correct AND-OR-network, therefore pole $y$ is connected to the output of some such $L$-gate $z$ which is a gate AND. In accordance with item «**i**», pole $y$ is $C^{\text{mpl}}$-intact. So gate $x$ undoubtedly keeps state «1» during whole interval $[t', t_z^+]$, where $t_z^+$ is the first such moment of interval $[t', t'']$ at which gate $z$ is in state «0». We see that $t' < t_z^- < t_z^+ < t_x^-$, therefore $t_z^- \in (t', t_x^-)$, but this contradicts definition of gate $x$;
- — gate $x$ is a gate AND. Let $Y_x$ denote the set of those $L$-gates whose outputs are connected to those inputs of gate $x$, which are $L$-poles. We see that no gate $y \in Y_x$ is in state «0» during interval $(t', t_x^-)$, otherwise $t' < t_y^- < t_y^+ < t_x^-$, entailing $t_y^- \in (t', t_x^-)$ in contradiction to definition of gate $x$. Taking into account this fact, as well as item «**iii₁**», we see that only passing of $[t', t'']$-step «$1 \to 0$» through such path, which goes through some $L$-gate $z$, can serve as reason for gate $x$ to switch into state «0» at some moment $t_x^+ \in (t', t'']$. It means $t' < t_z^- < t_z^+ < t_x^-$, but this entails $t_z^- \in (t', t_x^-)$ in contradiction to definition of gate $x$.

The lemma is proved.

* * *

**LEMMA 2.** Let some loop $L$ go within some correct direct-current AND-OR-network $S$. Let $t'$ and $t''$ (where $t' < t''$) be moments of $S$-halt. Let following conditions be met:

- **i**) each $L$-pole is $C^{\text{mpl}}$-intact;
- **ii**) the output of each $L$-gate OR has value «0» at moment $t'$;
- **iii**) for each such input of each $L$-gate OR, which is not $L$-pole, there are only 2 alternatives:
  - **iii$_1$**) this input keeps constant value during whole interval $[t', t'']$;
  - **iii$_2$**) each such $S$-path, which both leads towards this input and can be passed through by

$[t', t'']$-step «$0 \to 1$», goes through $L$-gate.

Then there is true

**Statement** $\Lambda_2(L, [t', t''])$: outputs of all $L$-gates keep value «0» during whole interval $[t', t'']$.

One can easily see that Lemma 2 is obtained from Lemma 1 by means of a complementary transforming:

- gates AND and OR are referred to instead of gates OR and AND, respectively;
- values «0» and «1» are referred to instead of values «1» and «0», respectively, including cases when they appear in denotations of $[t', t'']$-steps and stuck-at faults.

This complementary transforming can be also used to obtain *a proof of Lemma 2* from the proof of Lemma 1.

# A2. Some definitions relating towards «widened long flip-flop»

---

## I.

Let´s denote by

$$1 : 1_x \overset{1,\,2}{\longrightarrow} \mathrm{O} : 1_y \qquad (1)$$

for case $x = \overline{1,\ N+1}$ and $y = \overline{x,\ N+1}$ the path which leads from input 1 of gate $1_x$ towards output of gate $1_y$ and goes through no other gates except $1_1 \ldots 1_{N+1}$ and $2_1 \ldots 2_N$.

One can easily see that such path is single and in case $x = y$ this path goes only through such poles as the input 1 and the output of gate $1_x$, whereas in case $x < y$ this path goes only through such poles as the input 1 and the output of each of gates $1_x \ldots 1_y$ and $2_x \ldots 2_{y-1}$, and as outputs $7_x \ldots 7_{y-1}$.

> F o r  e x a m p l e , $1 : 1_2 \overset{1,\,2}{\longrightarrow} \mathrm{O} : 1_4$ denotes path «The input 1 of gate $1_2$ → The output of gate $1_2$ → The input 1 of gate $2_2$ → The output of gate $2_2$ → The output $7_2$ → The input 1 of gate $1_3$ → The output of gate $1_3$ → The input 1 of gate $2_3$ → The output of gate $2_3$ → The output $7_3$ → The input 1 of gate $1_4$ → The output of gate $1_4$».
>
> Moreover, $1 : 1_3 \overset{1,\,2}{\longrightarrow} \mathrm{O} : 1_3$ denotes path «The input 1 of gate $1_3$ → The output of gate $1_3$».

## II.

Let´s denote by

$$2 : 1_y \overset{1,\,3}{\rightarrow} \mathrm{O} : 1_x \qquad (2)$$

for case $x = \overline{1,\ N+1}$ and $y = \overline{x,\ N+1}$ the path which leads from input 2 of gate $1_y$ towards output of gate $1_x$ and goes through no other gates except $1_1 \ldots 1_{N+1}$ and $3_1 \ldots 3_N$.

One can easily see that such path is single and in case $x = y$ this path goes only through such poles as input 2 and the output of gate $1_x$, whereas in case $x < y$ this path goes only through such poles as input 2 and the output of each of gates $1_x \ldots 1_y$, as the input 1 and the output of each of gates $3_x \ldots 3_{y-1}$, and as outputs $8_x \ldots 8_{y-1}$.

> F o r  e x a m p l e , $2 : 1_4 \overset{1,\,3}{\rightarrow} \mathrm{O} : 1_2$ denotes path «The input 2 of gate $1_4$ → The output of gate $1_4$ → The input 1 of gate $3_3$ → The output of gate $3_3$ → The output $8_3$ → The input 2 of gate $1_3$ → The output of gate $1_3$ → The input 1 of gate $3_2$ → The output of gate $3_2$ → The output $8_2$ → The input 2 of gate $1_2$ → The output of gate $1_2$».
>
> Moreover, $1 : 1_3 \overset{1,\,3}{\rightarrow} \mathrm{O} : 1_3$ denotes path «The input 2 of gate $1_3$ → The output of gate $1_3$».

## III.

Let´s denote by

$$\mathrm{O} : 3_x \overset{5}{\rightarrow} 2 : 3_y \qquad (3)$$

for case $x = \overline{1,\ N–1}$ and $y = \overline{x+1,\ N}$ the path which leads from the output of gate $3_x$ towards the output 2 of gate $3_y$ and goes through no other gates except $5_1 \ldots 5_N$.

One can easily see that such path if exists represents path «The output of gate $3_x$ → Some input of gate $5_y$ → The output of gate $5_y$ → The input 2 of gate $3_y$».

## IV.

Let´s denote by

$$\text{O} : 2_y \xrightarrow{4} 2 : 2_x \quad (4)$$

for case $x = \overline{1, N-1}$ and $y = \overline{x+1, N}$ the path which leads from the output of gate $2_y$ towards the output 2 of gate $2_x$ and goes through no other gates except $4_1 \ldots 4_N$.

One can easily see that such path if exists represents path «The output of gate $2_y$ → Some input of gate $4_x$ → The output of gate $4_x$ → The input 2 of gate $2_x$».

As is easy to see, in each of expressions (1)—(4) there is «stenographic» information about the first pole of denoted path (see before long arrow), about the last pole of denoted path (see behind long arrow) and about group or groups of those gates (see above long arrow) which can be passed through by denoted path.

> In expressions (1)—(4) $1 : 1_x$ denotes pole «The input 1 of gate $1_x$»; $\text{O} : 1_y$ — pole «The output of gate $1_y$»; $2 : 1_y$ — pole «The input 2 of gate $1_y$»; $\text{O} : 1_x$ — pole «The output of gate $1_x$»; $\text{O} : 3_x$ — pole «The output of gate $3_x$»; $2 : 3_y$ — pole «The input 2 of gate $3_y$»; $\text{O} : 2_y$ — pole «The output of gate $2_y$»; $2 : 2_x$ — pole «The input 2 of gate $2_x$».

## V.

Let´s denote by name

$$\Psi_x$$

for case $x = \overline{1, N}$ the loop (in other words — closed path), which goes only through such poles as input 1 and the output of each of gates $1_{x+1}$, $2_x$ and $3_x$, as the input 2 and the output of gate $1_x$, and as outputs $7_x$ and $8_x$.

> For example, $\Psi_3$ denotes loop «The input 2 of gate $1_3$ → The output of gate $1_3$ → The input 1 of gate $2_3$ → The output of gate $2_3$ → The output $7_3$ → The input 1 of gate $1_4$ → The output of gate $1_4$ → The input 1 of gate $3_3$ → The output of gate $3_3$ → The output $8_3$ → The input 2 of gate $1_3$».

## VI.

Let´s denote by

$$T_\Delta$$

a static test of network $\Delta$ presented in Table III-1.

# A3. Proof of Theorem 2

---

We´ll divide the proof into steps **I—XIV**.

## I.

Let´s prove

> Statement $B_0(x)$: if path $1: 1_1 \overset{1,2}{\to} O: 1_x$ is $C^{mpl}$-intact, then $[t_4, t_5]$-step «$1 \to 0$» goes by path $2: 1_x \overset{1,3}{\to} O: 1_1$ for each $x \in \overline{1, N+1}$.

We´ll use an induction for $\gamma = \overline{1, x{-}1}$. *The inductive assumption:* $[t_4, t_5]$-step «$1 \to 0$» goes by path $2: 1_\gamma \overset{1,3}{\to} O: 1_1$. *The base of the induction:* the inductive assumption is true for case $\gamma = 1$ owing to items 3°, 5° and 8°. *The consequence from the inductive assumption:* only following 3 alternatives are possible:

- a) the value «0» remains on input 1 of gate $3_\gamma$ during whole interval $[t_4, t_5]$ whereas $[t_4, t_5]$-step «$1 \to 0$» goes by path «The input 2 of gate $3_\gamma \to$ The output of gate $3_\gamma \to$ The output $8_\gamma \to$ $|2: 1_\gamma \overset{1,3}{\to} O: 1_1|$»;
- b) the value «1» remains on input 2 of gate $1_{\gamma+1}$ during whole interval $[t_4, t_5]$ whereas $[t_4, t_5]$-step «$1 \to 0$» goes by path «The input 1 of gate $1_{\gamma+1} \to$ The output of gate $1_{\gamma+1} \to$ The input 1 of gate $3_\gamma \to$ The output of gate $3_\gamma \to$ The output $8_\gamma \to |2: 1_\gamma \overset{1,3}{\to} O: 1_1|$»;
- c) $[t_4, t_5]$-step «$1 \to 0$» goes by path $2: 1_{\gamma+1} \overset{1,3}{\to} O: 1_1$.

In case «*a*» we have 2 remarks:

- 1) during whole interval $[t_4, t_6]$ a constant value remains on input 1 of gate $5_\gamma$ (owing to item 4°) and also on each such input of gate $5_\gamma$ which is connected to one of inputs $6_1 \dots 6_K$ (owing to item 3°);
- 2) for case $\gamma = 1$ the alternative «*a*» is impossible, whereas for case $\gamma > 1$ there is such path $O: 3_j \overset{5}{\to} 2: 3_\gamma$ (where $j \in \overline{1, \gamma{-}1}$) which is passed through by $[t_4, t_5]$-step «$1 \to 0$» (This results from remark 1.).

In this connexion let´s denote by $L_{«a»}$ for case $\gamma > 1$ the loop which goes through each pole of each of loops $\Psi_1 \dots \Psi_{\gamma-1}$ (see A2) and goes through no other poles. We see that in accordance with Lemma 1 conditions of true of statement $A_1(L_{«a»}, [t_4, t_5])$ are met:

- condition «**i**» — because poles of path $1: 1_1 \overset{1,2}{\to} O: 1_x$ are $C^{mpl}$-intact (according to the condition of true of statement $B_0(x)$) and poles of path $2: 1_\gamma \overset{1,3}{\to} O: 1_1$ are $C^{mpl}$-intact (according to the inductive assumption),
- condition «**ii**» — because outputs of gates $1_1 \dots 1_\gamma$ have value «1» at moment $t_4$ (This results from the inductive assumption.),
- condition «**iii**» — because
  - — during whole interval $[t_4, t_5]$ the input 1 of gate $1_1$ keeps a constant value (owing to item

5°) and those inputs of gates $1_1 \dots 1_\gamma$, which are connected to any of inputs $6_1 \dots 6_K$, keep constant values (owing to item 3°),

- — any such path from input *10* towards input 2 of gate $1_\gamma$, which can be passed through by $[t_4,\ t_5]$-step «$1 \to 0$», passes through some $L_{«a»}$-gate $q \in \{3_1,\ 3_2,\ \dots,\ 3_{\gamma-1}\}$ (This results from remarks 1 and 2.).

In accordance with statement $A_1(L_{«a»},\ [t_4,\ t_5])$, the output of gate $1_\gamma$ keeps a constant value during whole interval $[t_4,\ t_5]$, but this contradicts the inductive assumption. I.e. the alternative «*a*» is impossible also in case $\gamma > 1$.

In case of the alternative «*b*», let´s denote by $L_{«b»}$ the loop, which goes through each pole of each of loops $\Psi_1 \dots \Psi_\gamma$ (see A2) and goes through no other poles. We see that in accordance with Lemma 1 conditions of true of statement $A_1(L_{«b»},\ [t_4,\ t_5])$ are met:

- condition «**i**» — because poles of path $1 : 1_1 \overset{1,\,2}{\longrightarrow} O : 1_x$ are $C^{mpl}$-intact (according to the condition of true of statement $B_0(x)$) and poles of path «The input 1 of gate $1_{\gamma+1}$ → The output of gate $1_{\gamma+1}$ → The input 1 of gate $3_\gamma$ → The output of gate $3_\gamma$ → The output $8_\gamma$ → $|2 : 1_\gamma \overset{1,\,3}{\longrightarrow} O : 1_1|$» are $C^{mpl}$-intact (according to the definition of alternative «*b*»),
- condition «**ii**» — because outputs of gates $1_1 \dots 1_{\gamma+1}$ have value «1» at moment $t_4$ (This results from the definition of alternative «*b*».),
- condition «**iii**» — because during whole interval $[t_4,\ t_5]$ the input 1 of gate $1_1$ keeps a constant value (owing to item 5°), the input 2 of gate $1_{\gamma+1}$ keeps a constant value (according to definition of alternative «*b*»), and those inputs of gates $1_1 \dots 1_{\gamma+1}$, which are connected to any of inputs $6_1 \dots 6_K$, keep constant values (owing to item 3°).

In accordance with statement $A_1(L_{«b»},\ [t_4,\ t_5])$, the output of gate $1_\gamma$ keeps a constant value during whole interval $[t_4,\ t_5]$, but this contradicts the inductive assumption. Therefore alternative «*b*» is impossible. Only alternative «*c*» remains possible, i.e. *the induction is confirmed.*

## II.

Let´s prove

> S t a t e m e n t $B_1$: there is no such $\alpha \in \overline{1,\ N}$ that $[t_2,\ t_3]$-step «$0 \to 1$» goes through the output of gate $3_\alpha$.

Let´s assume the opposite. Let´s denote by $Q$ the set, whose each element is some such $\alpha \in \overline{1,\ N}$ that $[t_2,\ t_3]$-step «$0 \to 1$» goes through output of gate $3_\alpha$. Let

$$\hat{\alpha} = \min_{\alpha \in Q}(\alpha).$$

We have 3 remarks:

- 1) during whole interval $[t_2,\ t_3]$ each input of gate $5_{\hat{\alpha}}$ keeps a constant value (the input 1 — owing to item 4°, each input connected to any of inputs $6_1 \dots 6_K$ — owing to item 3°, and each input connected to the output of any gate $3_x$ for $x \in \overline{1,\ \hat{\alpha}-1}$ — according to the definition of $\hat{\alpha}$);
- 2) $[t_2,\ t_3]$-step «$0 \to 1$» goes by path «$|1 : 1_1 \overset{1,\,2}{\longrightarrow} O : 1_{\hat{\alpha}+1}|$ → The input 1 of gate $3_{\hat{\alpha}}$ → The output of gate $3_{\hat{\alpha}}$ » (This results from remark 1 and from the impossibility of going of $[t_2,\ t_3]$-step «$0 \to 1$» through outputs of gates $3_1 \dots 3_{\hat{\alpha}-1}$ in accordance with the definition of $\hat{\alpha}$.);
- 3) $[t_4,\ t_5]$-step «$1 \to 0$» goes by path $2 : 1_{\hat{\alpha}+1} \overset{1,\,3}{\longrightarrow} O : 1_1$ (This results from statement $B_0(\hat{\alpha}+1)$, which is true because path $1: 1_1 \xrightarrow{1,2} \text{O}: 1_{\hat{\alpha}+1}$ is $C^{\text{mpl}}$-intact in accordance with remark 2.).

In this connexion, we see that according to Lemma 2 conditions of true of statement $A_2(\Psi_{\hat{\alpha}}, [t_2, t_3])$ are met:

- condition «**i**» — because input 2 of gate $1_{\hat{\alpha}}$ is $C^{\text{mpl}}$-intact (according to remark 3) and other poles of loop $\Psi_{\hat{\alpha}}$ (see A2) are $C^{\text{mpl}}$-intact (according to the remark 2),
- condition «**ii**» — because outputs of gates $2_{\hat{\alpha}}$ and $3_{\hat{\alpha}}$ have value «0» at moment $t_2$ (This results from remark 2.),
- condition «**iii**» — because
  - — during whole interval $[t_2, t_3]$ the input 2 of gate $3_{\hat{\alpha}}$ keeps a constant value (This results from remark 1.),
  - — any possible path, being passed through by $[t_2, t_3]$-step «$0 \to 1$» from the input $9$ towards input 2 of gate $2_{\hat{\alpha}}$, passes through $\Psi_{\hat{\alpha}}$-gate $1_{\hat{\alpha}}$ (This results from the impossibility of going this step through outputs of gates $3_1 \ldots 3_{\hat{\alpha}}$ in accordance with the definition of $\hat{\alpha}$.).

In accordance with statement $A_2(\Psi_{\hat{\alpha}}, [t_2, t_3])$, the output of gate $3_{\hat{\alpha}}$ keeps a constant value during whole interval $[t_2, t_3]$, but this contradicts the definition of $\hat{\alpha}$. Therefore $\hat{\alpha}$ doesn´t exist, i.e. $Q = \varnothing$.

## III.

Let´s prove

> S t a t e m e n t $B_2$: $[t_2, t_3]$-step «$0 \to 1$» goes by path $1: 1_1 \xrightarrow{1,2} \text{O}: 1_{N+1}$.

Let´s use an induction for $x = N+1, N, \ldots, 2$. *The inductive assumption:* $[t_2, t_3]$-step «$0 \to 1$» goes by path $1: 1_x \xrightarrow{1,2} \text{O}: 1_{N+1}$. *The base of the induction:* the inductive assumption is true for case $x = N+1$ owing to items 3°, 6° and 7°. *The consequence from the inductive assumption:* there are only 3 alternatives:

- a) during whole interval $[t_2, t_3]$ the input 1 of gate $2_{x-1}$ keeps value «0», whereas $[t_2, t_3]$-step «$0 \to 1$» goes by path «The input 2 of gate $2_{x-1}$ → The output of gate $2_{x-1}$ → The output $7_{x-1}$ → $|1: 1_x \xrightarrow{1,2} \text{O}: 1_{N+1}|$»;
- b) during whole interval $[t_2, t_3]$ the input 1 of gate $1_{x-1}$ keeps value «1», whereas $[t_2, t_3]$-step «$0 \to 1$» goes by path «The output of gate $3_{x-1}$ → The output $8_{x-1}$ → The input 2 of gate $1_{x-1}$ → The output of gate $1_{x-1}$ → The input 1 of gate $2_{x-1}$ → The output of gate $2_{x-1}$ → Output $7_{x-1}$ → $|1: 1_x \xrightarrow{1,2} \text{O}: 1_{N+1}|$»;
- c) $[t_2, t_3]$-step «$0 \to 1$» goes by path $1: 1_{x-1} \xrightarrow{1,2} \text{O}: 1_{N+1}$.

In case of alternative «*a*» we have 2 remarks:

- 1) during whole interval $[t_2, t_3]$ the input 1 of gate $4_{x-1}$ keeps a constant value (owing to item 4°) and each such input of gate $4_{x-1}$, which is connected to one of inputs $6_1 \ldots 6_K$, keeps a constant value (owing to item 3°);
- 2) for case $x = N+1$ the alternative «*a*» is impossible, whereas for case $x \le N$ there is some path $\text{O}: 2_\varepsilon \xrightarrow{4} 2: 2_{x-1}$ (where $\varepsilon \in \overline{x, N}$) being passed through by $[t_2, t_3]$-step «$0 \to 1$» (This results from remark 1.).

In this connexion let´s denote by $L$ for case $x \le N$ the loop «The input 2 of gate $2_{x-1}$ → The output of gate

$2_{x-1}$ → The output $7_{x-1}$ → $|1 : 1_x \xrightarrow{1,\,2} \mathrm{O} : 1_\varepsilon|$ → The input 1 of gate $2_\varepsilon$ → $|\mathrm{O} : 2_\varepsilon \xrightarrow{4} 2 : 2_{x-1}|$».

We see that in accordance with Lemma 2 conditions of true of statement $A_2(L,\ [t_2,\ t_3])$ are met:

- condition «**i**» — because poles of path «The input 2 of gate $2_{x-1}$ → The output of gate $2_{x-1}$ → The output $7_{x-1}$ → $|1 : 1_{x+1} \xrightarrow{1,\,2} \mathrm{O} : 1_{N+1}|$» are $C^{\mathrm{mpl}}$-intact (according to the definition of alternative «*a*») and poles of path $|\mathrm{O} : 2_\varepsilon \xrightarrow{4} 2 : 2_{x-1}|$ are $C^{\mathrm{mpl}}$-intact (according to remark 2),
- condition «**ii**» — because outputs of gates $2_{x-1}\ ...2_\varepsilon$ have value «0» at moment $t_2$ (This results from the definition of alternative «*a*».),
- condition «**iii**» — because
  - — the input 1 of gate $2_{x-1}$ keeps a constant value during whole interval $[t_2,\ t_3]$ (This results from the definition of alternative «*a*».),
  - — any such path from the input *9* towards input 2 of any of gates $2_x\ ...2_\varepsilon$, which is being passed through by $[t_2,\ t_3]$-step «$0 \to 1$», goes through $L$-gate $2_{x-1}$ (This results from the impossibility of going of this step through outputs of gates $3_1\ ...3_N$ in accordance with statement $B_1$.).

In accordance with statement $A_2(L,\ [t_2,\ t_3])$, the output of gate $1_x$ keeps a constant value during whole interval $[t_2,\ t_3]$, but this contradicts the inductive assumption. Therefore alternative «*a*» is impossible also in case $x \leq N$.

The alternative «*b*» is impossible in accordance with statement $B_1$.

Only the alternative «*c*» remains possible, i.e. *the induction is confirmed.*

## IV.

Statement $B_2$ directly entails

- — statement $B_3$: path $1 : 1_1 \xrightarrow{1,\,2} \mathrm{O} : 1_{N+1}$ is $C^{\mathrm{mpl}}$-intact;
- — statement $B_4$: the input *9* is $C^{\mathrm{mpl}}$-intact.

Statement $B_3$, in particular, entails

- — statement $B_5$: inputs 1 and outputs of gates $1_1\ ...1_{N+1}$ and $2_1\ ...2_N$, as well as outputs $7_1\ ...7_N$, are $C^{\mathrm{mpl}}$-intact;
- — statement $B_6$: inputs 2 of gates $2_1\ ...2_N$ don´t have faults « $\equiv 1$»;
- — statement $B_7$: each input $y \in \overline{3,\ \infty}$ of each gate $x \in \{1_1,\ 1_2,\ ...,\ 1_{N+1}\}$ doesn´t have fault « $\equiv 0$»;
- — statement $B_8$: each such input $x \in \{6_1,\ 6_2,\ ...,\ 6_K\}$, which is connected to $C^{\mathrm{mpl}}$-intact input of any of gate $1_1\ ...1_{N+1}$, doesn´t have fault « $\equiv 0$»;
- — statement $B_9$: for each $x = \overline{1,\ N}$ the output of gate $4_x$ doesn´t have fault « $\equiv 1$» in case the input 2 of gate $2_x$ is $C^{\mathrm{mpl}}$-intact.

Item 7° directly entails

Statement $B_{10}$: the output *14* is $C^{\mathrm{mpl}}$-intact.

In accordance with statement $B_3$ the path $1 : 1_1 \xrightarrow{1,\,2} \mathrm{O} : 1_{N+1}$ is $C^{\mathrm{mpl}}$-intact, therefore statement $B_0(N+1)$

directly entails

> S t a t e m e n t $B_{11}$: $[t_4,\ t_5]$-step «$1 \to 0$» goes by path $2:\ 1_{N+1} \overset{1,\,3}{\to} \mathrm{O}:\ 1_1$.

Statement $B_{11}$ directly entails

> - — s t a t e m e n t $B_{12}$: path $2:\ 1_{N+1} \overset{1,\,3}{\to} \mathrm{O}:\ 1_1$ is $C^{\mathrm{mpl}}$-intact;
> - — s t a t e m e n t $B_{13}$: the input *10* is $C^{\mathrm{mpl}}$-intact.

Statement $B_{12}$, in particular, entails

> - — s t a t e m e n t $B_{14}$: inputs 2 of gates $1_1 \ldots 1_{N+1}$, inputs 1 and outputs of gates $3_1 \ldots 3_N$, as well as outputs $8_1 \ldots 8_N$, are $C^{\mathrm{mpl}}$-intact;
> - — s t a t e m e n t $B_{15}$: inputs 2 of gates $3_1 \ldots 3_N$ don´t have faults «$\equiv 1$»;
> - — s t a t e m e n t $B_{16}$: for each $y = \overline{1,\ N}$ the output of gate $5_y$ doesn´t have fault «$\equiv 1$» in case the input 2 of gate $3_y$ is $C^{\mathrm{mpl}}$-intact.

Item 8° directly entails

> S t a t e m e n t $B_{17}$: the output *13* is $C^{\mathrm{mpl}}$-intact.

Statements $B_3$ and $B_{12}$ directly entail

> S t a t e m e n t $B_{18}$: loops $\Psi_1 \ldots \Psi_N$ are $C^{\mathrm{mpl}}$-intact.

## V.

Let´s prove

> S t a t e m e n t $B_{19}$: there is no such $\eta \in \overline{1,\ N}$ that $[t_5,\ t_6]$-step «$0 \to 1$» goes through the output of gate $2_\eta$.

Let´s assume the opposite. Let´s denote by $Q$ the set whose each element is some such $\eta \in \overline{1,\ N}$ that $[t_4,\ t_5]$-step «$0 \to 1$» goes through the output of gate $2_\eta$. Let

$$\hat{\eta} = \max_{\eta \in Q}(\eta).$$

We have 2 remarks:

- 1) during whole interval $[t_5,\ t_6]$ each input of gate $4_{\hat{\eta}}$ keeps a constant value (the input 1 — owing to item 4°, each such input which is connected to any of inputs $6_1 \ldots 6_K$ — owing to item 3°, and each such input which is connected to the output of some gate $2_x$ (where $x \in \overline{\hat{\eta}+1,\ N}$) — in accordance with the definition of $\hat{\eta}$);
- 2) $[t_5,\ t_6]$-step «$0 \to 1$» goes by path «$|2:\ 1_{N+1} \overset{1,\,3}{\to} \mathrm{O}:\ 1_{\hat{\eta}}| \to$ The input 1 of gate $2_{\hat{\eta}} \to$ The output of gate $2_{\hat{\eta}}$» (This results from remark 1 and from the impossibility of going of $[t_5,\ t_6]$-step «$0 \to 1$» through outputs of gates $2_{\hat{\eta}+1} \ldots 2_N$ in accordance with definition of $\hat{\eta}$.).

We see that in accordance with Lemma 2 conditions of true of statement $A_2\left(\Psi_{\hat{\eta}},\ [t_5,\ t_6]\right)$ are met:

- condition «**i**» — because loop $\Psi_{\hat{\eta}}$ (see A2) is $C^{\mathrm{mpl}}$-intact (This results from statement $B_{18}$.),
- condition «**ii**» — because outputs of gates $2_{\hat{\eta}}$ and $3_{\hat{\eta}}$ have value «0» at moment $t_5$ (This results from remark 2.),
- condition «**iii**» — because

- ○ — during whole interval $[t_5,\ t_6]$ the input 2 of gate $2_{\hat\eta}$ keeps a constant value (This results from remark 1.),
- ○ — any such path from the input *10* towards the input 2 of gate $3_{\hat\eta}$, which is being passed through by $[t_5,\ t_6]$-step «$0 \to 1$», goes through $\Psi_{\hat\eta}$-gate $1_{\hat\eta+1}$ (This results from the impossibility of going of this step through outputs of gates $2_{\hat\eta+1} \ldots 2_N$ in accordance with definition $\hat\eta$.).

According to statement $\Lambda_2\left(\Psi_{\hat\eta},\ [t_5,\ t_6]\right)$, the output of gate $2_{\hat\eta}$ keeps a constant value during whole interval $[t_5,\ t_6]$, but this contradicts the definition of $\hat\eta$. Therefore $\hat\eta$ doesn´t exist, i.e. $Q = \varnothing$.

## VI.

Let´s prove

> S t a t e m e n t $B_{20}$: $[t_1,\ t_6]$-step «$0 \to 1$» goes by path $2 : 1_{N+1} \overset{1,\,3}{\to} \mathrm{O} : 1_1$.

Let´s use an induction for $\gamma = \overline{1,\ N}$. *The inductive assumption:* $[t_5,\ t_6]$-step «$0 \to 1$» goes by path $2 : 1_\gamma \overset{1,\,3}{\to} \mathrm{O} : 1_1$. *The base of the induction:* the inductive assumption is true for case $\gamma = 1$ owing to items 3°, 5° and 8°. *The consequence from the inductive assumption:* only 3 alternatives are possible:

- a) during whole interval $[t_5,\ t_6]$ the input 1 of gate $3_\gamma$ keeps value «0», whereas $[t_5,\ t_6]$-step «$0 \to 1$» goes by path «The input 2 of gate $3_\gamma \to$ The output of gate $3_\gamma \to$ The output $8_\gamma \to$ $|2 : 1_\gamma \overset{1,\,3}{\to} \mathrm{O} : 1_1|$»;
- b) during whole interval $[t_5,\ t_6]$ the input 2 of gate $1_{\gamma+1}$ keeps value «1», whereas $[t_5,\ t_6]$-step «$0 \to 1$» goes by path «The output of gate $2_\gamma \to$ The output $7_\gamma \to$ The input 1 of gate $1_{\gamma+1} \to$ The output of gate $1_{\gamma+1} \to$ The input 1 of gate $3_\gamma \to$ The output of gate $3_\gamma \to$ The output $8_\gamma \to$ $|2 : 1_\gamma \overset{1,\,3}{\to} \mathrm{O} : 1_1|$»;
- c) $[t_5,\ t_6]$-step «$0 \to 1$» goes by path $|2 : 1_{\gamma+1} \overset{1,\,3}{\to} \mathrm{O} : 1_1|$.

In case of alternative «*a*» we have 2 remarks:

- 1) during whole interval $[t_5,\ t_6]$ the input 1 of gate $5_\gamma$ keeps a constant value (owing to item 4°) and each such input of gate $5_\gamma$, which is connected to any of inputs $6_1 \ldots 6_K$, keeps a constant value (owing to item 3°);
- 2) for case $\gamma = 1$ the alternative «*a*» is impossible, whereas for case $\gamma > 1$ there is some path $\mathrm{O} : 3_x \overset{5}{\to} 2 : 3_\gamma$ (where $x \in \overline{1,\ \gamma{-}1}$) which is being passed through by $[t_5,\ t_6]$-step «$0 \to 1$» (This results from remark 1.).

In this connexion let´s denote by $L$ for case $\gamma > 1$ the loop «The input 2 of gate $3_\gamma \to$ The output of gate $3_\gamma \to$ The output $8_\gamma \to |2 : 1_\gamma \overset{1,\,3}{\to} \mathrm{O} : 1_x| \to |\mathrm{O} : 3_x \overset{5}{\to} 2 : 3_\gamma|$».

We see, that in accordance with Lemma 2 conditions of true of statement $\Lambda_2(L,\ [t_5,\ t_6])$ are met:

- condition «**i**» — because poles of path «The output of gate $3_\gamma \to$ The output $8_\gamma \to$ $|2 : 1_\gamma \overset{1,\,3}{\to} \mathrm{O} : 1_1|$» are $C^{\mathrm{mpl}}$-intact (according to statement $B_{17}$) and poles of path $\mathrm{O} : 3_x \overset{5}{\to} 2 : 3_\gamma$ are $C^{\mathrm{mpl}}$-intact (according to remark 2),
- condition «**ii**» — because outputs of gates $3_\gamma \ldots 3_x$ have value «0» at moment $t_5$ (This results from the inductive assumption.),
- condition «**iii**» — because

- — the input 1 of gate $3_\gamma$ keeps a constant value during whole interval $[t_5,\ t_6]$ (This results from the definition of alternative «*a*».),
- — any such path from input *10* towards input 2 of any of gates $3_{\gamma-1}$ ...$3_x$, which is being passed through by $[t_5,\ t_6]$-step «$0 \to 1$», goes through $L$-gate $3_\gamma$ (This results from the impossibility of going of this step through outputs of gates $2_1$ ...$2_N$ in accordance with statement $B_{19}$.).

In accordance with statement $A_2(L,\ [t_5,\ t_6])$, the output of gate $1_\gamma$ keeps a constant value during whole interval $[t_5,\ t_6]$, but this contradicts the inductive assumption. Therefore alternative «*a*» is impossible also in case $\gamma > 1$.

The alternative «*b*» is impossible according to statement $B_{19}$.

Only alternative «*c*» remains possible, i.e. *the induction is confirmed.*

## VII.

Let´s prove

> S t a t e m e n t $B_{21}$: $[t_1,\ t_2]$-step «$1 \to 0$» goes by path $1 : 1_1 \overset{1,\,2}{\longrightarrow} O : 1_{N+1}$.

Let´s use an induction for $\gamma = N+1,\ N,\ \ ...,\ 2$. *The inductive assumption:* $[t_1,\ t_2]$-step «$1 \to 0$» goes by path $1 : 1_\gamma \overset{1,\,2}{\longrightarrow} O : 1_{N+1}$. *The base of the induction:* the inductive assumption is true for case $\gamma = N+1$ owing to items 3°, 6° and 7°. *The consequence from the inductive assumption:* there are only 3 alternatives:

- a) during whole interval $[t_1,\ t_2]$ the input 1 of gate $2_{\gamma-1}$ keeps value «0», whereas $[t_1,\ t_2]$-step «$1 \to 0$» goes by path «The input 2 of gate $2_{\gamma-1}$ → The output of gate $2_{\gamma-1}$ → The output $7_{\gamma-1}$ → $|1 : 1_\gamma \overset{1,\,2}{\longrightarrow} O : 1_{N+1}|$»;
- b) during whole interval $[t_1,\ t_2]$ the input 1 of gate $1_{\gamma-1}$ keeps value «1», whereas $[t_1,\ t_2]$-step «$1 \to 0$» goes by path «The input 2 of gate $1_{\gamma-1}$ → The output of gate $1_{\gamma-1}$ → The input 1 of gate $2_{\gamma-1}$ → The output of gate $2_{\gamma-1}$ → The output $7_{\gamma-1}$ → $|1 : 1_\gamma \overset{1,\,2}{\longrightarrow} O : 1_{N+1}|$»;
- c) $[t_1,\ t_2]$-step «$1 \to 0$» goes by path $1 : 1_{\gamma-1} \overset{1,\,2}{\longrightarrow} O : 1_{N+1}$.

In case of alternative «*a*» we have 2 remarks:

- 1) during whole interval $[t_1,\ t_2]$ the input 1 of gate $4_{\gamma-1}$ keeps a constant value (owing to item 4°) and each such input of gate $4_{\gamma-1}$, which is connected to one of inputs $6_1$ ...$6_K$, keeps a constant value (owing to item 3°);
- 2) the alternative «*a*» is impossible for $\gamma = N$, whereas for $\gamma < N$ there is some such path $O : 2_j \overset{4}{\to} 2 : 2_{\gamma-1}$ (where $j \in \overline{\gamma,\ N}$) which is being passed through by $[t_1,\ t_2]$-step «$1 \to 0$» (This results from remark 1.).

In this connexion, let´s denote by $L_{«a»}$ for case $\gamma < N$ the loop which goes through each pole of each of loops $\Psi_\gamma \ldots \Psi_N$ (see A2) and goes through no other poles. We see that according to Lemma 1 conditions of true of statement $A_1(L_{«a»},\ [t_1,\ t_2])$ are met:

- condition «**i**» — in accordance with statement $B_{18}$,
- condition «**ii**» — because outputs of gates $1_\gamma \ldots 1_{N+1}$ have value «1» at moment $t_1$ (This results from the inductive assumption.),
- condition «**iii**» — because

- ○ — during whole interval $[t_1,\ t_2]$ the input 2 of gate $1_{N+1}$ keeps a constant value (owing to item 6°) and those inputs of gates $1_\gamma \ldots 1_{N+1}$, which are connected with any of inputs $6_1 \ldots 6_K$, keep constant values (owing to item 3°),
- ○ — any such path from the input $9$ towards input 1 of gate $1_\gamma$, which is being passed by $[t_1,\ t_2]$-step «$1 \to 0$», goes through some $L_{«a»}$-gate $q \in \{2_\gamma,\ 2_{\gamma+1},\ \ldots,\ 2_N\}$ (This results from remarks 1 and 2.).

In accordance with statement $A_1(L_{«a»},\ [t_1,\ t_2])$, the output of gate $1_{N+1}$ keeps a constant value during whole interval $[t_1,\ t_2]$, but this contradicts the inductive assumption. Therefore the alternative «*a*» is impossible also in case $\gamma < N$.

In case of the alternative «*b*» let´s denote by $L_{«b»}$ the loop which goes through each pole of each of loops $\Psi_{\gamma-1} \ldots \Psi_N$ (see A2) and goes through no other poles. We see that in accordance with Lemma 1 conditions of true of statement $A_1(L_{«b»},\ [t_1,\ t_2])$ are met:

- condition «**i**» — in accordance with statement $B_{18}$,
- condition «**ii**» — because outputs of gates $1_{\gamma-1} \ldots 1_{N+1}$ have value «1» at moment $t_1$ (This results from the definition of alternative «*b*».),
- condition «**iii**» — because during whole interval $[t_1,\ t_2]$ the input 2 of gate $1_{N+1}$ keeps a constant value (owing to item 6°), the input 1 of gate $1_{\gamma-1}$ keeps a constant value (according to the definition of alternative «*b*»), and each such input of any gate $q \in \{1_{\gamma-1},\ 1_\gamma,\ \ldots,\ 1_{N+1}\}$, which is connected to any of inputs $6_1 \ldots 6_K$, keeps a constant value (owing to item 3°).

In accordance with statement $A_1(L_{«b»},\ [t_1,\ t_2])$, the output of gate $1_{N+1}$ keeps a constant value during whole interval $[t_1,\ t_2]$, but this contradicts the inductive assumption. Therefore the alternative «*b*» is impossible.

Only the alternative «*c*» remains possible, i.e. *the induction is confirmed.*

## VIII.

Let´s prove

> S t a t e m e n t $B_{22}$: there is no such $\chi \in \overline{1,\ N}$ that the input 2 of gate $3_\chi$ keeps value «0» during whole interval $[t_2,\ t_3]$.

Let´s assume the opposite.Statement $B_{18}$ entails that path «The output of gate $1_{\chi+1}$ → The input 1 of gate $3_\chi$ → The output of gate $3_\chi$» is $C^{mpl}$-intact, whereas statement $B_2$ entails that $[t_2,\ t_3]$-step «$0 \to 1$» goes through output of gate $1_{\chi+1}$. Therefore our assumption entails that $[t_2,\ t_3]$-step «$0 \to 1$» goes through output of gate $3_\chi$, but this contradicts statement $B_1$.

## IX.

Let´s prove

> S t a t e m e n t $B_{23}$: there is no such $\nu \in \overline{1,\ N}$ that the input 2 of gate $2_\nu$ keeps value «0» during whole interval $[t_5,\ t_6]$.

Let´s assume the opposite. Statement $B_{18}$ entails that path «The output of gate $1_\nu$ → The input 1 of gate $2_\nu$ → The output of gate $2_\nu$» is $C^{mpl}$-intact, whereas statement $B_{20}$ entails that $[t_5,\ t_6]$-step «$0 \to 1$» goes through output of gate $1_\nu$. Therefore our assumption entails that $[t_5,\ t_6]$-step «$0 \to 1$» goes through output of gate $2_\nu$, but this contradicts statement $B_{19}$.

## X.

Statement $B_{23}$ directly entails

> Statement $B_{24}$: inputs 2 of gates $2_1 \ldots 2_N$ don't have faults «$\equiv 0$».

Statements $B_6$ and $B_{24}$ directly entail

> Statement $B_{25}$: inputs 2 of gates $2_1 \ldots 2_N$ are $C^{mpl}$-intact.

Statements $B_9$ and $B_{25}$ directly entail

> Statement $B_{26}$: the outputs of gates $4_1 \ldots 4_N$ don't have faults «$\equiv 1$».

Statements $B_{23}$ and $B_{25}$ directly entail

> Statement $B_{27}$: the outputs of gates $4_1 \ldots 4_N$ don't have faults «$\equiv 0$».

Statements $B_{26}$ and $B_{27}$ directly entail

> Statement $B_{28}$: the outputs of gates $4_1 \ldots 4_N$ are $C^{mpl}$-intact.

Statements $B_{23}$, $B_{25}$ and $B_{28}$ directly entail

> - — statement $B_{29}$: the inputs of gates $4_1 \ldots 4_N$ don't have faults «$\equiv 0$»,
> - — statement $B_{30}$: each such input $x \in \{6_1, 6_2, \ldots, 6_K, 11\}$, which is connected to any $C^{mpl}$-intact input of any of gates $4_1 \ldots 4_N$, doesn't have fault «$\equiv 0$».

Statement $B_{22}$ directly entails

> Statement $B_{31}$: inputs 2 of gates $3_1 \ldots 3_N$ don't have faults «$\equiv 0$».

Statements $B_{15}$ and $B_{31}$ directly entail

> Statement $B_{32}$: inputs 2 of gates $3_1 \ldots 3_N$ are $C^{mpl}$-intact.

Statements $B_{16}$ and $B_{32}$ directly entail

> Statement $B_{33}$: the outputs of gates $5_1 \ldots 5_N$ don't have faults «$\equiv 1$».

Statements $B_{22}$ and $B_{31}$ directly entail

> Statement $B_{34}$: the outputs of gates $5_1 \ldots 5_N$ don't have faults «$\equiv 0$».

Statements $B_{33}$ and $B_{34}$ directly entail

> Statement $B_{35}$: outputs of gates $5_1 \ldots 5_N$ are $C^{mpl}$-intact.

Statements $B_{22}$, $B_{31}$ and $B_{35}$ directly entail

> - — statement $B_{36}$: the inputs of gates $5_1 \ldots 5_N$ don't have faults «$\equiv 0$»,
> - — statement $B_{37}$: each such input $x \in \{6_1, 6_2, \ldots, 6_K, 12\}$, which is connected to any $C^{mpl}$-intact input of any of gates $5_1 \ldots 5_N$, doesn't have fault «$\equiv 0$».

Statements $B_4$, $B_5$, $B_7$, $B_8$, $B_{10}$, $B_{13}$, $B_{14}$, $B_{17}$, $B_{25}$, $B_{28} \ldots B_{30}$, $B_{32}$, $B_{35} \ldots B_{37}$ directly entail

> Statement $B_{38}$ : the set of those elementary stuck-at-faults, which are present in network $\Delta$, may include only such elements as:
>
> - — the fault «$\equiv 1$» of some input $x \in \{6_1, 6_2, \ldots, 6_K, 11, 12\}$,

- — the fault « $\equiv 1$» of some input of some gate $x \in \{4_1, 4_2, ..., 4_N, 5_1, 5_2, ..., 5_N\}$,
- — the fault « $\equiv 1$» of some input $y \in \overline{3, \infty}$ of some gate $x \in \{1_1, 1_2, ..., 1_{N+1}\}$,
- — the fault « $\equiv 0$» of some input $x \in \{6_1, 6_2, ..., 6_K\}$, if there is no such $y \in \{1_1, 1_2, ..., N+1, 4_1, 4_2, ..., 4_N, 5_1, 5_2, ..., 5_N\}$ that the input $x$ is connected with $C^{\text{mpl}}$-intact input of gate $y$,
- — the fault « $\equiv 0$» of the input $11$, if each input 1 of each of gates $4_1 ... 4_N$ has fault « $\equiv 1$»,
- — the fault « $\equiv 0$» of the input $12$, if each input 1 of each of gates $5_1 ... 5_N$ has fault « $\equiv 1$».

Statements $B_{23}$, $B_{25}$ and $B_{28}$ directly entail

> S t a t e m e n t $B_{39}$: each such $C^{\text{mpl}}$-intact input $x \in \{6_1, 6_2, ..., 6_K\}$, which is connected to any $C^{\text{mpl}}$-intact input of any of gates $4_1 ... 4_N$, has value «1» at some moment of interval $[t_5, t_6]$.

Statement $B_{39}$ and item 3° directly entail

> S t a t e m e n t $B_{40}$: the value of each such input $x \in \{6_1, 6_2, ..., 6_K\}$, which is connected to any $C^{\text{mpl}}$-intact input of any of gates $4_1 ... 4_N$, conforms to Table III-1.

Statements $B_{22}$, $B_{32}$ and $B_{35}$ directly entail

> S t a t e m e n t $B_{41}$: each such $C^{\text{mpl}}$-intact input $x \in \{6_1, 6_2, ..., 6_K\}$, which is connected to any $C^{\text{mpl}}$-intact input of any of gates $5_1 ... 5_N$, has value «1» at some moment of interval $[t_2, t_3]$.

Statement $B_{41}$ and item 3° directly entail

> S t a t e m e n t $B_{42}$: the value of each such input $x \in \{6_1, 6_2, ..., 6_K\}$, which is connected to any $C^{\text{mpl}}$-intact input of any of gates $5_1 ... 5_N$, conforms to Table III-1.

## XI.

Let´s prove

> S t a t e m e n t $B_{43}$: there is no such $\nu \in \overline{1, N}$ that the input 1 of gate $4_\nu$ keeps value «1» during whole interval $[t_1, t_2]$.

Let´s assume the opposite. We have 4 remarks:

- 1) gate $4_\nu$ doesn´t have faults « $\equiv 0$» (This results from statements $B_{28}$ and $B_{29}$.);
- 2) each such input of gate $4_\nu$, which is connected to any input $x \in \{6_1, 6_2, ..., 6_K\}$, keeps value «1» during whole interval $[t_1, t_6]$
  (I n d e e d , if this input of gate $4_\nu$ has a fault, then in accordance with statement $B_{29}$ this fault is « $\equiv 1$»; if this input of gate $4_\nu$ is $C^{\text{mpl}}$-intact, then there are 2 alternatives: or input $x$ has a fault (then the input $x$ has fault « $\equiv 1$» in accordance with statement $B_{30}$), or the input $x$ is $C^{\text{mpl}}$-intact (then the input $x$ keeps value «1» during whole interval $[t_1, t_6]$ in accordance with statement $B_{40}$).);
- 3) the outputs of gates $1_1 ... 1_{N+1}$ have values «1» at moment $t_1$ (This results from statement $B_{21}$.);
- 4) the outputs of gates $2_1 ... 2_N$ have values «1» at moment $t_1$ (This results from remark 3 and from the fact that gates $2_1 ... 2_N$ are $C^{\text{mpl}}$-intact in accordance with statement $B_{38}$.).

We have only 2 alternatives:

- a) each such input of gate $4_\nu$, which is connected to the output of any of gates $2_{\nu+1} ... 2_N$, has fault « $\equiv 1$»,

- b) gate $4_\nu$ has such $C^{mpl}$-intact inputs which are connected to the outputs of gates belonging to some subset $Q_\nu$ of gates $2_{\nu+1} \ldots 2_N$, whereas all those inputs of gate $4_\nu$, which are connected to the outputs of gates belonging to the set $\overline{Q_\nu} = \{2_{\nu+1},\ 2_{\nu+2},\ \ldots,\ 2_N\} \backslash Q_\nu$, have fault « $\equiv 1$».

In case of alternative «*a*» the remarks 1 and 2 entail, that the output of gate $2_\nu$ keeps a constant value during whole interval $[t_1,\ t_2]$, but this contradicts statement $B_{21}$. Therefore alternative «*a*» is impossible.

In case of alternative «*b*» let´s denote by $L$ the loop which goes through each pole of each of loops $\Psi_{\nu+1} \ldots \Psi_N$ (see A2), through the output of gate $2_\nu$, through each pole of each path $\mathrm{O} : 2_x \xrightarrow{4} 2 : 2_\nu$ (where $2_x \in Q_\nu$), and through no other poles. We see that in accordance with Lemma 1 conditions of true of statement $A_1(L,\ [t_1,\ t_2])$ are met:

- condition «**i**» — in accordance with statement $B_{38}$,
- condition «**ii**» — because at moment $t_1$ the outputs of gates $1_{\nu+1} \ldots 1_{N+1}$ have values «1» (according to remark 3) and the output of gate $4_\nu$ has value «1» (in accordance with the definition of $\nu$, with remarks 1, 2, 4, and with the definition of alternative «*b*»),
- condition «**iii**» — because during whole interval $[t_1,\ t_2]$ the input 2 of gate $1_{N+1}$ keeps a constant value (owing to item 6°), the input 1 of gate $4_\nu$ keeps a constant value (according to definition of $\nu$), those inputs of gate $4_\nu$ which are connected to any of inputs $6_1 \ldots 6_K$ keep constant values (according to remark 2), and those inputs of gate $4_\nu$ which are connected to the outputs of gates belonging to the set $\overline{Q_\nu}$ keep constant values (according to the definition of alternative «*b*»).

In accordance with statement $A_1(L,\ [t_1,\ t_2])$ the output of gate $1_{N+1}$ keeps a constant value during whole interval $[t_1,\ t_2]$, but this contradicts statement $B_{21}$. I.e. the alternative «*b*» is also impossible.

## XII.

Let´s prove

> S t a t e m e n t $B_{44}$: there is no such $\chi \in \overline{1,\ N}$ that the input 1 of gate $5_\chi$ keeps value «1» during whole interval $[t_4,\ t_5]$.

Let´s assume the opposite. We have 4 remarks:

- 1) gate $5_\chi$ doesn´t have fault « $\equiv 0$» (This results from statements $B_{35}$ and $B_{36}$.);
- 2) each such input of gate $5_\chi$, which is connected to any input $x \in \{6_1,\ 6_2,\ \ldots,\ 6_K\}$, keeps value «1» during whole interval $[t_1,\ t_6]$
(I n d e e d , if this input of gate $5_\chi$ has a fault, then in accordance with statement $B_{36}$ this fault is « $\equiv 1$»; if this input of gate $5_\chi$ is $C^{mpl}$-intact, then there are 2 alternatives: or the input $x$ has a fault (then the input $x$ has fault « $\equiv 1$» in accordance with statement $B_{37}$), or the input $x$ is $C^{mpl}$-intact (then during whole interval $[t_1,\ t_6]$ the input $x$ keeps value «1» in accordance with statement $B_{42}$).);
- 3) the outputs of gates $1_1 \ldots 1_{N+1}$ have value «1» at moment $t_4$ (This results from statement $B_{11}$.);
- 4) the outputs of gates $3_1 \ldots 3_N$ have value «1» at moment $t_4$ (This results from remark 3 and from the fact that gates $3_1 \ldots 3_N$ are $C^{mpl}$-intact in accordance with statement $B_{38}$.).

We have only 2 alternatives:

- a) each such input of gate $5_\chi$, which is connected to the output of any of gates $3_1 \ldots 3_{\chi-1}$, has fault « $\equiv 1$»,
- b) gate $5_\chi$ has those $C^{mpl}$-intact inputs, which are connected to the outputs of gates belonging to some subset $Q_\chi$ of gates $3_1 \ldots 3_{\chi-1}$, whereas all those inputs of gate $5_\chi$, which are connected to the outputs of gates belonging to the set $\overline{Q_\chi} = \{3_1, 3_2, ..., 3_{\chi-1}\} \setminus Q_\chi$, have fault « $\equiv 1$».

In case of alternative «*a*» the remarks 1 and 2 entail, that the output of gate $3_\chi$ keeps a constant value during whole interval $[t_4, t_5]$, but this contradicts statement $B_{11}$. Therefore the alternative «*a*» is impossible.

In case of alternative «*b*» let´s denote by $L$ the loop which goes through each pole of each of loops $\Psi_1 ... \Psi_{\chi-1}$ (see A2), through the output of gate $3_\chi$, through each pole of each path $O : 3_\nu \xrightarrow{5} 2 : 3_\chi$ (where $3_\nu \in Q_\chi$), and through no other poles.

We see that in accordance with Lemma 1 conditions of true of statement $A_1(L, [t_4, t_5])$ are met:

- condition «**i**» — in accordance with statement $B_{38}$,
- condition «**ii**» — because at moment $t_4$ the outputs of gates $1_1 ... 1_\chi$ have value «1» (according to remark 3) and the output of gate $5_\chi$ has value «1» (in accordance with the definition of $\chi$, with remarks 1, 2, 4, and with the definition of alternative «*b*»),
- condition «**iii**» — because during whole interval $[t_4, t_5]$ the input 1 of gate $1_1$ keeps a constant value (owing to item 5°), the input 1 of gate $5_\chi$ keeps a constant value (according to the definition of $\chi$), those inputs of gate $5_\chi$, which are connected to any of inputs $6_1 ... 6_K$, keep constant values (according to remark 2), and those inputs of gate $5_\chi$, which are connected to the outputs of gates belonging to the set $\overline{Q_\chi}$, keep constant values (according to the definition of alternative «*b*»).

In accordance with statement $A_1(L, [t_4, t_5])$ the output of gate $1_1$ keeps a constant value during whole interval $[t_4, t_5]$, but this contradicts statement $B_{11}$. Therefore alternative «*b*» is also impossible.

## XIII.

Statement $B_{43}$ directly entails

> S t a t e m e n t $B_{45}$: inputs 1 of gates $4_1 ... 4_N$ don´t have faults « $\equiv 1$».

Statements $B_{29}$ and $B_{45}$ directly entail

> S t a t e m e n t $B_{46}$: inputs 1 of gates $4_1 ... 4_N$ are $C^{mpl}$-intact.

Statements $B_{43}$ and $B_{46}$ directly entail

> S t a t e m e n t $B_{47}$: the input *11* doesn´t have fault « $\equiv 1$».

Statements $B_{30}$ and $B_{46}$ directly entail

> S t a t e m e n t $B_{48}$: the input *11* doesn´t have fault « $\equiv 0$».

Statements $B_{47}$ and $B_{48}$ directly entail

> S t a t e m e n t $B_{49}$: the input *11* is $C^{mpl}$-intact.

Statements $B_{43}$, $B_{46}$ and $B_{49}$ directly entail

> S t a t e m e n t $B_{50}$: the input *11* has value «1» at some moment of interval $[t_1, t_2]$.

Statement $B_{50}$ and item 4° directly entail

> S t a t e m e n t $B_{51}$: the value of the input *11* conforms to Table III-1 during whole interval $[t_1, t_3]$.

Statements $B_{23}$, $B_{25}$, $B_{28}$, $B_{46}$ and $B_{49}$ directly entail

S t a t e m e n t $B_{52}$: the input *11* has value «1» at some moment of interval $[t_5, t_6]$.

Statement $B_{52}$ and item 4° directly entail

S t a t e m e n t $B_{53}$: the value of the input *11* conforms to Table III-1 during whole interval $[t_4, t_6]$.

Statements $B_{51}$ and $B_{53}$ directly entail

S t a t e m e n t $B_{54}$: the value of the input *11* conforms to Table III-1.

Statement $B_{44}$ directly entails

S t a t e m e n t $B_{55}$: inputs 1 of gates $5_1 \ldots 5_N$ don´t have faults « $\equiv 1$».

Statements $B_{36}$ and $B_{55}$ directly entail

S t a t e m e n t $B_{56}$: inputs 1 of gates $5_1 \ldots 5_N$ are $C^{\mathrm{mpl}}$-intact.

Statements $B_{44}$ and $B_{56}$ directly entail

S t a t e m e n t $B_{57}$: the input *12* doesn´t have fault « $\equiv 1$».

Statements $B_{37}$ and $B_{56}$ directly entail

S t a t e m e n t $B_{58}$: the input *12* doesn´t have fault « $\equiv 0$».

Statements $B_{57}$ and $B_{58}$ directly entail

S t a t e m e n t $B_{59}$: the input *12* is $C^{\mathrm{mpl}}$-intact.

Statement $B_{44}$, $B_{56}$ and $B_{59}$ directly entail

S t a t e m e n t $B_{60}$: the input *12* has value «0» at some moment of interval $[t_4, t_5]$.

Statement $B_{60}$ and item 4° directly entail

S t a t e m e n t $B_{61}$: the value of the input *12* conforms to Table III-1 during whole interval $[t_4, t_6]$.

Statement $B_{22}$, $B_{32}$, $B_{35}$, $B_{56}$ and $B_{56}$ directly entail

S t a t e m e n t $B_{62}$: the input *12* keeps value «1» during whole interval $[t_2, t_3]$.

Statement $B_{62}$ and item 4° directly entail

S t a t e m e n t $B_{63}$: the value of the input *12* conforms to Table III-1 during whole interval $[t_1, t_3]$.

Statements $B_{61}$ and $B_{63}$ directly entail

S t a t e m e n t $B_{64}$: the value of the input *12* conforms to Table III-1.

Statement $B_{21}$ and item 5° directly entail

S t a t e m e n t $B_{65}$: the value of the input *9* conforms to Table III-1 during whole interval $[t_1, t_2]$.

Statement $B_2$ and item 5° directly entail

S t a t e m e n t $B_{66}$: the value of the input *9* conforms to Table III-1 during whole interval $[t_2, t_3]$.

Statements $B_4$ and $B_{20}$ and item 5° directly entail

> S t a t e m e n t $B_{67}$: the value of the input *9* conforms to Table III-1 during whole interval $[t_4, t_6]$.

Statements $B_{65} \ldots B_{67}$ directly entail

> S t a t e m e n t $B_{68}$: the value of the input *9* conforms to Table III-1.

Statement $B_{11}$ and item 6° directly entail

> S t a t e m e n t $B_{69}$: the value of the input *10* conforms to Table III-1 during whole interval $[t_4, t_5]$.

Statement $B_{20}$ and item 6° directly entail

> S t a t e m e n t $B_{70}$: the value of the input *10* conforms to Table III-1 during whole interval $[t_5, t_6]$.

Statements $B_2$ and $B_{13}$ as well as item 6° directly entail

> S t a t e m e n t $B_{71}$: the value of the input *10* conforms to Table III-1 during whole interval $[t_1, t_3]$.

Statements $B_{69} \ldots B_{71}$ directly entail

> S t a t e m e n t $B_{72}$: the value of the input *10* conforms to Table III-1.

Statement $B_2$ and item 3° directly entail

> S t a t e m e n t $B_{73}$: the value of each such $C^{\text{mpl}}$-intact input $x \in \{б_1, б_2, \ldots, б_K\}$, which is connected to any $C^{\text{mpl}}$-intact input of any of gates $1_1 \ldots 1_{N+1}$, conforms to Table III-1.

## XIV.

Statements $B_{38}$, $B_{46}$, $B_{49}$, $B_{56}$, and $B_{59}$ directly entail item 1°°.

Statements $B_{54}$, $B_{64}$, $B_{68}$, and $B_{72}$ directly entail item 2°°.

Statements $B_{40}$, $B_{42}$, and $B_{73}$ directly entail item 3°°.

Theorem 2 is proved.

# A4. Proof of Theorem 3

## I.

Let´s show, that during interval $[t_1,\ t_3]$

- a) the value «0» goes by path $1\ :\ 1_1 \xrightarrow{1,\,2} \mathrm{O}\ :\ 1_{N+1}$ at moment $t_2$,
- b) the value «1» goes by path $1\ :\ 1_1 \xrightarrow{1,\,2} \mathrm{O}\ :\ 1_{N+1}$ at moments $t_1$ and $t_3$.

Item «*a*» is obvious if taking into account 2 facts:

- — in accordance with item 1°°, path $1\ :\ 1_1 \xrightarrow{1,\,2} \mathrm{O}\ :\ 1_{N+1}$ is $C^{\mathrm{mpl}}$-intact,
- — in accordance with items 1°° and 2°°, the value «0», going from input *11* through gates $4_1\ ...4_N$, remains on inputs 2 of gates $2_1\ ...2_N$ during whole interval $[t_1,\ t_3]$.

Item «*b*» can be proved by induction for $j = \overline{1,\ N}$. We have:

- $\alpha_0$) in accordance with items 1°° and 2°°, the value «1» comes onto input 1 of gate $1_1$ from input *9* at moments $t_1$ and $t_3$;
- $\alpha_1$) the value «1» comes onto input 2 of gate $1_1$ through gate $3_1$ from the output of gate $5_1$ during whole interval $[t_1,\ t_3]$
(Let´s p r o v e item «$\alpha_1$». In accordance with item 1°°, gate $3_1$, as well as input 1 and the output of gate $5_1$, are $C^{\mathrm{mpl}}$-intact. In accordance with items 1°° and 2°°, the value «1», coming from input *12*, remains on input 1 of gate $5_1$ during whole interval $[t_1,\ t_3]$. In accordance with item 1°°, any input $x \in \overline{2,\ \infty}$ of gate $5_1$ either has fault « $\equiv 1$», or is $C^{\mathrm{mpl}}$-intact. In latter case the value «1» comes onto input $x$ of gate $5_1$ at moments $t_1$ and $t_3$, because this input is connected to some such input $y \in \{6_1,\ 6_2,\ ...,\ 6_K\}$, which in accordance with items 1°° and 3°° either has fault « $\equiv 1$», or, being $C^{\mathrm{mpl}}$-intact, keeps value «1» during whole interval $[t_1,\ t_3]$.);
- $\alpha_2$) according to item 1°°, any input $x \in \overline{3,\ \infty}$ of gate $1_1$ either has fault « $\equiv 1$», or is $C^{\mathrm{mpl}}$-intact. In latter case the value «1» comes onto input $x$ of gate $1_1$ at moments $t_1$ and $t_3$, because this input is connected to some such input $y \in \{6_1,\ 6_2,\ ...,\ 6_K\}$, which in accordance with items 1°° and 3°° either has fault « $\equiv 1$», or, being $C^{\mathrm{mpl}}$-intact, keeps value «1» during whole interval $[t_1,\ t_3]$;
- $\alpha_3$) in accordance with item 1°°, inputs 1 and 2 and the output of gate $1_1$ are $C^{\mathrm{mpl}}$-intact.

Items «$\alpha_0$»—«$\alpha_3$» entail *the base of induction*: the value «1» goes by path $1\ :\ 1_1 \xrightarrow{1,\,2} \mathrm{O}\ :\ 1_j$ at moments $t_1$ and $t_3$ for case $j = 1$.

*Inductive assumption:* the value «1» goes by path $1\ :\ 1_1 \xrightarrow{1,\,2} \mathrm{O}\ :\ 1_j$ at moments $t_1$ and $t_3$. We have:

- $\beta_0$ ) in order that inductive assumption were true, outputs of gates $3_1\ ...3_j$ must have value «1» at moments $t_1$ and $t_3$, because in accordance with item 1°° the inputs 2 of gates $1_1\ ...1_j$ are $C^{\mathrm{mpl}}$-intact,
- $\beta_1$) the value «1» comes onto input 1 of gate $1_{j+1}$ through input 1 and the output of gate $2_j$ at moments $t_1$ and $t_3$

(Let´s p r o v e item «$\beta_1$»: in accordance with item 1°° gate $2_j$ is $C^{mpl}$-intact, besides in accordance with inductive assumption the value «1» comes onto input 1 of gate $2_j$ at moments $t_1$ and $t_3$.);

- $\beta_2$) when $j = N$, in accordance with items 1°° and 2°° the value «1», coming from input *10*, remains on input 2 of gate $1_{j+1}$ during whole interval $[t_1, t_3]$;
- $\beta_3$) when $j \in \overline{1, N-1}$, the value «1» comes onto input 2 of gate $1_{j+1}$ from output of gate $5_{j+1}$ through gate $3_{j+1}$ at moments $t_1$ and $t_3$
  (Let´s p r o v e item «$\beta_3$». In accordance with item 1°°, gate $3_{j+1}$, as well as input 1 and the output of gate $5_{j+1}$, are $C^{mpl}$-intact. In accordance with items 1°° and 2°°, the value «1» comes onto input 1 of gate $5_{j+1}$ from input *12* during whole interval $[t_1, t_3]$. In accordance with item 1°°, any input $x \in \overline{2, \infty}$ of gate $5_{j+1}$ either has fault «$\equiv 1$», or is $C^{mpl}$-intact. In latter case the value «1» comes onto input $x$ of gate $5_{j+1}$ at moments $t_1$ and $t_3$, because this input is connected to one of following two poles:
  - — some input $y \in \{6_1, 6_2, \ldots, 6_K\}$, which in accordance with items 1°° and 3°° either has fault «$\equiv 1$», or, being $C^{mpl}$-intact, keeps value «1» during whole interval $[t_1, t_3]$,
  - — the output of one of gates $3_1 \ldots 3_j$, which in accordance with item «$\beta_0$» has value «1» at moments $t_1$ and $t_3$.);
- $\beta_4$) in accordance with item 1°°, any input $x \in \overline{3, \infty}$ of gate $1_{j+1}$ either has fault «$\equiv 1$», or is $C^{mpl}$-intact. In latter case the value «1» comes onto input $x$ of gate $1_{j+1}$ at moments $t_1$ and $t_3$, because this input is connected to some input $y \in \{6_1, 6_2, \ldots, 6_K\}$, which in accordance with items 1°° and 3°° either has fault «$\equiv 1$», or, being $C^{mpl}$-intact, keeps value «1» during whole interval $[t_1, t_3]$;
- $\beta_5$) in accordance with item 1°°, inputs 1 and 2 and the output of gate $1_{j+1}$ are $C^{mpl}$-intact.

Thus at moments $t_1$ and $t_3$ the value «1» goes

- — in accordance with inductive assumption — by path $1 : 1_1 \xrightarrow{1,\,2} \mathrm{O} : 1_j$,
- — in accordance with item «$\beta_1$» — through input 1 and the output of gate $2_j$,
- — in accordance with items «$\beta_1$»—«$\beta_5$» — through input 1 and the output of gate $1_{j+1}$.

I.e. at moments $t_1$ and $t_3$ the value «1» goes by path $1 : 1_1 \xrightarrow{1,\,2} \mathrm{O} : 1_{j+1}$. *The induction is confirmed.*

## II.

Let´s show, that during interval $[t_4, t_6]$

- a) the value «0» goes by path $2 : 1_{N+1} \xrightarrow{1,\,3} \mathrm{O} : 1_1$ at moment $t_5$,
- b) the value «1» goes by path $2 : 1_{N+1} \xrightarrow{1,\,3} \mathrm{O} : 1_1$ at moments $t_4$ and $t_6$ during whole interval $[t_4, t_6]$.

Item «*a*» is obvious if taking into account 2 facts:

- — in accordance with item 1°°, path $2 : 1_{N+1} \xrightarrow{1,\,3} \mathrm{O} : 1_1$ is $C^{mpl}$-intact,
- — in accordance with items 1°° and 2°°, the value «0», coming from input *12* through gates $5_1 \ldots 5_N$, remains on inputs 2 of gates $3_1 \ldots 3_N$.

Item «$b$» can be proved by induction for $j = N+1,\ N,\ \ldots,\ 2$. We have:

- $\gamma_0$) in accordance with items 1°° and 2°°, the value «1» comes onto input 2 of gate $1_{N+1}$ through input *10* at moments $t_4$ and $t_6$;
- $\gamma_1$) the value «1» comes onto input 1 of gate $1_{N+1}$ through gate $2_N$ from the output of gate $4_N$ during whole interval $[t_4,\ t_6]$
(Let´s prove item «$\gamma_1$». In accordance with item 1°°, gate $2_N$, as well as input 1 and the output of gate $4_N$, are $C^{\mathrm{mpl}}$-intact. In accordance with items 1°° and 2°°, the value «1», coming from input *11*, remains on input 1 of gate $4_N$ during whole interval $[t_4,\ t_6]$. In accordance with item 1°°, any input $x \in \overline{2,\ \infty}$ of gate $4_N$ either has fault «$\equiv 1$», or is $C^{\mathrm{mpl}}$-intact. In latter case the value «1» comes onto input $x$ of gate $4_N$ at moments $t_4$ and $t_6$, because this input is connected to some input $y \in \{\sigma_1,\ \sigma_2,\ \ldots,\ \sigma_K\}$, which in accordance with items 1°° and 3°° either has fault «$\equiv 1$», or, being $C^{\mathrm{mpl}}$-intact, keeps value «1» during whole interval $[t_4,\ t_6]$.);
- $\gamma_2$) according to item 1°°, any input $x \in \overline{3,\ \infty}$ of gate $1_{N+1}$ either has fault «$\equiv 1$», or is $C^{\mathrm{mpl}}$-intact. In latter case the value «1» comes onto input $x$ of gate $1_{N+1}$ at moments $t_4$ and $t_6$, because this input is connected to some such input $y \in \{\sigma_1,\ \sigma_2,\ \ldots,\ \sigma_K\}$, which in accordance with items 1°° and 3°° either has fault «$\equiv 1$», or, being $C^{\mathrm{mpl}}$-intact, keeps value «1» during whole interval $[t_4,\ t_6]$;
- $\gamma_3$) in accordance with item 1°°, the input 1 and 2 and the output of gate $1_{N+1}$ are $C^{\mathrm{mpl}}$-intact.

Items «$\gamma_0$»—«$\gamma_3$» entail *the base of induction:* the value «1» goes by path $2 : 1_{N+1} \overset{1,\,3}{\to} \mathrm{O} : 1_j$ at moments $t_4$ and $t_6$ for case $j = N+1$.

*Inductive assumption:* the value «1» goes by path $2 : 1_{N+1} \overset{1,\,3}{\to} \mathrm{O} : 1_j$ at moments $t_4$ and $t_6$. We have:

- $\delta_0$ ) in order that inductive assumption were true, outputs of gates $2_{j-1} \ldots 2_N$ must have value «1» at moments $t_4$ and $t_6$, because in accordance with item 1°° the inputs 1 of gates $1_j \ldots 1_{N+1}$ are $C^{\mathrm{mpl}}$-intact;
- $\delta_1$) at moments $t_4$ and $t_6$ the value «1» comes onto input 2 of gate $1_{j-1}$ through input 1 and the output of gate $3_{j-1}$
(Let´s prove item «$\delta_1$»: in accordance with item 1°° gate $3_{j-1}$ is $C^{\mathrm{mpl}}$-intact, besides in accordance with inductive assumption the value «1» comes onto input 1 of gate $3_{j-1}$ at moments $t_4$ and $t_6$.);
- $\delta_2$) when $j = 2$, in accordance with items 1°° and 2°° the value «1», coming from input *9*, remains on input 1 of gate $1_{j-1}$ during whole interval $[t_4,\ t_6]$;
- $\delta_3$) when $j > 2$, at moments $t_4$ and $t_6$ the value «1» comes onto input 1 of gate $1_{j-1}$ from output of gate $4_{j-2}$ through gate $2_{j-2}$
(Let´s prove item «$\delta_3$». In accordance with item 1°°, gate $2_{j-2}$, as well as input 1 and the output of gate $4_{j-2}$, are $C^{\mathrm{mpl}}$-intact. In accordance with items 1°° and 2°°, the value «1», coming from input *11*, remains on the input 1 of gate $4_{j-2}$ during whole interval $[t_4,\ t_6]$. In accordance with item 1°°, any input $x \in \overline{2,\ \infty}$ of gate $4_{j-2}$ either has fault «$\equiv 1$», or is $C^{\mathrm{mpl}}$-intact. In latter case the value «1» comes onto input $x$ of gate $4_{j-2}$ at moments $t_4$ and $t_6$, because this input is connected to one of following two poles:
  - — some input $y \in \{\sigma_1,\ \sigma_2,\ \ldots,\ \sigma_K\}$, which in accordance with items 1°° and 3°° either has fault «$\equiv 1$», or, being $C^{\mathrm{mpl}}$-intact, keeps value «1» during whole interval $[t_4,\ t_6]$,
  - — the output of one of gates $2_{j-1} \ldots 2_N$, which in accordance with item «$\delta_0$» has value «1» at moments $t_4$ and $t_6$.);

- $\delta_4$) in accordance with item 1°°, any input $x \in \overline{3, \infty}$ of gate $1_{j-1}$ eithwer has fault « $\equiv 1$», or is $C^{\text{mpl}}$-intact. In latter case the value «1» comes onto input $x$ of gate $1_{j-1}$ at moments $t_4$ and $t_6$, because this input is connected to some input $y \in \{6_1, 6_2, ..., 6_K\}$, which in accordance with items 1°° and 3°° either has fault « $\equiv 1$», or, being $C^{\text{mpl}}$-intact, keeps value «1» during whole interval $[t_4, t_6]$;
- $\delta_5$) in accordance with item 1°°, inputs 1 and 2 and the output of gate $1_{j-1}$ are $C^{\text{mpl}}$-intact.

Thus at moments $t_4$ and $t_6$ the value «1» goes

- — in accordance with inductive assumption — by path $2 : 1_{N+1} \overset{1,\,3}{\to} \text{O} : 1_j$,
- — in accordance with item «$\delta_1$» — through input 1 and the output of gate $3_{j-1}$,
- — in accordance with items «$\delta_1$»—«$\delta_5$» — through input 2 and the output of gate $1_{j-1}$.

I.e. at moments $t_4$ and $t_6$ the value «1» goes by path $2 : 1_{N+1} \overset{1,\,3}{\to} \text{O} : 1_{j-1}$. *The induction is confirmed.*

## III.

Considering items I«$a$» and I«$b$» and taking into account the fact that in accordance with item 1°° the output *14* is $C^{\text{mpl}}$-intact, we see that the value «0» at moment $t_2$ and the value «1» at moments $t_1$ and $t_3$ go through output *14*.

Besides, considering items II«$a$» and II«$b$» and taking into account the fact that in accordance to item 1°° the output *13* is $C^{\text{mpl}}$-intact, we see that the value «0» at moment $t_5$ and the value «1» at moments $t_4$ and $t_6$ go through output *13*.

Theorem 3 is proved.